\documentclass[12pt,tightenlines,showpacs,preprintnumbers,%
superscriptaddress,nofootinbib]{revtex4-1}
\newcommand{\PRE}[1]{{#1}} 
\usepackage{bm}
\usepackage{subfigure}
\usepackage{slashed}
\usepackage{epsfig}
\usepackage{hyperref}
\usepackage{color}
\usepackage{amsmath}
\usepackage{bbold}

\newcommand{\OmegaDM}{\Omega_{\text{DM}}}

\newcommand{\gev}{\text{GeV}}
\newcommand{\tev}{\text{TeV}}

\renewcommand{\eqref}[1]{Eq.~(\ref{#1})}

\newcommand{\secref}[1]{Sec.~\ref{sec:#1}}

\newcommand{\figref}[1]{Fig.~\ref{fig:#1}}
\newcommand{\figsref}[2]{Figs.~\ref{fig:#1} and \ref{fig:#2}}

\newcommand{\gsim}{\lower.7ex\hbox{$\;\stackrel{\textstyle>}{\sim}\;$}}
\newcommand{\lsim}{\lower.7ex\hbox{$\;\stackrel{\textstyle<}{\sim}\;$}}

\newcommand{\bea}{\begin{eqnarray}}
\newcommand{\eea}{\end{eqnarray}}

\newcommand{\Sec}[1]{Sec.~\ref{#1}}

\newcommand{\mchi}{M_{\chi}}

\newcommand{\Omegachi}{\Omega_{\chi}}

\newcommand{\sigmaSI}{\sigma^{\text{SI}}}
\newcommand{\zb}{\text{zb}}
\newcommand{\chxx}{c_{h \chi\chi}}
\newcommand{\ahxx}{a_{h \chi\chi}}
\newcommand{\xiwt}{\xi_{\text{WT}}}
\newcommand{\xibs}{\xi_{\text{BS}}}
\DeclareMathOperator{\tr}{tr}
\DeclareMathOperator{\eqomega}{\Omegachi = \OmegaDM}

\begin{document}

\preprint{CALT 68-2870}

\title{ \PRE{\vspace*{1.5in}}
\Large Simplified Models of Mixed Dark Matter
\PRE{\vspace*{0.3in}} }

\author{Clifford Cheung} 
\email{clifford.cheung@caltech.edu}
\affiliation{California Institute of Technology, Pasadena, CA 91125,
  USA \PRE{\vspace*{.5in}}
}

\author{David Sanford\PRE{\vspace*{.3in}}}
\email{dsanford@caltech.edu}
\affiliation{California Institute of Technology, Pasadena, CA 91125,
  USA \PRE{\vspace*{.5in}}
}

\begin{abstract}
\PRE{\vspace*{.3in}} We explore simplified models of mixed dark matter
(DM), defined here to be a stable relic composed of a singlet and an
electroweak charged state.  Our setup describes a broad spectrum of
thermal DM candidates that can naturally accommodate the observed DM
abundance but are subject to substantial constraints from current and
upcoming direct detection experiments.  We identify ``blind spots'' at
which the DM-Higgs coupling is identically zero, thus nullifying
direct detection constraints on spin independent scattering.
Furthermore, we characterize the fine-tuning in mixing angles,
{\it i.e.}~well-tempering, required for thermal freeze-out to
accommodate the observed abundance.  Present and projected limits from
LUX and XENON1T force many thermal relic models into blind spot
tuning, well-tempering, or both.  This simplified model framework
generalizes bino-Higgsino DM in the MSSM, singlino-Higgsino DM in the
NMSSM, and scalar DM candidates that appear in models of extended
Higgs sectors.
\end{abstract}


\maketitle

\newpage

\tableofcontents

\newpage

\section{Introduction}

The gravitational evidence for dark matter (DM) is very strong, but
its precise particle properties remain elusive.  Long ago, laboratory
experiments excluded the simplest models of weakly interacting massive
particle (WIMP) DM which predicted DM-nucleon scattering via spin
independent $Z$ boson exchange.  Today, direct detection experiments,
particularly XENON100~\cite{Aprile:2011hi,Aprile:2012nq} and
LUX~\cite{Akerib:2013tjd}, have become sensitive to the large class of
theories that predict spin independent (SI) DM-nucleon scattering
mediated by the Higgs boson.  Future experiments such as
XENON1T~\cite{Aprile:2012zx} and LZ~\cite{Malling:2011va} will have
improved sensitivities and their results will have even stronger
implications. In light of the discovery of the Higgs boson at the
LHC~\cite{:2012gk,:2012gu}, present and future limits on
Higgs-mediated scattering can be recast in terms of the effective
parameter space defined by the DM mass, $\mchi$, and its coupling to
the Higgs boson, $c_{h\chi\chi}$.

What are the natural values for $\mchi$ and $c_{h\chi\chi}$?  In the
absence of additional theory input, these parameters are arbitrary --
there is simply no reason why DM should be accessible through direct
detection.  For example, DM could be completely inert and thus
impervious to non-gravitational probes.  While many theories offer a
DM candidate as part of a new physics framework, the only general
impetus for couplings between DM and the Standard Model (SM) is
cosmological in nature: if DM is a thermal relic, then it is
reasonable for it be thermalized with the SM in the early universe.
In minimal extensions of the SM, DM couples via electroweak gauge
interactions and/or via the Higgs portal.  Going beyond this setup
requires more elaborate models that entail richer structures like dark
force carriers \cite{ArkaniHamed:2008qn, Baumgart:2009tn,
  Cheung:2009qd, Chen:2009ab, Katz:2009qq, Andreas:2011in,
  Feldman:2007wj, Cohen:2010kn} or other mediators
\cite{Finkbeiner:2008qu, Nomura:2008ru}.  Thus, an important question
for present and upcoming experiments is the status of thermal relic
DM, broadly defined.

The literature provides a litany of well-motivated theories of DM,
both within and outside of broader new physics frameworks, though by
far the most popular is neutralino DM in supersymmetry (SUSY).  While
supersymmetric theories are a useful benchmark for models, analyses of
SUSY DM are often colored by theory biases and disparate connections
to unrelated experimental data.  For instance, under specific model
assumptions, issues of naturalness are still taken as a hard
constraint on the parameter space of SUSY DM models.  Another example
is the discrepancy in $g-2$ of the muon~\cite{Bennett:2006fi,
  Davier:2010nc, Jegerlehner:2011ti}, which prefers certain signs of
the $\mu$ parameter, influencing the perceived viability of SUSY DM.
However, given the current sensitivity of experiments, overarching
theory assumptions like SUSY are not required to pare down the
parameter space -- experiments will do so.

Instead it can be fruitful to take the approach of simplified models:
effective theories that describe a broad class of theories but are
tailored to extract maximal information from experimental results.
Simplified models of DM have appeared in the literature in a number of
guises.  In the case of minimal DM~\cite{Cirelli:2005uq,
  Cirelli:2007xd}, pure gauge representations were considered.  Others
have studied simplified models of a singlet and colored particle, {\it
  a.k.a.} the effective ``bino-squark" system~\cite{DiFranzo:2013vra,
  Chang:2013oia, Bai:2013iqa}.  For thermal relics, it was found that
many of these models are bowing under the weight of present
experimental constraints from direct detection and the LHC.  Recently,
there has also been growing interest in effective operator
descriptions of DM~\cite{Goodman:2010yf, Fox:2011pm,
  Fox:2011fx,Essig:2013vha, Buckley:2011kk, Chae:2012bq}.  Modulo the
well-known limits of their validity~\cite{Fox:2011pm, Essig:2013vha,
  Shoemaker:2011vi, Busoni:2013lha, Buchmueller:2013dya}, these
effective theories have been used to determine quite general bounds on
DM from colliders.

In the present work, we consider simplified models of mixed DM,
defined here as renormalizable theories of fermion or scalar DM
comprised of a singlet and an additional electroweak charged state.
Generically, the singlet and charged states will mix after electroweak
symmetry breaking.  As a consequence, the DM possesses annihilation
channels inherited from its electroweak charged component, and thermal
relic DM can be achieved with an appropriate degree of mixing.  In a
sense, this simplified model is a generalization of the
``well-tempered'' neutralino~\cite{ArkaniHamed:2006mb, Baer:2006te,
  Feng:2010ef} found most commonly in focus point SUSY
scenarios~\cite{Feng:2000gh, Baer:2005ky, Feng:2011aa, Draper:2013cka}
to a more diverse set of DM charges and spins.  By enumeration, there
exist three renormalizable, gauge invariant simplified models of mixed
dark matter: fermion singlet-doublet ({\it Model A}), scalar
singlet-doublet ({\it Model B}), and scalar singlet-triplet ({\it
  Model C}).  More complicated models necessarily include additional
degrees of freedom or higher dimension operators to induce mixing.  We
evaluate the viability of models based upon current limits at
LUX~\cite{Akerib:2013tjd} and the expected reach of
XENON1T~\cite{Aprile:2012zx}.  Our main conclusions are:

\begin{itemize}

\item In light of current LUX limits and the projected reach of
  XENON1T, we have determined the viable parameter space of thermal
  relic DM in \figref{Plot2c} (singlet-doublet fermion),
  \figref{Plot6} (singlet-doublet scalar), and \figref{PlotViableST}
  (singlet-triplet scalar).  We have cast our results in terms of the
  parameter space of physical quantities: the DM mass, $\mchi$, and
  the DM-Higgs coupling, $c_{h \chi\chi}$.

\item {\it Model A: Singlet-Doublet Fermion}. LUX stringently
  constrains this model except in regions with relative signs among
  the DM-Higgs Yukawa couplings.  Given the overall Yukawa coupling
  strength $y$ defined in \secref{sdfermionmodel}, XENON1T will
  eliminate all of the viable parameter space with $y\gtrsim 0.1$
  except near blind spots with tuning more severe than at least
  $\lesssim 10$\%.  XENON1T will allow for regions with $y\lesssim
  0.1$, but they require $\lesssim 10$\% tuning in order to
  accommodate the observed relic density.

\item {\it Model B: Singlet-Doublet Scalar}. LUX places modest limits
  on this model but leaves much of the parameter space still open.
  Given the overall DM-Higgs quartic coupling strength $\lambda$
  defined in \secref{sdscalar}, XENON1T will eliminate essentially all
  of the parameter space for $\lambda \leq 0$.  For $\lambda > 0$,
  blind spots appear but a typical tuning of $\lesssim 10$\% is still
  required to escape the projected reach of XENON1T.  Within the
  limits of our parameter scan, a similar tuning is required to obtain
  the observed relic abundance.

\item {\it Model C}: {\it Singlet-Triplet Scalar}.  LUX places
  relatively weak limits on the thermal relic parameter space because
  triplets annihilate very efficiently in the early universe.  Given
  the overall DM-Higgs quartic coupling strength $\lambda$ defined in
  \secref{stscalar}, XENON1T will strongly constrain models with
  $\lambda \leq 0$, requiring a tuning of $\lesssim 1$\% to match the
  observed DM abundance except for nearly pure triplet DM.  For
  $\lambda > 0$, the coupling is suppressed and there are regions with
  all fine-tunings alleviated to $\gtrsim 10$\%.

\end{itemize}

In \Sec{sec:model} we give a precise definition of {\it Models A,B,}
and {\it C}, and discuss general aspects of the thermal relic
abundance and the DM-nucleon scattering cross-section.  For the
latter, we derive analytic formulas indicating when the DM-Higgs
coupling identically vanishes.  In \secref{sdfermion_analysis},
\secref{sdscalar_analysis}, and \secref{stscalar_analysis}, we examine
the parameter spaces of {\it Models A,B,} and {\it C}, both in
generality and for the case of thermal relic DM which saturates the
observed abundance.  Current bounds from LUX and projected reach of
XENON1T bounds are shown throughout.  Finally, in \secref{conclusion}
we present a discussion of our results and concluding thoughts.

\section{Model Definitions}
\label{sec:model}

\label{sec:modelintro}

In this section we explicitly define our simplified models.
Throughout, we focus on the case of a singlet mixed with a
non-singlet, which is a natural generalization of many models of
theoretical interest.  Of course, mixing among non-singlet states is
also viable, but in this case the preferred mass range for DM is
typically in the multi-TeV range, with a lower bound of several
hundred GeV.

Furthermore, we restrict our discussion to models with renormalizable
interactions.  While non-renormalizable interactions are of course
allowed, they are competitive with renormalizable operators only when
the cutoff is so low that the effective theory is invalid.  Indeed,
mixing induced by higher dimension operators is highly suppressed by
the electroweak symmetry breaking scale divided by the cutoff scale.
Even in scenarios where such a theory is valid and produces the
appropriate relic density, a large degree of well-tempering is
required for even marginal mixing, disfavoring it for this study.
Restricting to renormalizable models limits us to three simplified
models of mixed DM:
\begin{itemize}
\item {\it Model A:} Majorana fermion DM composed of a Majorana
  fermion singlet and Dirac fermion doublet with hypercharge $Y =
  1/2$.
\item {\it Model B:} Real scalar DM composed of a real scalar singlet
  and complex scalar doublet with hypercharge $Y = 1/2$.
\item {\it Model C:} Real scalar DM composed of a real scalar singlet
  and real scalar triplet with hypercharge $Y = 0$.
\end{itemize}
In principle, one can consider singlets which are Dirac
fermions or complex scalars, but these theories have more degrees of
freedom and the analysis does not qualitatively change.

Throughout, we take $\Omega_\chi$ to be the relic abundance for the DM
predicted by a thermal history.  In all cases, the relic abundance and
direct detection cross-section are calculated with {\sc
  micrOMEGAs~2.4.5}~\cite{Belanger:2006is, Belanger:2008sj,
  Belanger:2010gh} using model files generated using {\sc
  FeynRules~1.6.11}~\cite{Christensen:2008py}.  We are predominantly
interested in the parameter space that saturates the DM abundance
observed by Planck~\cite{Ade:2013zuv},
\begin{eqnarray}
\Omega_{\rm DM}h^2 &\simeq & 0.1199 \pm 0.0027.
\end{eqnarray}
Obviously, the DM relic abundance will drastically vary if the
cosmological history is non-thermal, and in such cases there is no
requirement that DM couples to the SM at all.  Our initial analysis
for each model will highlight the location of the $\eqomega$ line in
parameter space, along with regions of $\Omegachi < \OmegaDM$ and
$\Omegachi > \OmegaDM$.  Our more detailed analysis will be restricted
to the $\eqomega$ region, examining the behavior of other observables
within the thermal relic context.

In all of our models, the DM particle is either a real scalar or
Majorana fermion.  Consequently, SI scattering through the $Z$ boson
is inelastic and can be ignored.  On the other hand, SI scattering
through the Higgs boson is mediated via mixing between the singlet and
and non-singlet components, though it is suppressed at direct
detection ``blind spots'': regions of parameter space at which the
coupling of DM to the Higgs boson vanishes identically.  As noted
in~\cite{Cheung:2012qy}, the existence of blind spots depend
sensitively on relative signs among the DM parameters.  In the blind
spot parameter space, the SI scattering DM-nucleon cross-section is
zero at tree-level.  Radiative corrections are typically sub-dominant
in the parameter space except very close to the blind spot
cancellation points.  However, a proper evaluation of these higher
order effects may become important for the status of DM if direct
detection experiments do not observe SI scattering.  We also neglect
radiative corrections to the masses, which are important for the
phenomenology of minimal DM~\cite{Cirelli:2005uq} but are sub-dominant
when large mixing effects are introduced.

In the spirit of low energy Higgs theorems, we can straightforwardly
compute the coupling of DM to the Higgs via a Taylor expansion of the
DM mass term with respect to the Higgs vacuum expectation value (VEV),
$v$.  For Majorana fermion DM, we obtain
\begin{eqnarray}
-{\cal L} &=& \frac{1}{2} \mchi(v + h) \chi \chi \\
&=& \frac{1}{2} \mchi(v) \chi \chi+ \frac{1}{2} \frac{\partial
  \mchi(v)}{\partial v} h \chi \chi + {\cal O}(h^2)\,,
  \label{eq:vshift}
\end{eqnarray}
where $v=246$ GeV.  \eqref{eq:vshift} implies a dimensionless DM-Higgs
boson coupling given by $c_{h \chi\chi} =\partial \mchi (v)/ \partial
v$.  For real scalar DM the same formula applies except with the
replacement $\mchi(v) \rightarrow \mchi^2(v)$.  In this case we define
a dimensionful coupling $\ahxx = \partial \left[\mchi^2 (v)\right]/
\partial v$ which is proportional to the DM mass, though for ease of
discussion we will sometimes use the effective dimensionless coupling,
$\chxx = \ahxx / \mchi$ instead.  As discussed in
Ref.~\cite{Cheung:2012qy}, the blind spot is defined by $c_{h
  \chi\chi} = 0$, computed by taking the $\partial /\partial v$
derivative of the characteristic eigenvalue equation for the DM mass.

The DM-Higgs coupling maps straightforwardly onto limits from direct
detection.  The spin independent DM-nucleon cross-section is mediated
by Higgs exchange and scales as
\begin{eqnarray}
\sigmaSI & \propto & \frac{\mu^2}{m_h^4} \times c_{h\chi\chi}^2 \,,
\label{eq:DDformula}
\end{eqnarray}
where we use $m_h = 125.6~\gev$ throughout~\cite{:2012gk,:2012gu}.
Strictly speaking, both $\mu$ and $\chxx$ vary with $\mchi$.  In our
region of interest, though, we require $\mchi \gtrsim 100~\gev$ to
avoid LEP constraints on additional charged states which accompany the
DM particle.  Thus, $\mu$ is approximately equal to the nucleon mass.
Meanwhile, we can compare the $\sigmaSI$ computed from theory with the
limits from LUX and XENON1T.  These limits have complicated mass
dependence at low mass due to reduced efficiency in observing low
energy events.  However, for $\mchi \gtrsim 100~\gev$, the
cross-section bounds $\sigmaSI_{\rm LUX}, \sigmaSI_{\rm X1T}$ rise
linearly with $\mchi$ because the event rates are proportional to the
DM density, which falls with $1/\mchi$.  Throughout, we use the lattice values for the quark content of the nucleon from \cite{Giedt:2009mr}.

We will be interested in models which evade present and projected
limits from direct detection while accommodating a thermal relic
abundance consistent with observation.  However, these theories may
require tuning for either or both of these aspects.  For direct
detection, the DM-Higgs coupling is of the schematic form
$c_{h\chi\chi} = a+b$, where $a$ and $b$ depend on different sets of
model parameters.  We can characterize the degree of tuning required
for a blind spot cancellation by
\begin{eqnarray}
\xibs &=& \frac{|a+b|}{|a|+|b|}\, .
\label{eq:BStuningdef}
\end{eqnarray}
Blind spot tuning grows more severe as $\xibs \rightarrow 0$.  This
effectively captures the tuning inherent in $\chxx \rightarrow 0$ when
individual Higgs couplings remain non-zero.  If $a$ and $b$ have the
same sign, no cancellation in $\chxx$ is possible, and $\xibs = 1$.

Meanwhile, achieving the correct thermal relic abundance may require
fine-tuning of mixing angles, or ``well-tempering'', since the DM must
be the appropriate admixture of singlet and non-singlet state.
Heuristically, well-tempering is correlated with the existence of
small mass splittings in the DM multiplet relative to the dimensionful
input parameters.  Concretely, if the mass squared matrix is an
$N\times N$ matrix ${\bf M}^2$, then the severity of well-tempering is
linked to the relative size of the traceless component of ${\bf M}^2$
(the mass splittings) relative to the trace of ${\bf M}^2$ (the
overall mass scale).  Hence, we define a parameter that describes the
well-tempering in the mixing angle,
\begin{eqnarray}
\xiwt &=& \left( \frac{N \, {\rm Tr}[ {\bf M}^4]}{{\rm Tr} [{\bf
      M}^2]^2} -1\right)^{\frac{1}{2}} \, ,
\label{eq:WTtuningdef}
\end{eqnarray}
which is related to the variance of the ${\bf M}^2 $ matrix.  Indeed,
$\xiwt$ is precisely the fractional standard deviation of the
eigenvalues of ${\bf M}^2$.  In the limit that the entire DM multiplet
is exactly degenerate, the mixing angle is very fine-tuned, ${\bf M}^2
\propto \mathbb{1}$ and $\xiwt \rightarrow 0$.

\subsection{Singlet-Doublet Fermion}
\label{sec:sdfermionmodel}

\begin{figure}[tb]
\subfigure[]{
  \includegraphics*[width=0.45\textwidth]{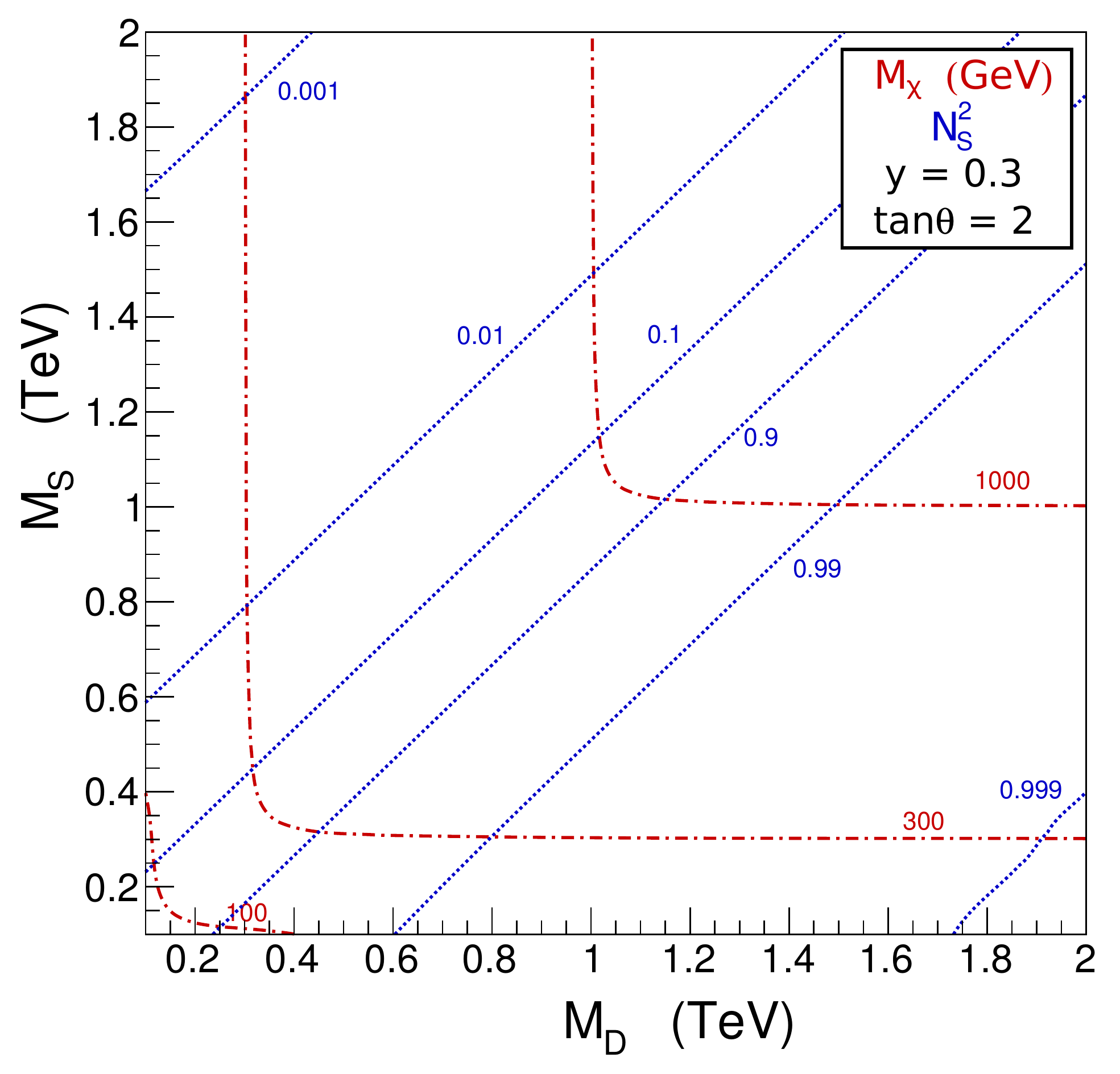}
  \label{fig:Plot0_y=0.3_tantheta=2}}
\subfigure[]{
  \includegraphics*[width=0.45\textwidth]{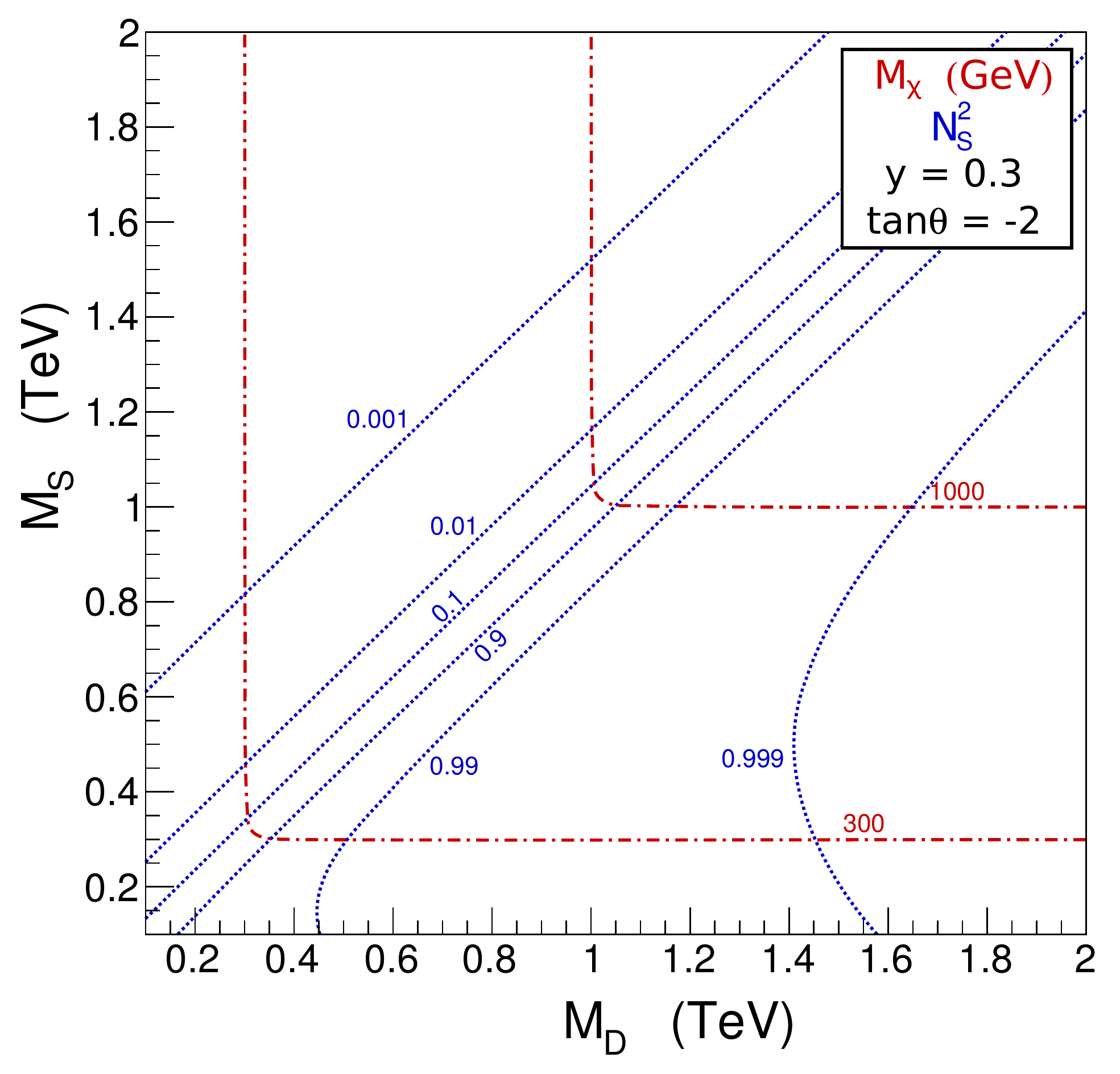}
  \label{fig:Plot0_y=0.3_tantheta=-2}}
\subfigure[]{
  \includegraphics*[width=0.45\textwidth]{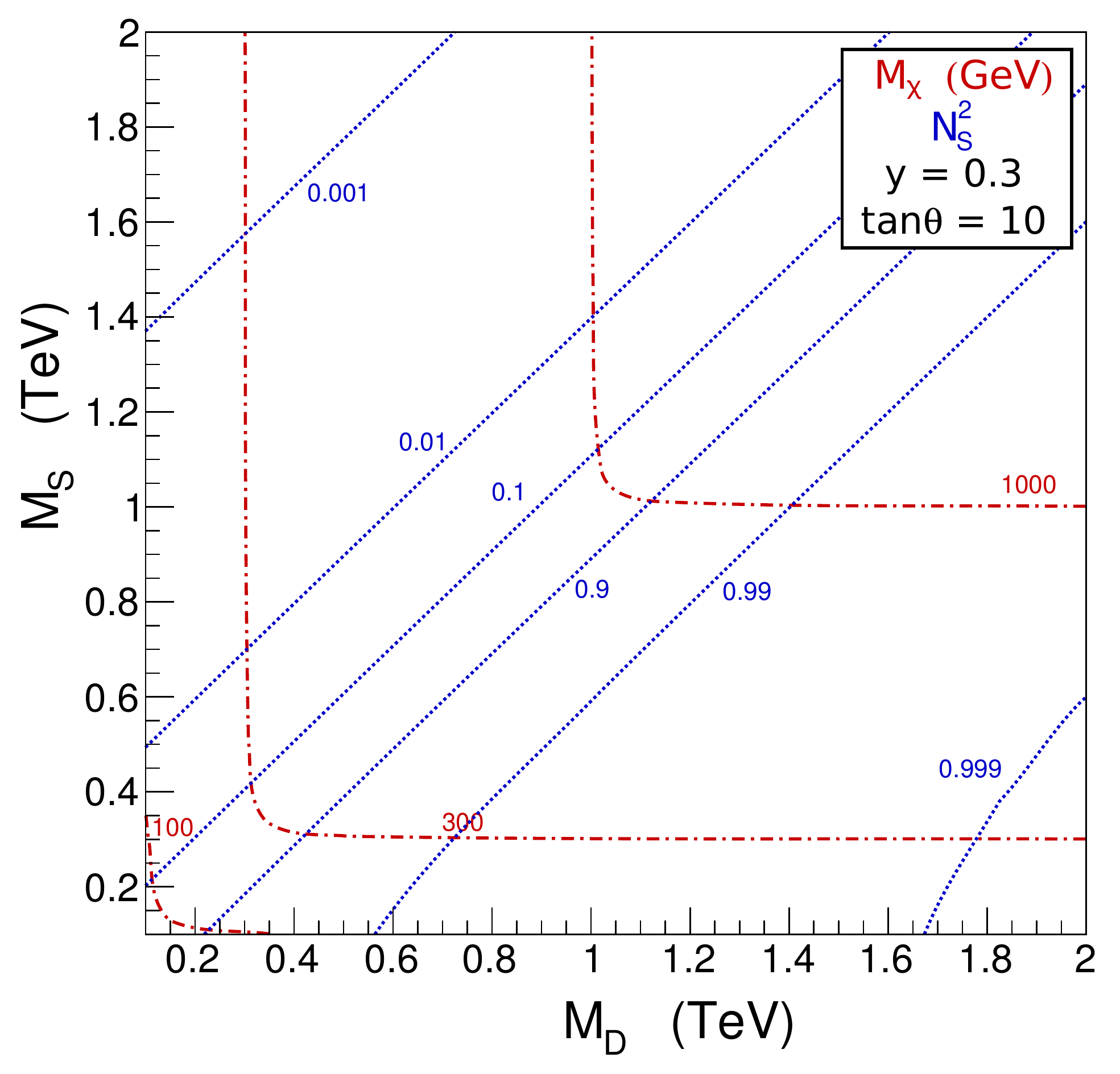}
  \label{fig:Plot0_y=0.3_tantheta=10}}
\subfigure[]{
  \includegraphics*[width=0.45\textwidth]{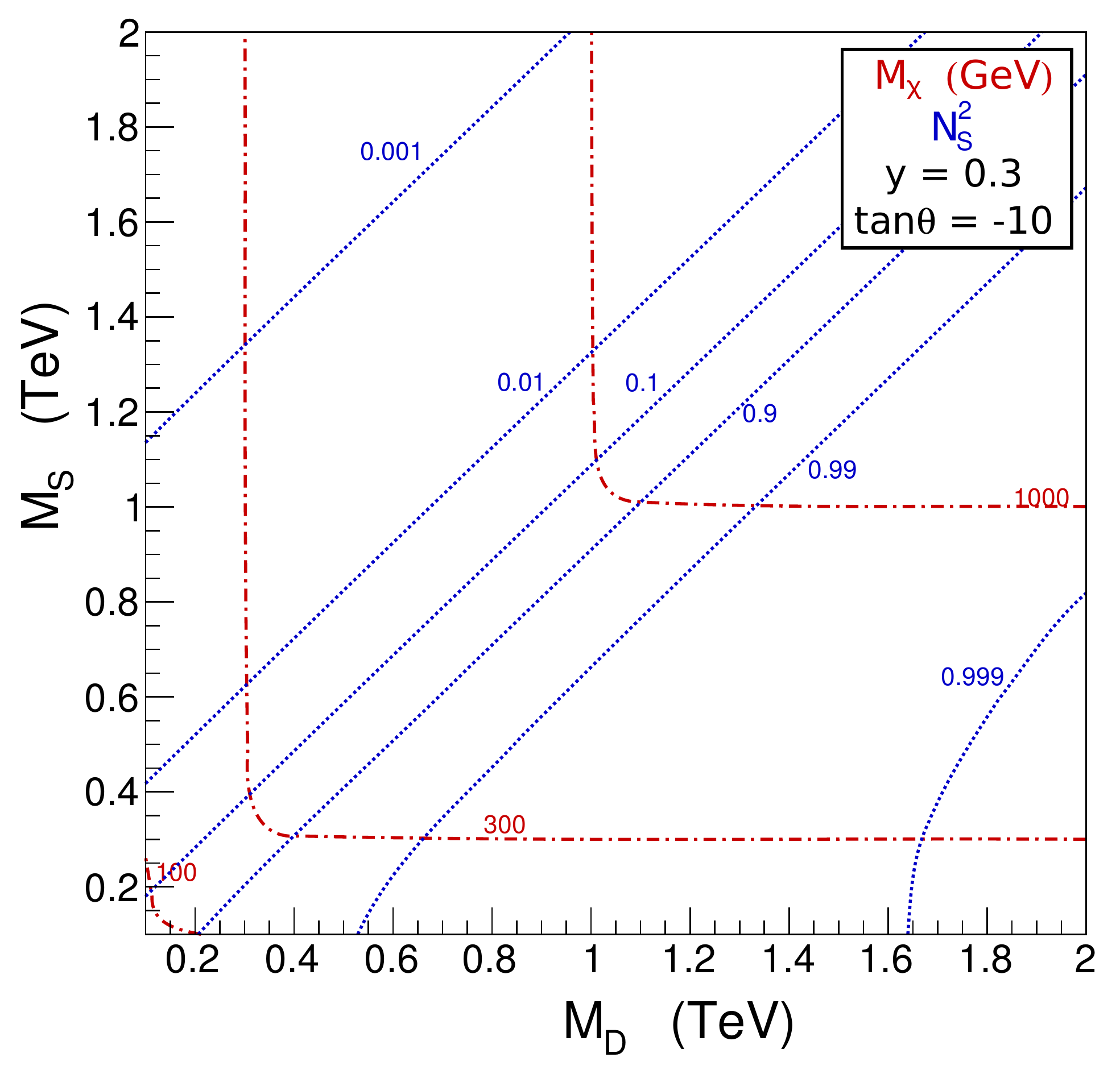}
  \label{fig:Plot0_y=0.3_tantheta=-10}}
\vspace*{-.1in}
\caption{\label{fig:Plot0_y=0.3} \textit{Mass and mixing angle
    contours in singlet-doublet fermion DM for $y=0.3$ and $\tan\theta
    = \pm 2,\pm 10 $.}  Shown are contours of the DM mass $\mchi$ in GeV (red
  dot-dashed) and the DM-singlet mixing angle squared $N_S^2$ (blue
  dotted).}
\end{figure}

\begin{figure}[tb]
\subfigure[]{
  \includegraphics*[width=0.45\textwidth]{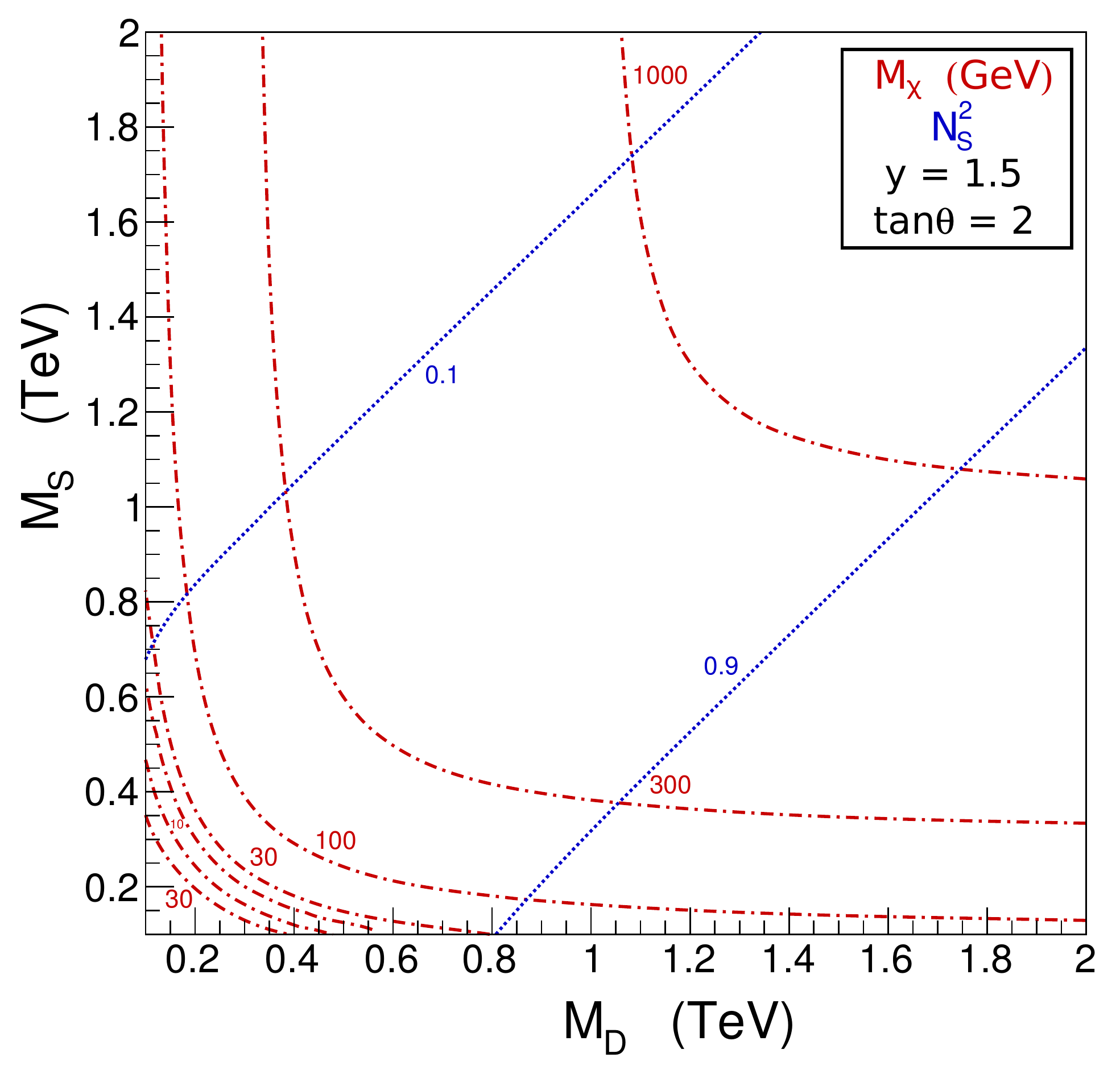}
  \label{fig:Plot0_y=1.5_tantheta=2}}
\subfigure[]{
  \includegraphics*[width=0.45\textwidth]{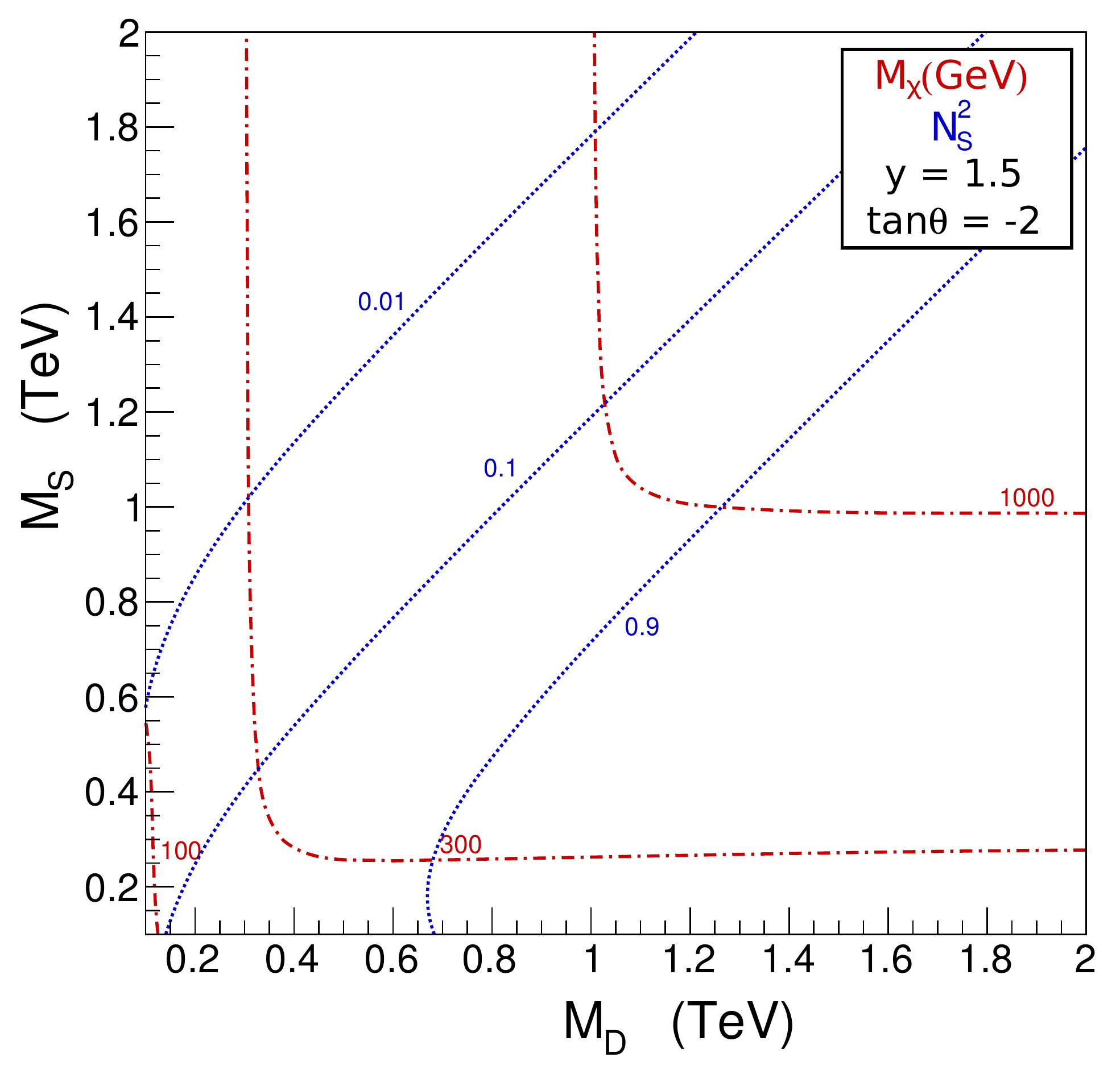}
  \label{fig:Plot0_y=1.5_tantheta=-2}}
\subfigure[]{
  \includegraphics*[width=0.45\textwidth]{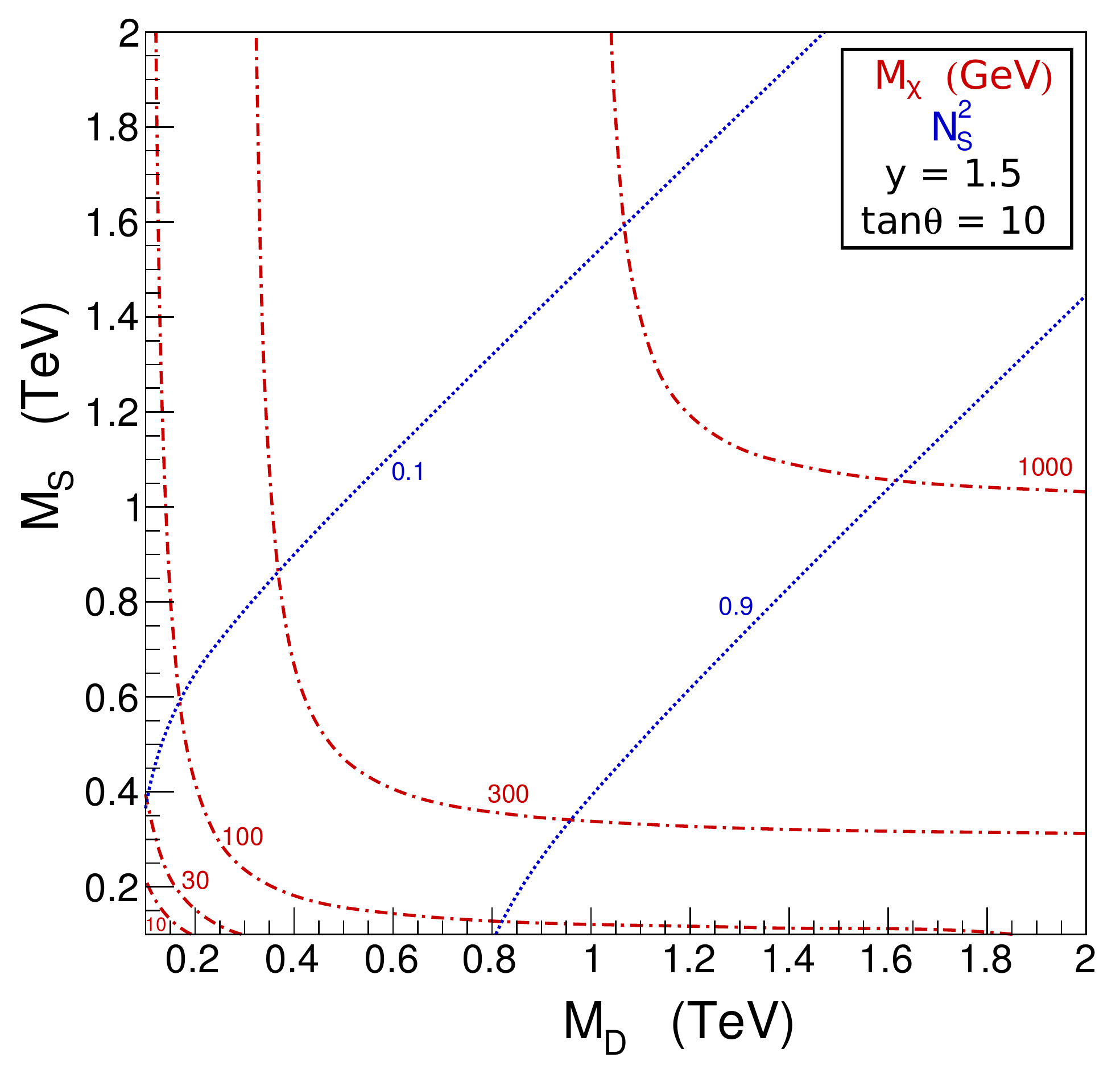}
  \label{fig:Plot0_y=1.5_tantheta=10}}
\subfigure[]{
  \includegraphics*[width=0.45\textwidth]{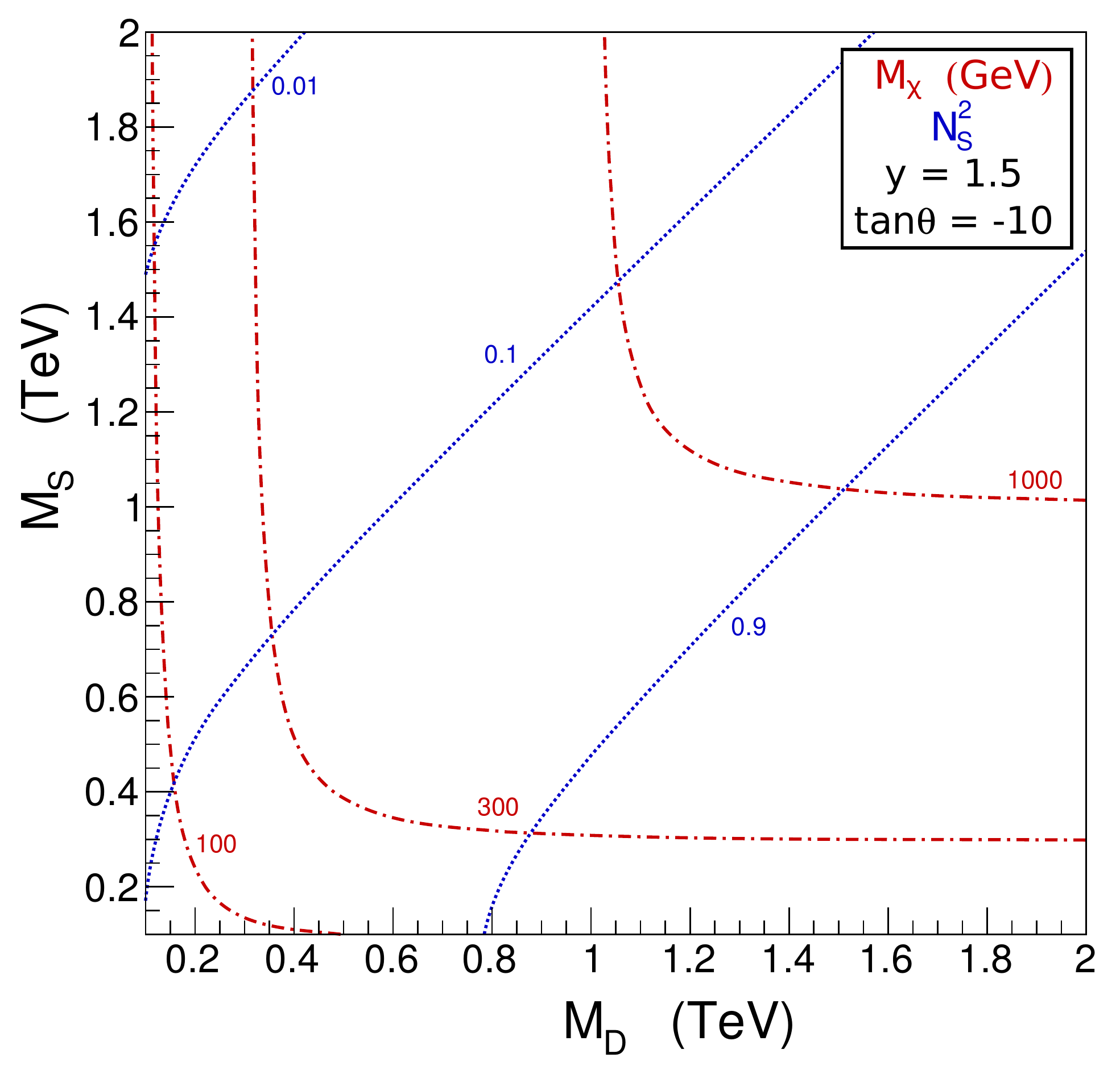}
  \label{fig:Plot0_y=1.5_tantheta=-10}}
\vspace*{-.1in}
\caption{\label{fig:Plot0_y=1.5} \textit{Mass and mixing angle
    contours in singlet-doublet fermion DM for $y=1.5$ and $\tan\theta
    = \pm 2,\pm 10 $.}  Contours shown are the same as in
  \figref{Plot0_y=0.3}.}
\end{figure}

In this section we define a simplified model for fermionic mixed DM
comprised of a Majorana singlet and Dirac doublet of $SU(2)_{W} \times
U(1)_Y$.  Here we have introduced a Dirac doublet in order to cancel
anomalies and allow for a bare mass term.  A priori, the Dirac doublet
can have arbitrary hypercharge, but to induce mixing via
renormalizable interactions, we demand that $Y = \pm 1/2$ so that
either the left-handed or right-handed doublet component can couple to
the Higgs and the singlet.  The field content of this simplified model
is
\begin{center}
\begin{tabular} {|c|c|c|}
\hline
\multicolumn{3}{|c|}{\it Model A} \\
\hline
Field & Charges & Spin \\
\hline
$S$ & $(\mathbf{1},0) $  & 1/2 \\
\hline
$D_1$ & $(\mathbf{2}, -1/2) $ &1/2 \\
\hline
$D_2$ & $(\mathbf{2}, 1/2) $ & 1/2 \\
\hline
\end{tabular}\,,
\end{center}
and the general renormalizable Lagrangian is
\begin{eqnarray}
-{\cal L}_\textrm{\it Model A} &=& \frac{1}{2} M_S S^2 + M_D D_1 D_2 +
y_{D_1} S H D_1+ y_{D_2} S H^\dagger D_2 + H.c. \, ,
\label{eq:SDfermion}
\end{eqnarray}
where we have dropped the kinetic terms for simplicity.  In our
notation, $D_1$ and $D_2$ have the same quantum numbers as
$\tilde{H}_d$ and $\tilde{H}_u$ of the MSSM, respectively. We will
sometimes parametrize the Yukawa couplings in polar coordinates,
\begin{eqnarray}
y_{D_1} &=& y \cos \theta \\
y_{D_2} &=& y \sin \theta\,.
\end{eqnarray}
Throughout, we work in a convention where $M_S$ and $M_D$ are positive
but $y_{D_1}$ and $y_{D_2}$ have indefinite sign.  Note, however, that
using a parity transformation $S \rightarrow -S$, we can
simultaneously flip the signs of $y_{D_1}$ and $y_{D_2}$, so only
their relative sign is physical.  Likewise, a parity transformation
$D_1 \rightarrow - D_1$ or $D_2 \rightarrow -D_2$ flips the signs of
$M_D$ and either $y_{D_1}$ or $y_{D_2}$, respectively, so only a
singlet sign among the three parameters is physical.  After
electroweak symmetry breaking, $S$ mixes with $D_1$ and $D_2$, and
this mixing simultaneously controls the thermal relic density of the
DM as well as its coupling to the Higgs boson.  The lightest neutral
state, $\chi$, is stable DM, and is defined as a linear combination of
the interaction eigenstates, $\chi = N_S S + N_{D_1} D_1 + N_{D_2}
D_2$, where $N_S^2 + N_{D_1}^2 + N_{D_2}^2 =1$.  There is also an
electrically charged state in the spectrum which we denote by
$\chi^\pm$.  Two heavier neutral states are also present, but they
have little effect except when coannihilation controls the thermal
relic density.

\eqref{eq:SDfermion} parametrizes a broad class of models which have
been discussed in the literature.  Most prominently, it describes
bino-Higgsino DM in the MSSM, with $y=g'/\sqrt{2}$ and $\theta =
\beta$.  It also describes singlino-Higgsino mixing in the NMSSM,
where mixing is controlled by the superpotential term $W = \lambda S
H_u H_d$, so $y = \lambda$ and $\theta$ and $ \beta$ are offset by
$\pi/2$.  Singlet-doublet Majorana DM has also been discussed in great
detail in Ref.~\cite{Cohen:2011ec}.

Next, we present an analytic derivation of the relevant properties of
$\chi$.  In the basis $(S, D_1, D_2)$, the neutral mass matrix is
\begin{eqnarray}
\mathbf{M} &=& \left(
\begin{array}{ccc}
M_S & \frac{1}{\sqrt{2}} y_{D_1} v & \frac{1}{\sqrt{2}} y_{D_2} v \\
\frac{1}{\sqrt{2}} y_{D_1} v & 0 & M_D \\
\frac{1}{\sqrt{2}} y_{D_2} v & M_D & 0
\end{array}
\right) \,.
\end{eqnarray}
The characteristic equation of the mass matrix is
\begin{eqnarray}
\nonumber 0 & = & \left( \mchi^2 - M_D^2 \right) \left( M_S - \mchi
\right) + M_D y_{D_1} y_{D_2} v^2 + \frac{1}{2} \mchi
\left( y_{D_1}^2 + y_{D_2}^2 \right) v^2 \\
& = & \left( \mchi^2 - M_D^2 \right) \left( M_S - \mchi \right) +
\frac{1}{2} y^2 v^2 \left(\mchi + M_D \sin 2 \theta \right)\,,
\label{eq:chareqSD}
\end{eqnarray}
where we are interested in the smallest eigenvalue, $\mchi$.  Since
$y$ labels the overall magnitude of the Yukawa couplings, its sign is
unphysical and the characteristic equation depends only on $y^2$.  On
the other hand, the sign of $y_{D_1} / y_{D_2}$ is physical, and is
represented by the sign of $\tan\theta$ (or equivalently $\sin
2\theta$).  The DM-Higgs coupling, $\chxx = \partial M_\chi(v) /
\partial v$, can be computed exactly by differentiating the
characteristic equation in \eqref{eq:chareqSD} with respect to
$\partial/\partial v$ and solving, yielding
\begin{eqnarray}
\chxx & = & - \frac{y^2 v \left(\mchi + M_D \sin 2\theta \right)}{
M_D^2+2 M_S M_\chi- 3 M_\chi^2 + y^2 v^2/2}\,.
\label{eq:sdfermionhiggs}
\end{eqnarray}
We define ``blind spot'' for spin independent direct detection by all
parameter points which satisfy $c_{h \chi\chi}=0$, so
\begin{eqnarray}
\mchi + M_D \sin 2\theta =0\,,
\label{eq:fermionSDblindspot}
\end{eqnarray}
as discussed in~\cite{Cheung:2012qy}.  Because $M_S$ and $M_D$ are
positive definite, a blind spot can only occurs when $\sin 2\theta <
0$.

In general, we will be interested in the amount of tuning required of
thermal relic DM permitted by current and future direct detection
sensitivity.  Using the general expressions for the tuning measures in
\eqref{eq:BStuningdef} and \eqref{eq:WTtuningdef}, we define
\begin{eqnarray}
\xiwt &=& \frac{\sqrt{2 \left( M_D^2 - M_S^2 \right)^2 + \frac{1}{2}
    y^4 v^4 + 2 y^2 v^2 \left[ M_D^2 + 2 M_S^2 + 3 M_D M_S \sin
      2\theta \right]}}{2 M_D^2 + M_S^2 + y^2 v^2} \\
\xibs &=& \left| \frac{\mchi + M_D \sin 2\theta}{\mchi + M_D |\sin
  2\theta |}\right| \,,
\end{eqnarray}
corresponding to the amount of tuning required for a properly
well-tempered thermal relic, and to the amount of tuning required to
reside sufficiently close to a blind spot cancellation, respectively.

Finally, let us consider the DM mass and mixing angles in the
parameter space of singlet-doublet DM.
\figsref{Plot0_y=0.3}{Plot0_y=1.5} show contours of $\mchi$ and
DM-singlet mixing angle squared, $N_S^2$, in the $(M_D, M_S)$ plane
for $\tan\theta = \pm 2, \pm 10$ for $y=0.3$ and $y=1.5$,
respectively.  For $y=0.3$, $\mchi$ is equal to the lower of $M_S$ or
$M_D$ in most of the parameter space, only deviating near $M_S \approx
M_D$.  Significant mixing occurs when $M_S \approx M_D$, and away from
this region the DM quickly approaches a pure singlet or pure doublet.
For $y=1.5$, $\mchi$ is significantly offset from both $M_S$ and $M_D$
throughout the plotted range for $\tan\theta > 0$, with a larger
offset around $M_S \approx M_D$.  A sizable region even exists with
$\mchi < 100~\gev$ for low values of $M_S$ and/or $M_D$.  For
$\tan\theta < 0$, however, the offset is more modest.  As indicated,
the degree of mixing typically much greater for $y=1.5$ than for
$y=0.3$.

\subsection{Singlet-Doublet Scalar}
\label{sec:sdscalar}

\begin{figure}[tb]
\subfigure[]{
  \includegraphics*[width=0.31\textwidth]{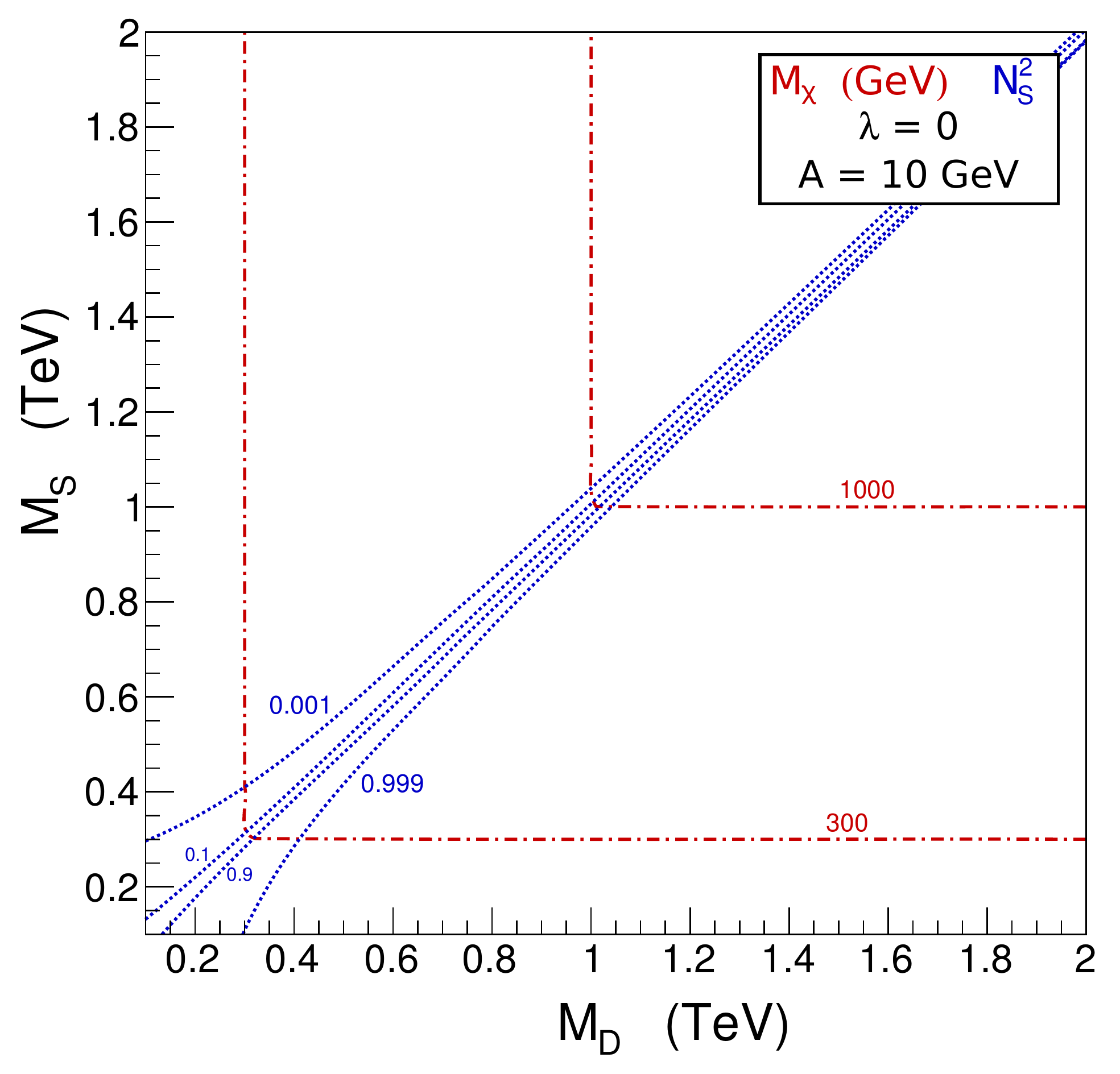}
  \label{fig:Plot0b_MSMD_A=10_y=0}}
\subfigure[]{
  \includegraphics*[width=0.31\textwidth]{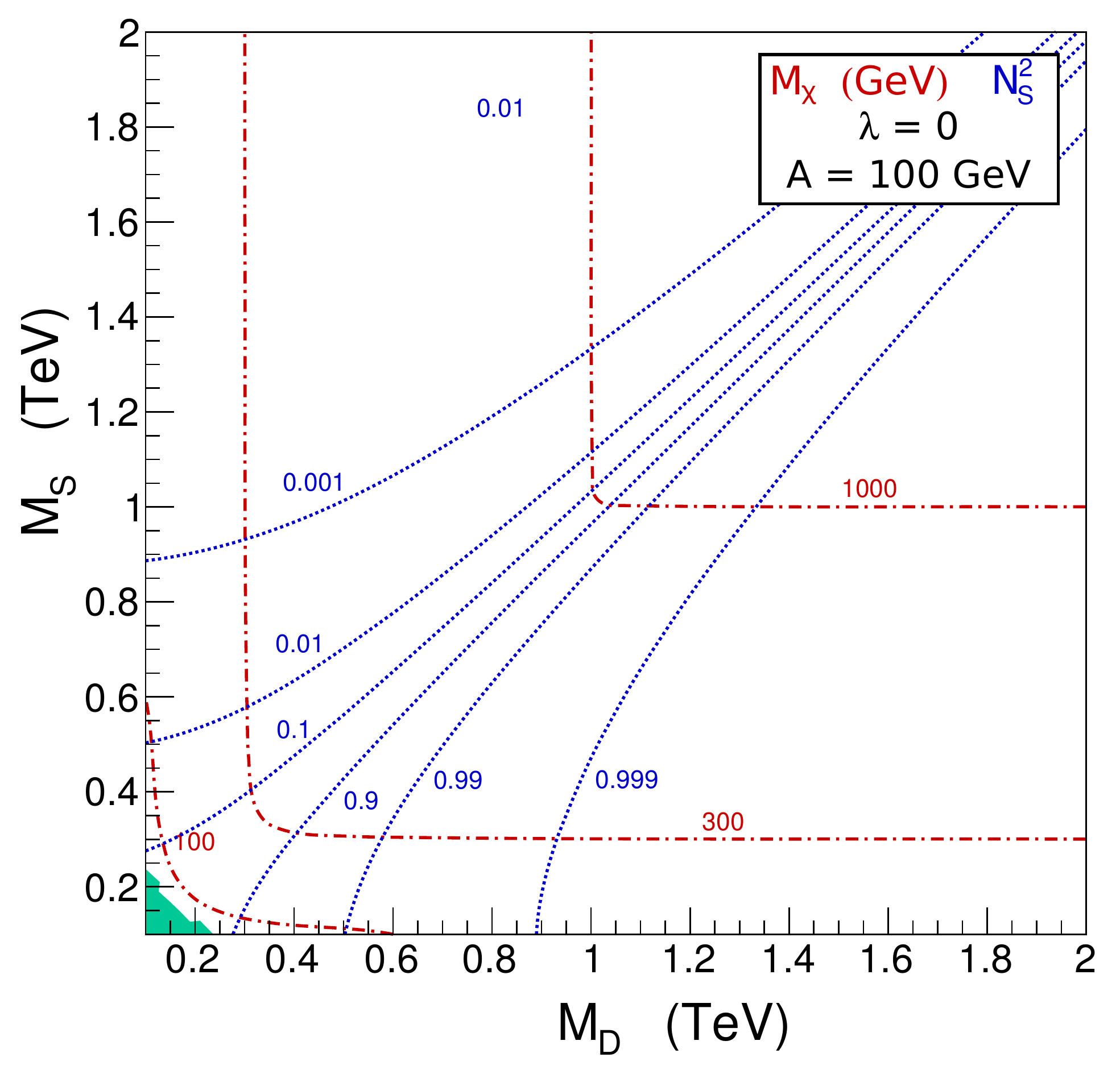}
  \label{fig:Plot0b_MSMD_A=100_y=0}}
\subfigure[]{
  \includegraphics*[width=0.31\textwidth]{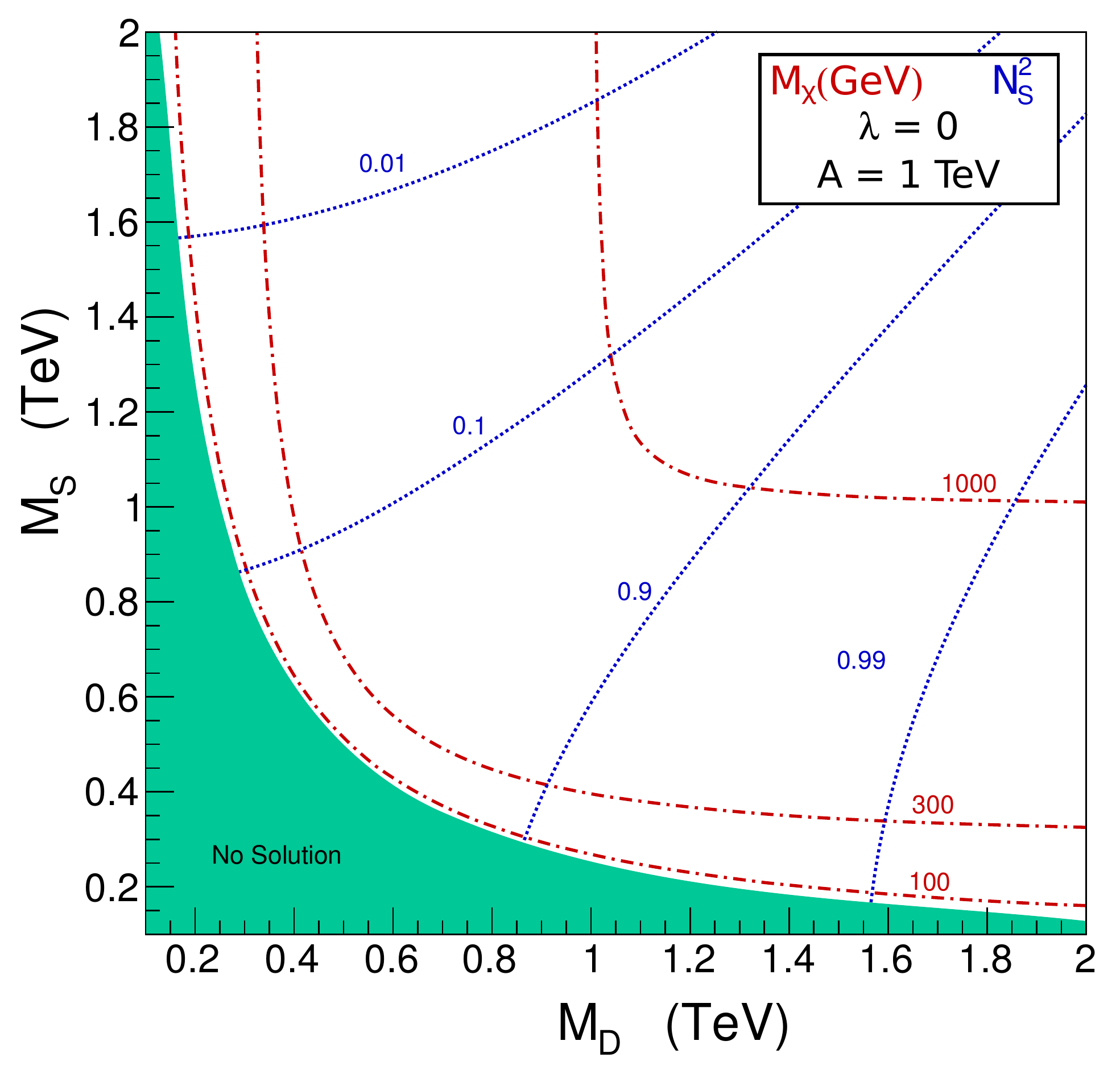}
  \label{fig:Plot0b_MSMD_A=1000_y=0}}
\vspace*{-.1in}
\caption{\label{fig:Plot0b} \textit{Mass and mixing angle contours for
    singlet-doublet scalar DM for $\lambda = 0$ and $A= 10, 100,
    1000~\gev$.}  Contours shown are the same as in
  \figref{Plot0_y=0.3}.}
\end{figure}

Next, we define another simplified DM model with singlet-doublet
mixing, only with scalars rather than fermions.  As in the case of
fermionic DM, singlet-doublet mixing for scalars requires that the
doublet have hypercharge $Y = 1/2$.  However, in this case a second
doublet is not required to either accommodate a bare doublet mass or
cancel anomalies.  The field content is
\begin{center}
\begin{tabular} {|c|c|c|}
\hline
\multicolumn{3}{|c|}{\it Model B} \\
\hline
Field & Charges & Spin \\
\hline
$S$ & $(\mathbf{1},0) $  & 0\\
\hline
$D$ & $(\mathbf{2},1/2) $ & 0 \\
\hline
\end{tabular}\,,
\end{center}
with a corresponding Lagrangian
\begin{eqnarray}
-\nonumber \mathcal{L}_\textrm{\it Model B} & = & \frac{1}{2} M_S^2 S^2
+ M_D^2 \left|D\right|^2 + \frac{1}{2} \lambda_S S^2 \left|H\right|^2
+ \lambda_D \left|D\right|^2 \left|H\right|^2 + \lambda_D' \left| H
D^\dagger \right|^2 \\
&& + \frac{1}{2} \lambda_D'' \left[ \left( H D^\dagger \right)^2 + h.c. \right] +
A \left[ S H D^\dagger + h.c. \right]\,.
\end{eqnarray}
Here we drop kinetic terms, along with all interactions involving only
$S$ and $D$, which are DM self-interactions which do not effect DM
annihilation or direct detection.  For simplicity, we assume positive
$M_S^2$ and $M_D^2$, although strictly speaking this is not necessary
because the singlet and doublet acquire masses after electroweak
symmetry breaking.  Note that by applying a parity transformation $S
\rightarrow -S$, we can flip the sign of $A$; consequently this sign
is unphysical.

Scalar dark matter theories have a long pedigree.  Scalar singlet DM
is often considered the most minimal DM candidate, and has been
studied extensively~\cite{Silveira:1985rk, McDonald:1993ex,
  Burgess:2000yq, Boehm:2003hm}.  Current bounds on scalar singlet
models with the correct relic density require $M_S \gtrsim 100~\gev$
except for a small region of viable parameter space for $50~\gev
\lesssim M_S \lesssim 65~\gev$~\cite{Cline:2013gha}, but XENON1T reach
expected to cover $M_S \gtrsim 10~\tev$.  Scalar doublet DM also has
been studied extensively, most often in the case of a two-Higgs
doublet model where only one Higgs receives a VEV~\cite{Ma:2006km},
often called the ``inert doublet model''~\cite{LopezHonorez:2006gr}.
Mixed singlet-doublet scalar models have also bee considered
previously~\cite{Kadastik:2009dj, Kadastik:2009cu, Cohen:2011ec},
though most such studies have considered a sub-set of the possible
phenomenology motivated by grand unification.

Paralleling the fermion case, singlet-doublet mixing for scalars
produces three real neutral scalars and one charged scalar.  Mixing
among states is induced by the $A$ term after electroweak
symmetry breaking.  We can work in  the basis of real neutral scalars, $( S, D_R, D_I )$, where  $D_R$ and $D_I$ are the real and imaginary components of the
neutral component of $D$.  The mass squared matrix is
\begin{equation}
\mathbf{M}^2 = \left(\begin{array}{ccc} M_S^2 + \frac{1}{2} v^2
  \lambda_S & A v & 0 \\
A v & M_D^2 + \frac{1}{2} v^2 \left( \lambda_D + \lambda_D' +
\lambda_D'' \right) & 0 \\
0 & 0 & M_D^2 + \frac{1}{2} v^2 \left( \lambda_D + \lambda_D' -
\lambda_D'' \right)
\end{array} \right).
\end{equation}
We assume the absence of CP violating
couplings, so $D_I$ cannot mix with either $D_R$ or $S$.  Focusing on
mixed DM, we choose a very tiny but negative value of $\lambda_D''$ to
ensure that the doublet which mixes has a smaller mass term.  With
this restriction we define the lightest mixed state as the DM particle
$\chi = N_S S + N_D D_R$, where $N_S^2 + N_D^2 = 1$.

DM mixing is induced by the upper left $2\times 2$ block of the full
mixing matrix, whose characteristic eigenvalue equation is
\begin{eqnarray}
0 & = & \left( \mchi^2 - \tilde{M}_S^2 \right) \left( \mchi^2 -
\tilde{M}_D^2 \right) - v^2 A^2 ,
\end{eqnarray}
where $\tilde{M}_S^2 = M_S^2 + \frac{1}{2} v^2 \lambda_S $ and
$\tilde{M}_D^2 = M_D^2 + \frac{1}{2} v^2 \left( \lambda_D + \lambda_D'
+ \lambda_D'' \right) $.  The two eigenvalues of the mass-squared
matrix are
\begin{eqnarray}
\frac{1}{2} \left[ \tilde{M}_S^2 + \tilde{M}_D^2 \pm \sqrt{\left(
    \tilde{M}_S^2 - \tilde{M}_D^2 \right)^2 + 4 v^2 A^2} \right]\,,
\end{eqnarray}
with the smaller eigenvalue corresponding to $\mchi^2$.  As shown
earlier, the associated DM-Higgs coupling is given by the derivative
of $\mchi^2$ with respect to $v$,
\begin{eqnarray}
\ahxx & = & \frac{1}{2} v \left( \lambda_S + \lambda_D + \lambda_D' +
\lambda_D'' \right) - \frac{2 v A^2 + \frac{1}{2}v \left(\tilde{M}_S^2
  - \tilde{M}_D^2 \right) \left(\lambda_S - \lambda_D - \lambda_D' -
  \lambda_D'' \right)}{\sqrt{\left(\tilde{M}_S^2 - \tilde{M}_D^2
    \right)^2 + 4 v^2 A^2}} \,.
\label{eq:sdscalarhiggs}
\end{eqnarray}
In our analysis we will make use of the simplifying limit $\lambda_S =
\lambda_D = \lambda$ and $ \lambda_D' = \lambda_D'' = 0$.  In this
limit, $\ahxx$ simplifies to
\begin{eqnarray}
\ahxx & = & \lambda v - \frac{2 v A^2}{\sqrt{\left(M_S^2 -M_D^2
    \right)^2 + 4 v^2 A^2}} \,.
\label{eq:ahxx_sdscalar_simplified}
\end{eqnarray}
In analogy with the case of fermion DM, we define the blind spot
region by the condition $a_{h\chi\chi} = 0$.  However, in the scalar
case it is complicated by the presence of Higgs couplings to the pure
states.  For fermionic DM, mixing is induced by Yukawa terms, leading
to a correlation between mixing strength and the Higgs coupling.  In
the scalar case, however, for any degree of mixing, the direct quartic
couplings can be modified to create a blind spot.  The enhancement or
suppression of $\ahxx$ depends on the sign of $\lambda$.  For positive
$\lambda$, a blind spot can occur, and interestingly, this is the sign
preferred in general by considerations of tree-level vacuum stability.
In particular, if $\lambda$ is too negative then the potential may
contain unbounded from below directions at large field values.

As in the case of singlet-doublet fermion DM, we can characterize the
amount of tuning required to accommodate a thermal relic abundance and
evade direct detection.  Again using \eqref{eq:BStuningdef} and
\eqref{eq:WTtuningdef}, we define (using only the $2\times2$ mixing
sub-matrix)
\begin{eqnarray}
\xiwt & = & \frac{\sqrt{\left( M_S^2 - M_D^2 \right)^2 + 4 v^2
    A^2}}{\left( M_S^2 + M_D^2 +\lambda v^2\right)} \\
\xibs & = & \frac{\left| \lambda v \sqrt{\left(M_S^2 -M_D^2 \right)^2 +
    4 v^2 A^2} - 2 v A^2 \right|}{\left| \lambda v
  \right|\sqrt{\left(M_S^2 -M_D^2 \right)^2 + 4 v^2 A^2} + 2 v A^2}\,,
\label{eq:sdscalar_tuning}
\end{eqnarray}
to be the tuning measures for well-tempering and blind spot
cancellations, respectively.

Last of all, we consider the DM mass and mixing angles in
singlet-doublet scalar DM.  \figref{Plot0b} shows contours of $\mchi$
and $N_S^2$ in the $(M_D, M_S)$ plane, with $A = 10, 100,
1000~\gev$ and $\lambda_S = \lambda_D = \lambda_D' = \lambda_D'' =
0$.  For sufficiently small masses in the $A=100~\gev$ and $A=1~\tev$
cases, one of the eigenstates becomes tachyonic and thus the region is
excluded.  The well-mixed scenario occurs near $M_S \approx M_D$, with
the degree of mixing dropping rapidly except for the very large value
of $A=1~\tev$.  Likewise, except for $A=1~\tev$ the mass contours are
restricted to $\mchi \approx M_S$ or $\mchi \approx M_D$ except for
very close to the $M_S \approx M_D$ line.  Allowing for non-zero
quartic DM-Higgs couplings does not qualitatively change the mass and
mixing contours, but simply shifts their positions.

\subsection{Singlet-Triplet Scalar}
\label{sec:stscalar}

\begin{figure}[tb]
\subfigure[]{
  \includegraphics*[width=0.48\textwidth]{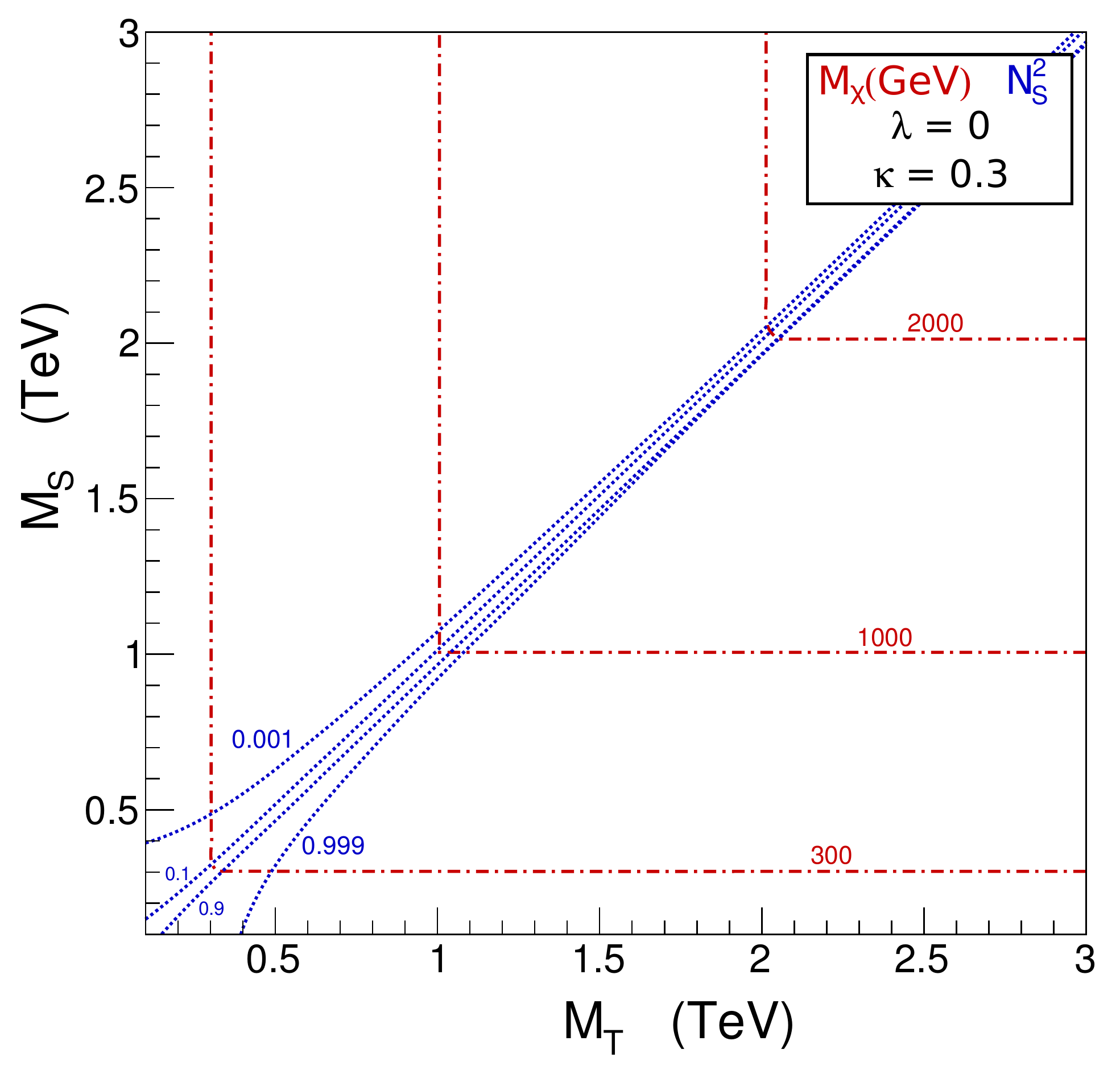}
  \label{fig:Plot0_MSMT_yST=0.3}}
\subfigure[]{
  \includegraphics*[width=0.48\textwidth]{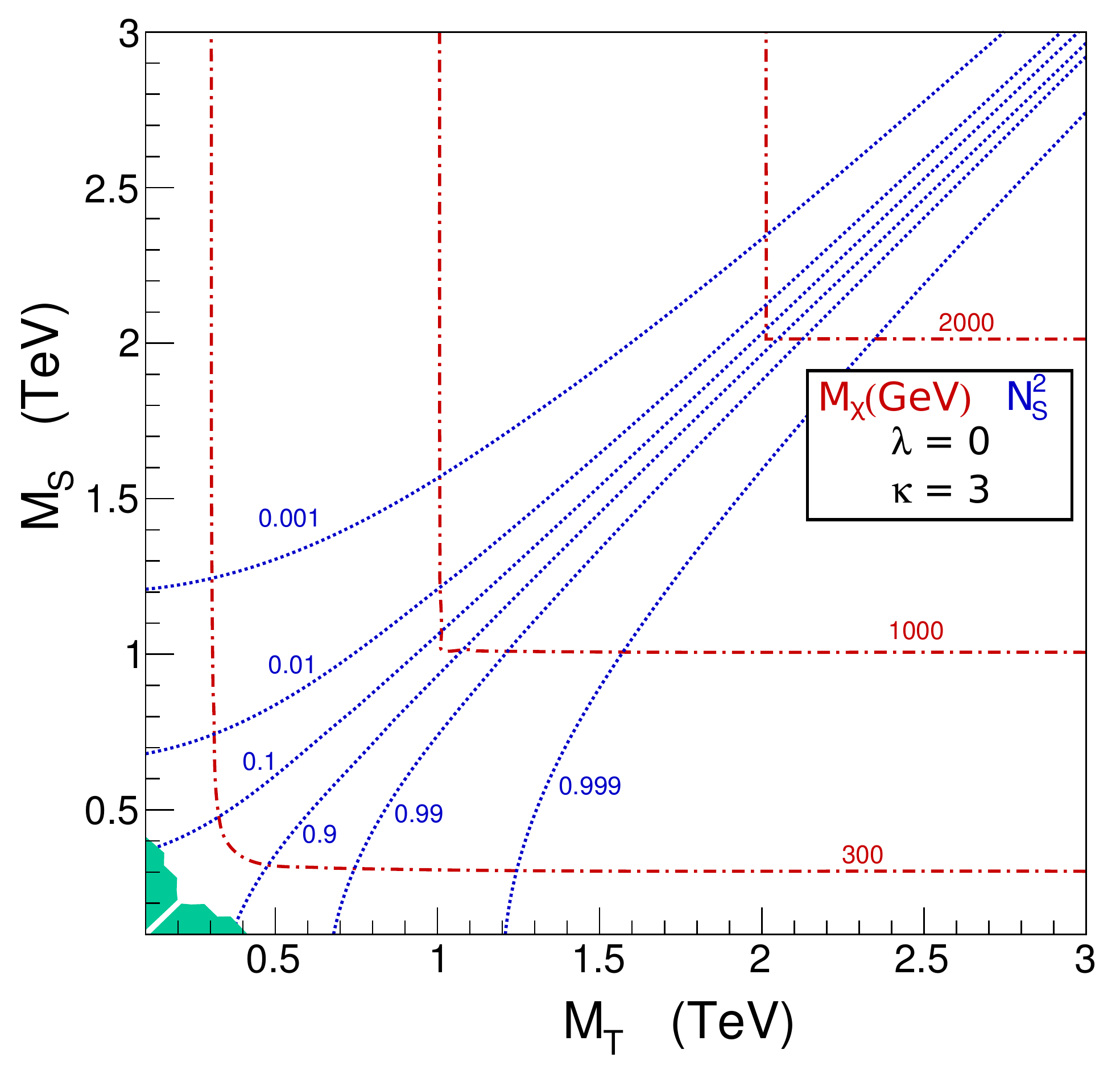}
  \label{fig:Plot0_MSMT_yST=3}}
\vspace*{-.1in}
\caption{\label{fig:Plot0c} \textit{Mass and mixing angle contours for
    singlet-triplet scalar DM for $\lambda = 0$ and $\kappa = 0.3,
    3$.}  Contours shown are the same as in \figref{Plot0_y=0.3}.}
\end{figure}

Lastly, we consider scalar DM comprised of a mixed singlet and
triplet.  For the sake of simplicity, we consider a real triplet,
which necessarily carries zero hypercharge.  The field content of this
model is
\begin{center}
\begin{tabular} {|c|c|c|}
\hline
\multicolumn{3}{|c|}{\it Model C} \\
\hline
Field & Charges & Spin \\
\hline
$S$ & $(\mathbf{1},0) $  & 0\\
\hline
$T$ & $(\mathbf{3},0) $ & 0 \\
\hline
\end{tabular}\,,
\end{center}
with a corresponding Lagrangian,
\begin{eqnarray}
-\nonumber \mathcal{L}_\textrm{\it Model C} & = & \frac{1}{2} M_S^2 S^2
+ M_T^2 \tr \left( T^2 \right) + \frac{1}{2} \lambda_S S^2
\left|H\right|^2 + \lambda_T \tr \left( T^2 \right)\left|H\right|^2 +
\kappa S H^\dagger T H \,,
\end{eqnarray}
where once again we have dropped kinetic terms and interactions
involving only $S$ and $T$.  After electroweak symmetry breaking, the
singlet and triplet mix via the dimensionless quartic interaction
$\kappa$.  Note that by applying the parity transformation,
$S\rightarrow -S$, we can freely flip the sign of $\kappa$, so its
sign is unphysical.

While models involving triplet scalar DM have not received the same
level of attention as those with singlet or doublet scalar DM, both
pure triplet~\cite{Cirelli:2005uq, FileviezPerez:2008bj} and mixed
singlet-triplet~\cite{Fischer:2011zz, Fischer:2013hwa}.  However, as
for mixed singlet-doublet scalar DM, previous studies have primarily
focused on a sub-set of parameter space motivated by grand
unification.

The mass matrix for the singlet-triplet scalar case resembles the
mixed sub-matrix for the singlet-doublet scalar, with the substitution
$A \rightarrow - \kappa v/4$ in the $(S, T)$ basis,
\begin{equation}
\mathbf{M}^2 = \left(\begin{array}{cc} M_S^2 + \frac{1}{2} v^2
  \lambda_S & - \frac{1}{4} \kappa v^2 \\
-\frac{1}{4} \kappa v^2 & M_T^2 + \frac{1}{2} v^2 \lambda_T
\end{array} \right)\,.
\end{equation}
After diagonalization, the DM particle is $\chi = N_S S + N_T T$,
where $N_S^2 + N_T^2 = 1$.  Likewise, the mass-squared eigenvalues
become
\begin{eqnarray}
\frac{1}{2} \left[ \tilde{M}_S^2 + \tilde{M}_T^2 \pm \sqrt{\left(
    \tilde{M}_S^2 - \tilde{M}_T^2 \right)^2 + \frac{1}{4} \kappa^2 v^4}
  \right]\,,
\end{eqnarray}
where $\tilde{M}_S^2 = M_S^2 + \lambda_S v^2 / 2$ and $\tilde{M}_T^2 =
M_T^2 + \lambda_T v^2 / 2$.  However, due to the $v$-dependence of the
mixing term for the singlet-triplet case, the same substitution does
not apply to $\ahxx$, which has the form
\begin{eqnarray}
\ahxx & = & \frac{1}{2} v \left( \lambda_S + \lambda_T \right) -
\frac{\frac{1}{4} \kappa^2 v^3 + \frac{1}{2}v \left(\tilde{M}_S^2 -
  \tilde{M}_T^2 \right) \left(\lambda_S - \lambda_T
  \right)}{\sqrt{\left(\tilde{M}_S^2 - \tilde{M}_T^2 \right)^2 +
    \frac{1}{4} \kappa^2 v^4}} \,.
\end{eqnarray}
The contribution to the Higgs coupling from the mixing term in the
singlet-triplet case is roughly twice as large
relative to the singlet-doublet case for models with an equivalent
mass spectrum and mixing.  For $\lambda_S = \lambda_D = \lambda$, this
reduces to
\begin{eqnarray}
\label{eq:ahchichiST}
\ahxx & = & \lambda v  - \frac{\frac{1}{4} \kappa^2
  v^3}{\sqrt{\left(M_S^2 - M_T^2 \right)^2 + \frac{1}{4} \kappa^2
    v^4}} \,.
\end{eqnarray}

As in the singlet-doublet case, cancellations in the DM-Higgs coupling
only occur if $\lambda$ is positive, with the same implication that
vacuum stability favors positive values of $\lambda$ and thus
cancellation.  From \eqref{eq:BStuningdef} and \eqref{eq:WTtuningdef},
we define tuning measures for blind spot cancellations and
well-tempering,
\begin{eqnarray}
\xiwt & = & \frac{\sqrt{\left( M_T^2 - M_D^2 \right)^2 + \frac{1}{4}
    \kappa^2 v^4}}{\left( M_S^2 + M_T^2 + \lambda v^2 \right)} \\
\xibs & = & \frac{\left| \lambda v \sqrt{\left(M_S^2 -M_T^2 \right)^2
    + \frac{1}{4} \kappa^2 v^2} - \frac{1}{4} \kappa^2 v^3
  \right|}{\left| \lambda v \right|\sqrt{\left(M_S^2 -M_T^2 \right)^2
    + \frac{1}{4} \kappa^2 v^2} + \frac{1}{4} \kappa^2 v^3}\,.
\label{eq:stscalar_tuning}
\end{eqnarray}

As before, we plot the DM mass, $\mchi$, and singlet mixing angle
squared, $N_S^2$, as a function of the parameter space in
\figref{Plot0c}.  Qualitatively, our results are similar to the
singlet-doublet case.  However, there is a major quantitative
difference, which is that mixing effects are minimal away from $M_S
\approx M_D$ even at very large values of the coupling, $\kappa = 3$.
As we will see in \secref{stscalar_analysis}, this implies
substantially different experimental constraints on singlet-doublet
versus singlet-triplet scalar DM.

\section{Model A: Singlet-Doublet Fermion DM}
\label{sec:sdfermion_analysis}

To begin, we will analyze the full four dimensional parameter space of
singlet-doublet fermion DM, $(M_S, M_D, y_{D_1}, y_{D_2})$, imposing
no constraints beyond the definition of the theory in
\secref{sdfermionmodel}.  We will display experimental constraints and
regions consistent with $\eqomega$ as a function of the bare masses
$M_S$ and $M_D$, fixing $(y, \theta)$ to several characteristic
values.  We will then focus on the subspace of thermal relic DM,
fixing one of the model parameters ($M_D$, $M_S$, or $\theta$,
depending on the plot) to accommodate $\eqomega$.  Lastly, we will
study the thermal relic scenario further restricted to the parameter
space residing exactly at the present (and future) limits of direct
detection experiments.  In particular, for this analysis we
will fix $y$ to $\sigmaSI = \sigmaSI_{\text{LUX}}$ or $\sigmaSI =
\sigmaSI_{\text{X1T}}$, corresponding to the space of models which are ``marginally excluded'' by LUX or XENON1T, respectively.

\subsection{Exclusion Plots (General)}

First, we consider the unconstrained parameter space, focusing on the
position of the $\eqomega$ line.  A pure singlet Majorana fermion does
not couple to the SM at the renormalizable level, so its thermal relic
abundance is typically very large.  Meanwhile, a pure doublet has
$\eqomega$ for $M_D \simeq 1~\tev$~\cite{Chattopadhyay:2005mv}, with
$\Omegachi > \OmegaDM$ for $M_D \gtrsim 1~\tev$ and $\Omegachi <
\OmegaDM$ for $M_D \lesssim 1~\tev$.  Annihilation of the DM into
gauge bosons occurs entirely through the electroweak charged component
of the DM.  Hence, this contribution to the annihilation cross-section
is suppressed by the doublet mixing angles, $N_{D_1}$ and $N_{D_2}$.
In principle, annihilation can also occur via Higgs exchange, and this
will be enhanced at large $y$.  However, annihilation via the Higgs is
a p-wave suppressed process for Majorana fermions, and thus
sub-dominant to gauge boson processes unless $y$ is very large.

\begin{figure}[tb]
\subfigure[]{
  \includegraphics*[width=0.45\textwidth]{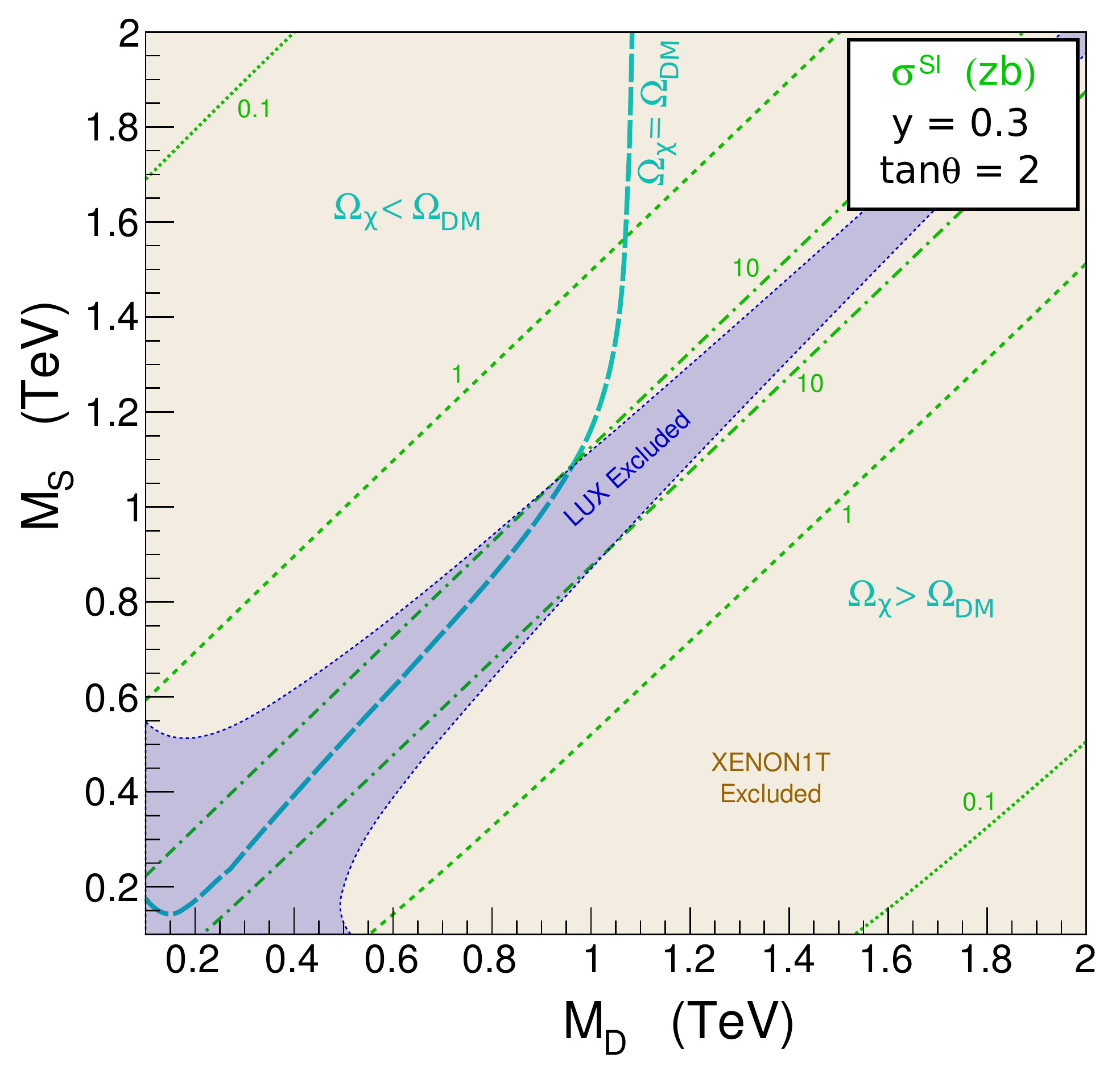}
  \label{fig:Plot1_y=0.3_tantheta=2}}
\subfigure[]{
  \includegraphics*[width=0.45\textwidth]{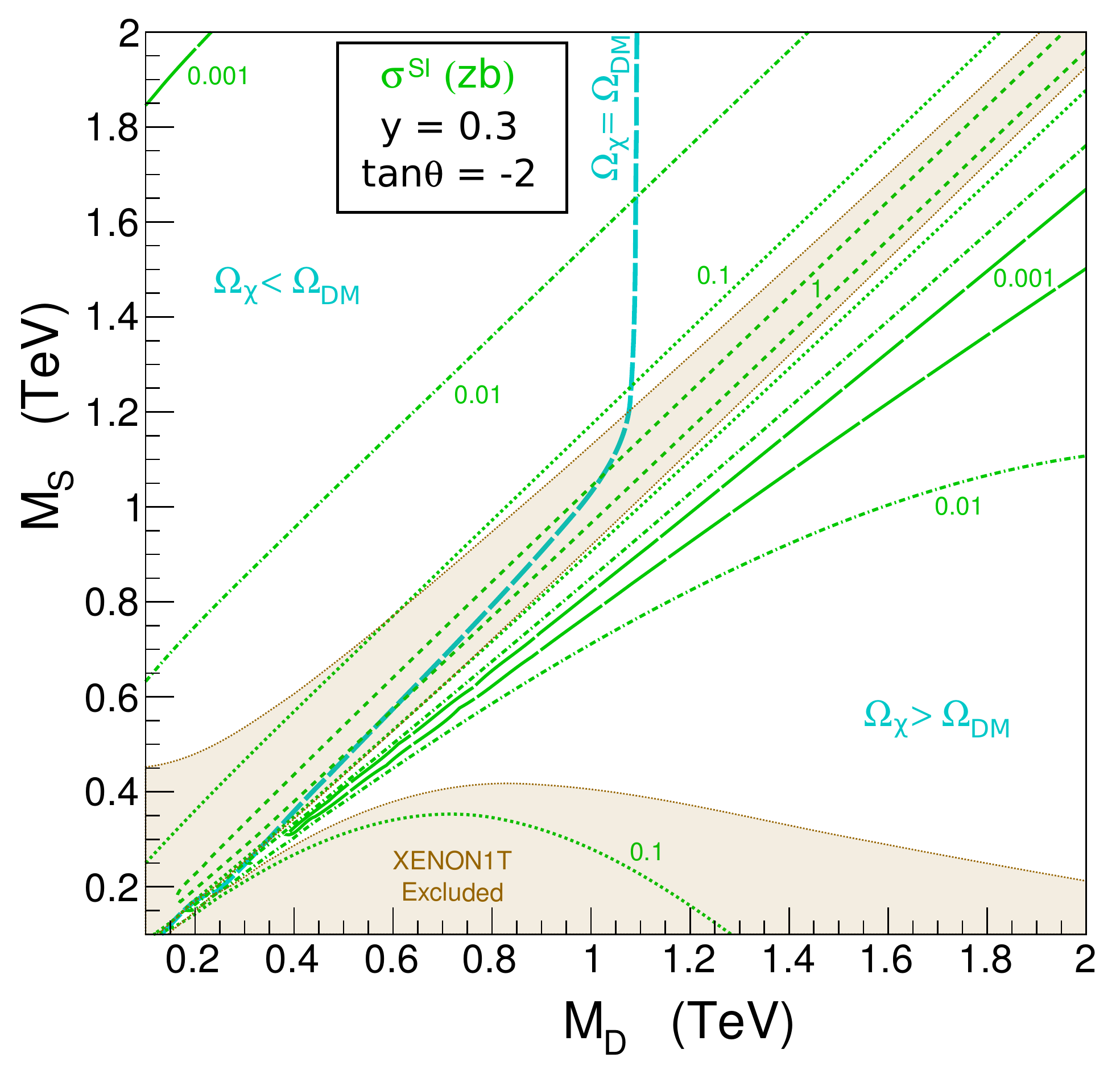}
  \label{fig:Plot1_y=0.3_tantheta=-2}}
\subfigure[]{
  \includegraphics*[width=0.45\textwidth]{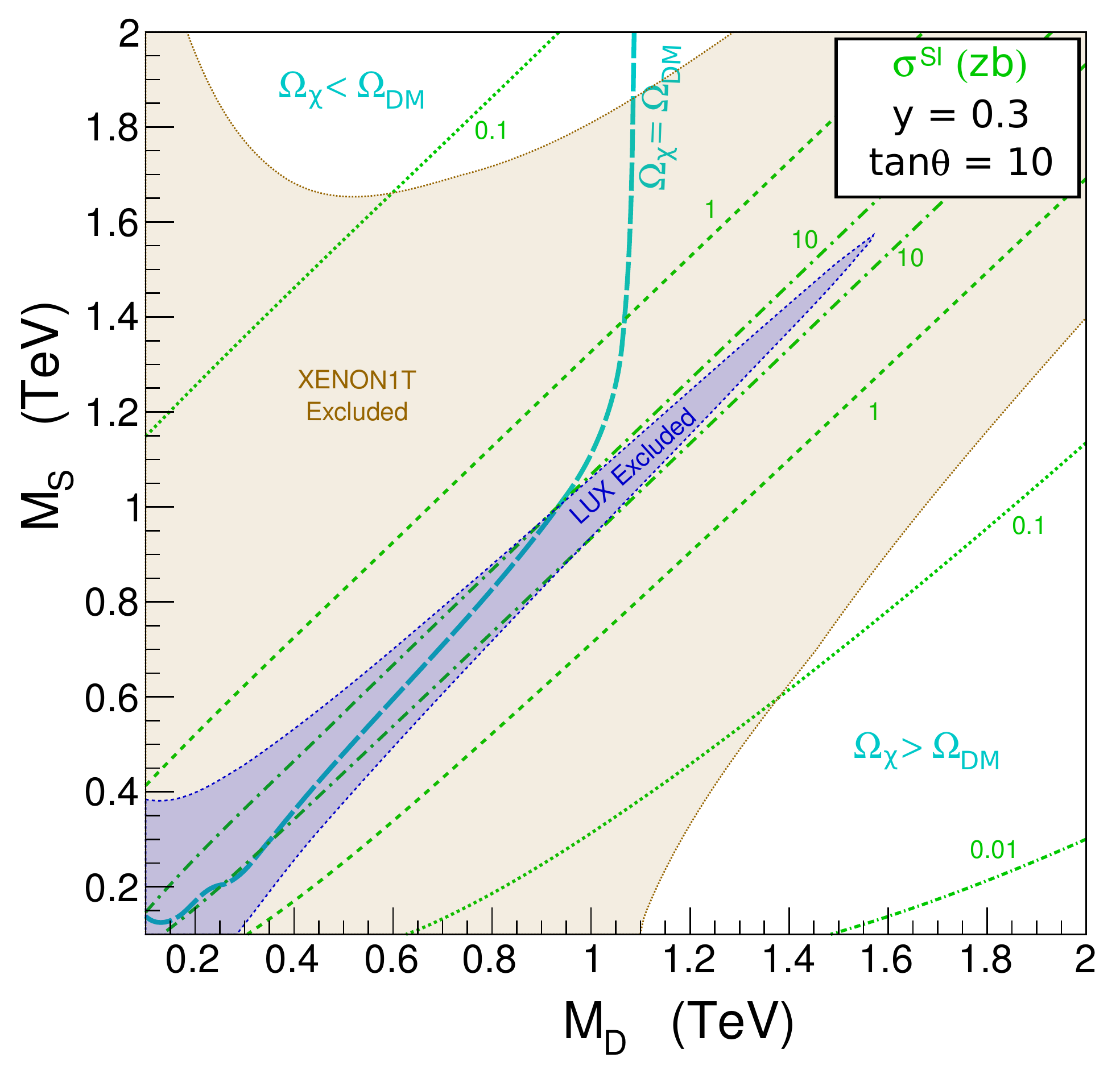}
  \label{fig:Plot1_y=0.3_tantheta=10}}
\subfigure[]{
  \includegraphics*[width=0.45\textwidth]{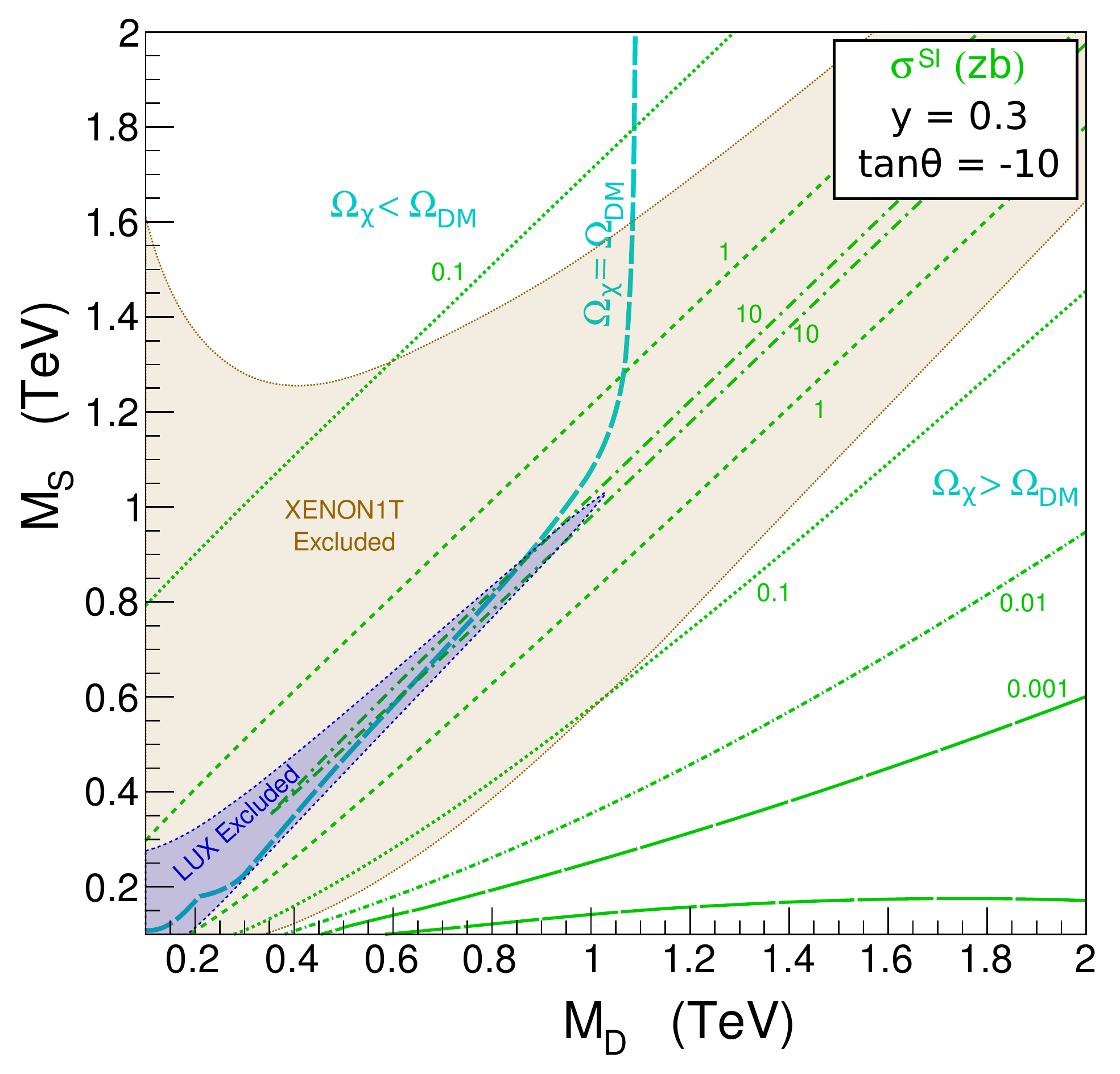}
  \label{fig:Plot1_y=0.3_tantheta=-10}}
\vspace*{-.1in}
\caption{\label{fig:Plot1_y=0.3} \textit{Direct detection prospects
    for $y=0.3$ and $\tan\theta = \pm 2, \pm 10$.}  Shown are contours
  of $\sigmaSI$ in zb (green, various styles) together with current
  bounds from LUX (blue shaded) and projected reach at XENON1T (gold
  shaded).  Away from the line consistent with the observed relic
  density (teal dashed), we compute LUX and XENON1T bounds assuming
  that the DM relic density is equal to the observed one due to a
  non-thermal cosmology.}
\end{figure}

In \figref{Plot1_y=0.3}, we have plotted constraints in the $(M_D,
M_S)$ plane, fixing $y=0.3$ and $\tan\theta = \pm 2$ (top) or $\pm 10$
(bottom).  For $M_S \lesssim 1~\tev$, significant well-tempering of
the DM mixing angles is required to produce $\eqomega$ for all
parameter combinations.  This is the case because $y v \ll M_{S,D}$,
requiring $M_S \approx M_D$ for any significant level of mixing.  For
$M_S \gtrsim 1~\tev$ the thermal relic line asymptotes to $M_D \simeq
1~\tev$, the mass at which pure doublet DM is a thermal relic with the
correct abundance.  The fact that $\eqomega$ is not possible for $M_D
> 1~\tev$ implies that the Higgs coupling $y=0.3$ affects the
annihilation cross-section primarily through mixing angles that
control gauge boson processes.  That is, processes involving the Higgs
boson directly do not strongly affect the DM annihilation
cross-section.
 
In terms of direct detection, the value of $\sigmaSI$ depends
sensitively on the sign of $\tan\theta$.  For $\tan\theta > 0$ it lies
in the range $0.01~\zb \lesssim \sigmaSI \lesssim 10~\zb$ throughout
most of the region shown, and exhibits no blind spot behavior.
Conversely, for $\tan\theta < 0$, the maximum value of $\sigmaSI$ is
reduced, and a blind spot occurs where $\sigmaSI$ vanishes (the
minimum value shown is $0.001~\zb$).  The position of the blind spot
changes for different values of $\tan\theta$, and is located at $M_S +
M_D \sin 2\theta\approx 0$, roughly consistent with the blind spot
condition of \eqref{eq:fermionSDblindspot}.

For positive $\tan\theta$, the thermal line is constrained by LUX for
$M_S\gtrsim 1.1~\tev$ for $\tan\theta=2$ and $M_S\gtrsim 1~\tev$ for
$\tan\theta=10$.  For negative $\tan\theta$, however, LUX provides no
bound for $\tan\theta=-2$ and bounds the thermal scenario only up to
$900~\gev$ for $\tan\theta=-10$.  Meanwhile, for both values of
$\tan\theta$, XENON1T constrains the thermal scenario well into the
nearly pure doublet region, as well as large swaths of non-thermal
scenarios.  Even after XENON1T, large swaths of parameter space will
still be allowed at small and negative $\tan\theta$, and for
$\tan\theta=-2$ a small portion of the thermal line with $M_S\sim
200~\gev$ remains viable.  The relatively low values of $\sigmaSI$ and
weak exclusion for $\tan\theta=-2$ are due to a blind spot near $\tan
\theta = -1$ for doublet-like DM~\cite{Cheung:2012qy}.  At large
values of $|\tan \theta|$ (corresponding to large $\tan \beta$ in the
MSSM), the sign of $\tan\theta$ becomes unphysical
\cite{Cheung:2012qy}, as shown by the relative similarity of the
contours for $\tan\beta=\pm 10$ as opposed to $\tan\theta=\pm 2$ in
\figref{Plot1_y=0.3}.

There are two major implications of \figref{Plot1_y=0.3}.  First, the
region consistent with a thermal relic is already quite constrained by
LUX for $y=0.3$.  XENON1T will further constrain this region,
excluding at least up to the point at which DM is nearly a pure
doublet except for a very finely tuned region of parameter space.
Second, the sensitivity of limits depends greatly on the sign of
$\tan\theta$, due to blind spot cancellation points, and in small
regions of parameter space even relatively light masses of a few
hundred GeV remain viable.

\begin{figure}[tb]
\subfigure[]{
  \includegraphics*[width=0.45\textwidth]{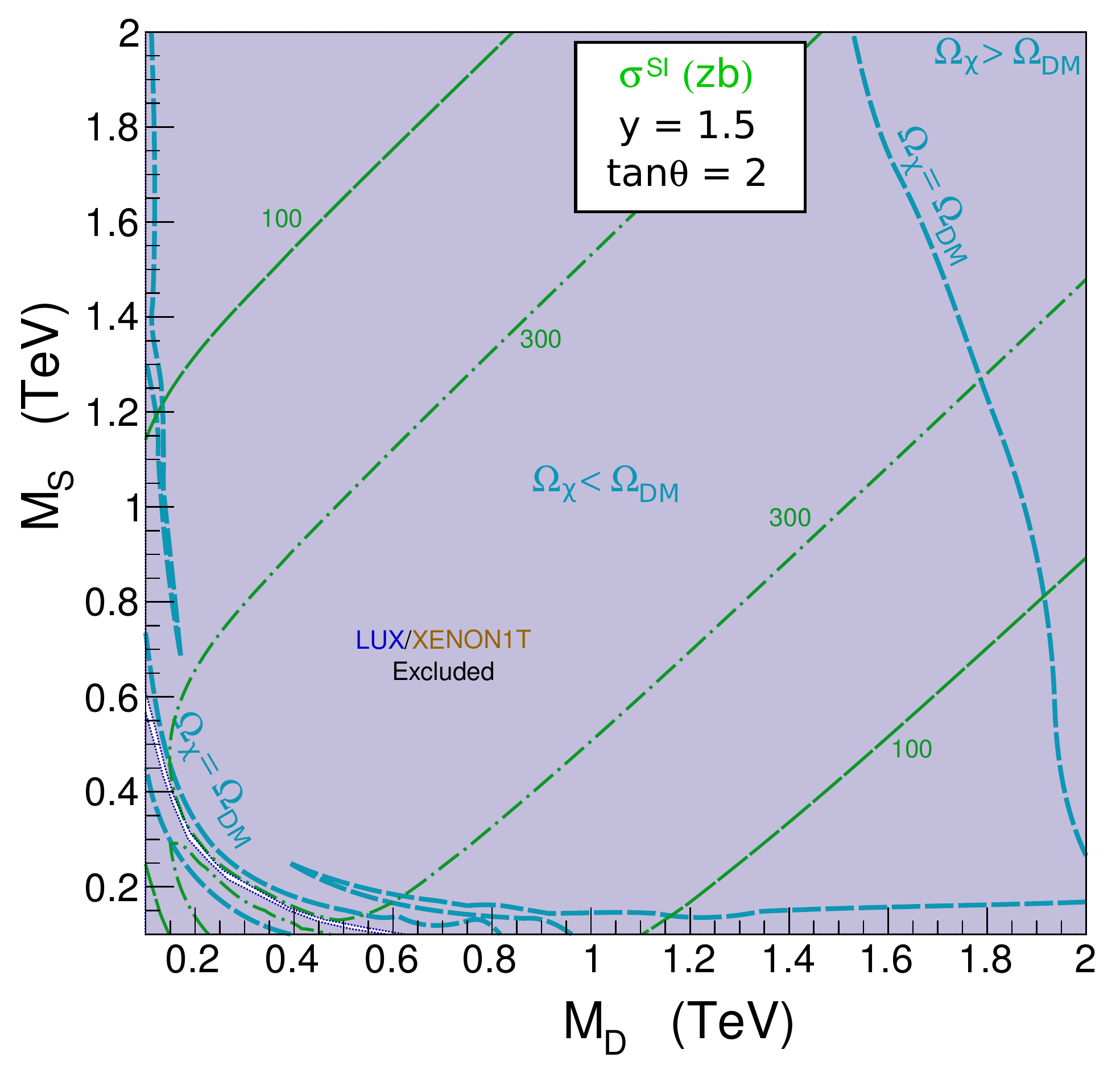}
  \label{fig:Plot1_y=1.5_tantheta=2}}
\subfigure[]{
  \includegraphics*[width=0.45\textwidth]{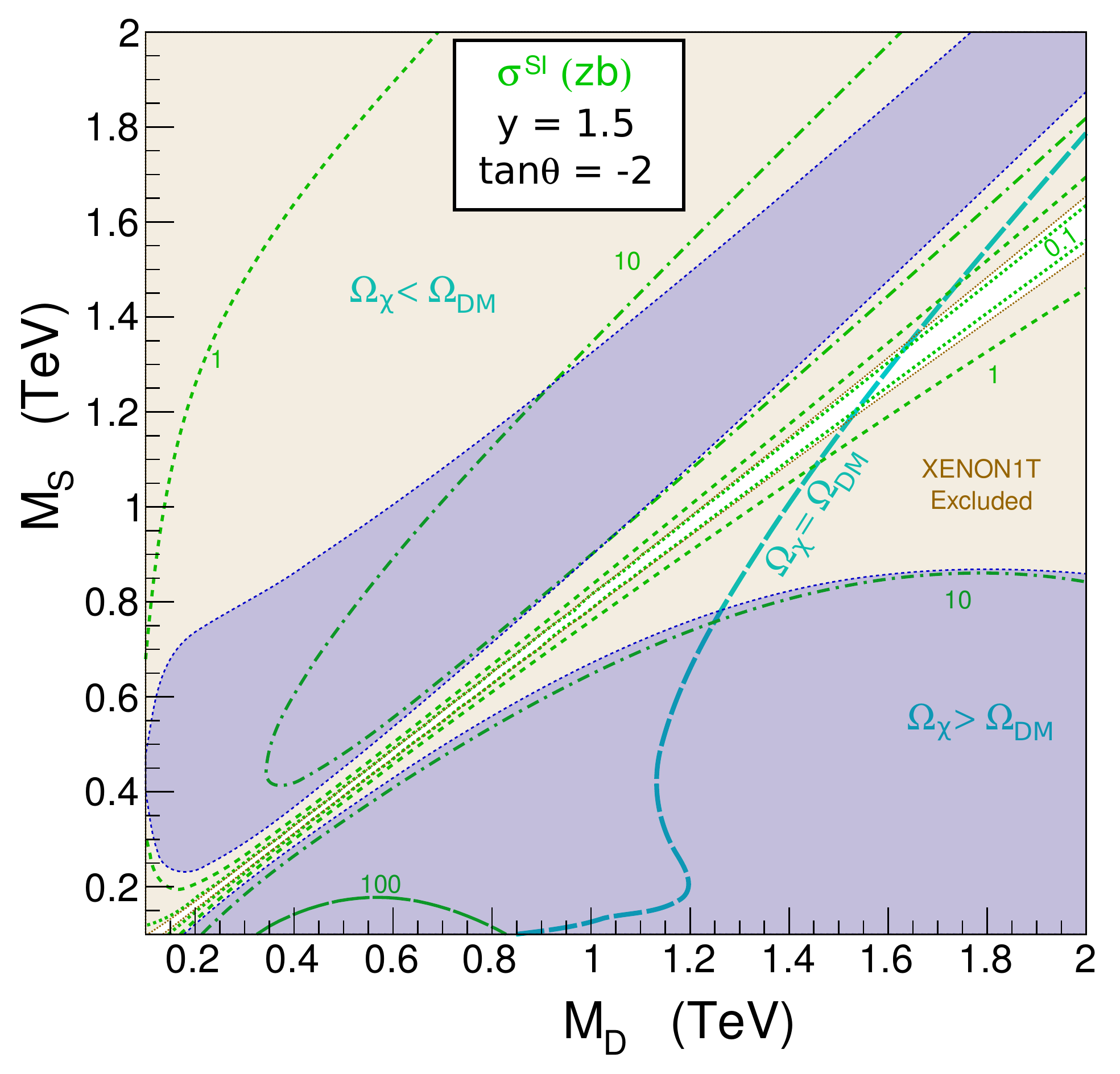}
  \label{fig:Plot1_y=1.5_tantheta=-2}}
\subfigure[]{
  \includegraphics*[width=0.45\textwidth]{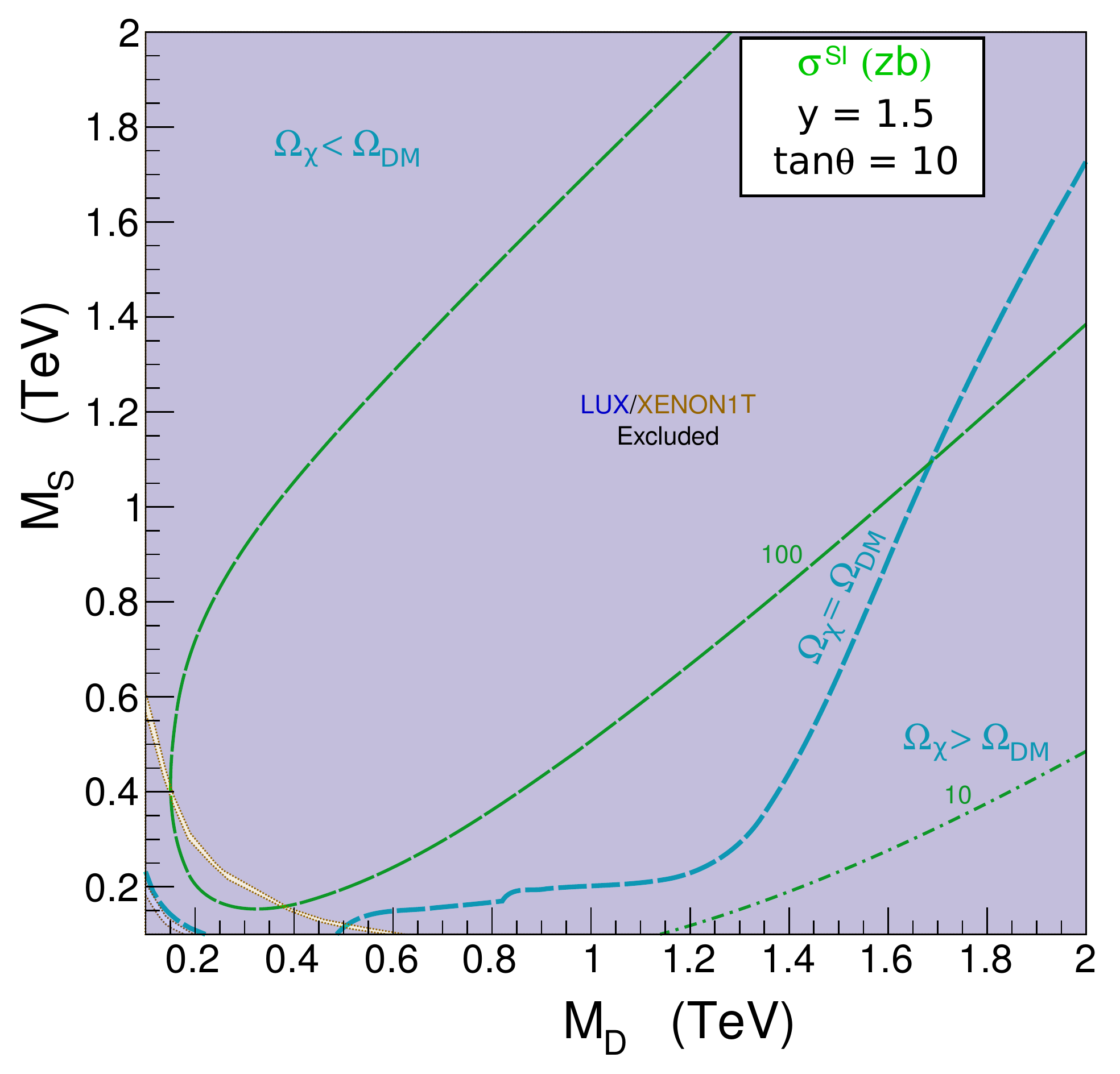}
  \label{fig:Plot1_y=1.5_tantheta=10}}
\subfigure[]{
  \includegraphics*[width=0.45\textwidth]{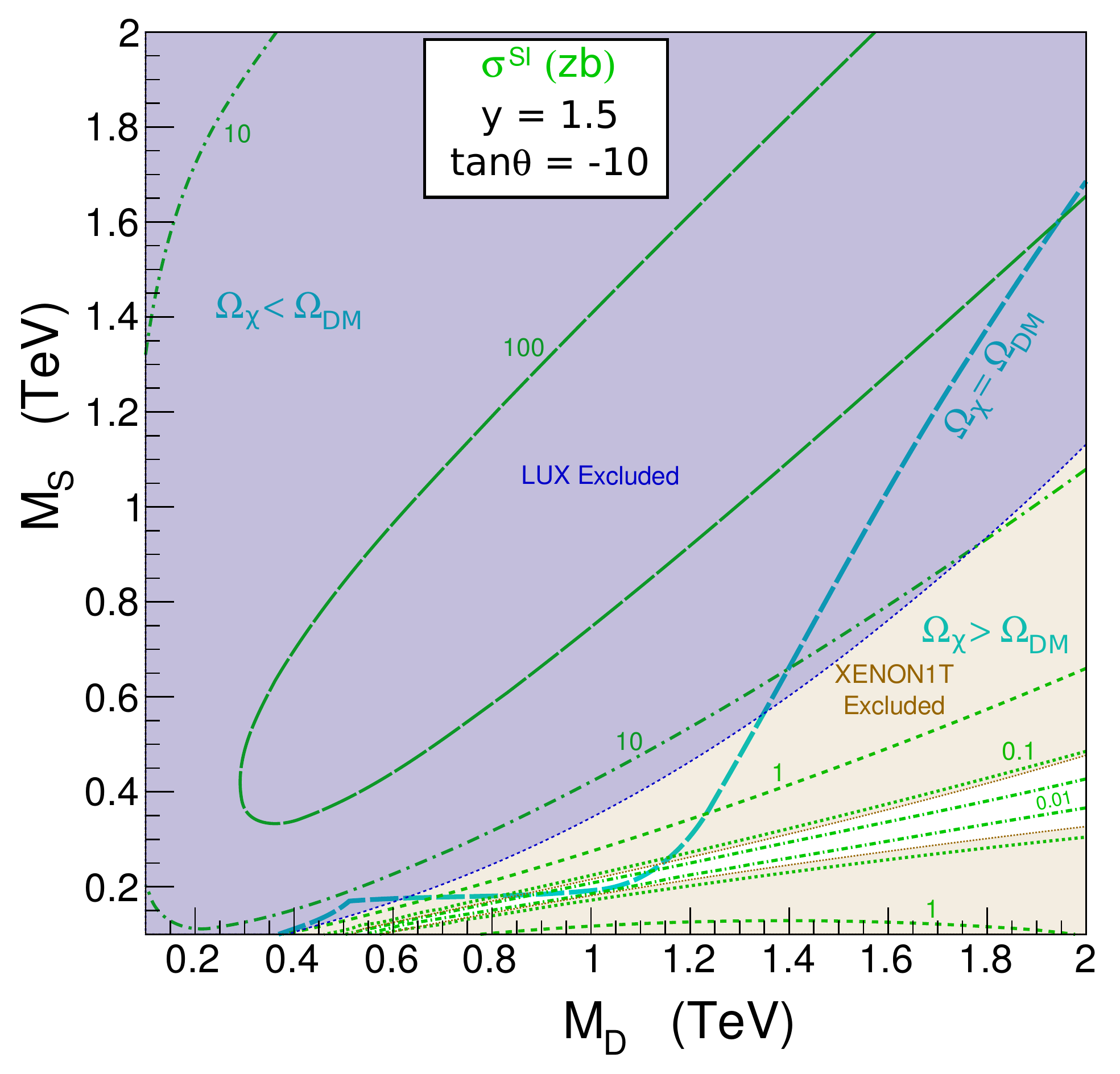}
  \label{fig:Plot1_y=1.5_tantheta=-10}}
\vspace*{-.1in}
\caption{\label{fig:Plot1_y=1.5} \textit{Direct detection prospects
    for $y=1.5$ and $\tan\theta = \pm 2, \pm 10$.}  Contours
  shown are the same as in \figref{Plot1_y=0.3}.}
\end{figure}

As shown in \figref{Plot1_y=1.5}, the situation changes drastically
when the Higgs-DM coupling is increased to $y=1.5$.  There is a
significant shift in the position of the thermal relic line, primarily
toward larger values of $M_D$ for a given value of $M_S$.  This
behavior stems from the increase in the size of the off-diagonal terms
in the mass matrix, resulting in $y v \sim M_{S,D}$ for a larger
portion of the space scanned in the figures.  As a result,
well-tempering is no longer required to produce significant mixing,
and so $M_D$ must be increased significantly to suppress DM
annihilation mediated through the doublet component.  Moreover, a
significant portion of the $\eqomega$ line is located at $M_D >
1~\tev$, implying that annihilation for large $y$ is substantially
stronger than for a pure doublet.   The dominant effect is
enhanced annihilation mediated by a $Z$-boson due to a modification of
the $\chi - Z$ coupling, $i c_{Z\chi\chi} \overline{\chi} \gamma^\mu
\gamma^5 \chi Z_\mu$.  For mixed scenarios
\begin{equation}
c_{Z\chi\chi} = \frac{g}{2 \cos\theta_W} \left( N_{D_1}^2 - N_{D_2}^2
\right)\,,
\end{equation}
where $g$ is the weak gauge coupling and $\theta_W$ is the Weinberg
angle.  In the $y \rightarrow 0$ limit, the singlet and doublet
partially decouple, resulting in $N_{D_1} = N_{D_2} \rightarrow 0$ for
$M_D > M_S$ or $\left|N_{D_1}\right| = \left|N_{D_2}\right|
\rightarrow 1 / \sqrt{2}$ for $M_D < M_S$.  In either case,
$c_{Z\chi\chi} \rightarrow 0$, so long as the very small splitting
necessary to break the doublet into two Majorana states remains.
However, for larger values of $y$ and $\tan\theta \neq \pm 1$, the
degree of $S-D_1$ mixing and $S-D_2$ mixing is different, resulting in
$\left|N_{D_1}\right| \neq \left|N_{D_2}\right|$ and thus
$c_{Z\chi\chi} \neq 0$.  The size of $c_{Z\chi\chi}$ grows with the
degree of mixing, and thus with $y$, though it of course can never
exceed $g / 2 \cos\theta_W$.  For large $y$, enhanced Higgs-mediated
annihilation also becomes important, though such processes remain
p-wave suppressed and are somewhat smaller than the $Z$-mediated
diagrams.

While stronger annihilation channels shifts the $\eqomega$ line to
larger values of $M_D > 1~\tev$ over most of the range shown, for
sufficiently large $M_S \sim 3-4~\tev$, the thermal relic line
asymptotes to the $M_D \approx 1~\tev$ line corresponding to
nearly-pure doublet DM.  In the right panels, $\tan\theta < 0$ and
$\chxx$ is suppressed, so $Z$-mediated processes dominate
annihilation.  As a result, annihilation becomes markedly stronger
approaching the $M_S \approx M_D$ line, and the $\eqomega$ line moves
to larger $M_D$ with increasing $M_S$.  The coupling $c_{Z\chi\chi}$
is marginally smaller for $\tan\theta = -2$ than for $\tan\theta =
-10$, but not sufficiently so to produce significantly different
results.  For $\tan\theta > 0$, however, the $\eqomega$ line shows
distinctly different behavior for $\tan\theta = 2$.  Because
$Z$-mediated processes vanish identically in the limit of
$c_{Z\chi\chi} \rightarrow 0$ for $\tan\theta \rightarrow \pm 1$, this
coupling is suppressed in the neighborhood of $\tan\theta \approx 1$.
As a result, the $\eqomega$ contour is located at a lower value of
$M_D$ for $\tan\theta = 2$.  Higgs-mediated annihilation compensates
somewhat at lower $M_S$, and indeed the $\tan\theta = 2$ case has the
largest Higgs coupling of the cases shown, but Higgs-mediated
annihilation does not exhibit the same degree of enhancement for $M_D
\approx M_S$.  As a result, the $\eqomega$ line is located at large
$M_D$ for small $M_S$ and shifts to smaller values of $M_D$ as $M_S$
increases.  The $\tan\theta =10$ case, however, is far enough from
$\tan\theta = 1$ that the behavior seen in $\tan\theta < 0$ is
recovered.

Moreover, the availability of annihilation channels is important for
low masses -- in the lower two panels of \figref{Plot1_y=1.5}, the
$\eqomega$ line is nearly horizontal over a range of several hundred
GeV for $M_D$ in multiple plots for $M_S \sim m_t$.  The increase of
$M_D$ for nearly fixed $M_S$ in this region effectively reduces the
mixing angle to compensate for an enhancement in annihilation from the
opening of the $\chi\chi \rightarrow t\bar{t}$ annihilation
channel. Likewise, the singlet-doublet mixing terms are so large that
$\mchi \rightarrow 0$ for $\tan\theta > 0$ and $M_{S,D}$ of order a
few hundred GeV.  For $\tan\theta < 0$ this does not occur due to the
structure of the mass matrix.  In the upper left panel of
\figref{Plot1_y=1.5}, $y=1.5, \tan\theta = 2$ and the $\eqomega$ line
is roughly parallel to the low mass DM contours.  This occurs because
the relic abundance is strongly controlled by the opening of
annihilation channels when the DM mass crosses the bottom quark, $W$
boson, and top quark thresholds.

Regarding direct detection, for $y=1.5$, the raw value of $\sigmaSI$
is increased by more than an order of magnitude relative to $y=0.3$.
LUX excludes the entire region shown for $\tan\theta > 0$, except for
a small region at low $M_{S,D}$ for which $\mchi \lesssim m_b$ is
below the experimental threshold.  However, this low mass region
retains a large Higgs coupling and will be excluded by constraints on
the invisible decay width of the Higgs~\cite{Espinosa:2012im}.  Blind
spot cancellations occur for $\tan\theta < 0$, however, with large
portions of the thermal relic line remaining viable given LUX limits.
XENON1T still has much greater reach for $\tan\theta < 0$, with the
increased Higgs coupling resulting in only the blind spot and an
associated small portion of the thermal relic line evading XENON1T
sensitivity.

\subsection{Exclusion Plots (Thermal Relic)}
\label{sec:sdfermionthermal}

Next, we consider singlet-doublet fermion DM in a reduced parameter
space.  We focus on thermal relic DM by restricting to regions that
saturate the observed DM abundance, defined by $\Omegachi = \OmegaDM$.
Concretely, we fix one of parameters $(M_S, M_D, y, \theta)$ to
saturate the thermal relic constraint.

\begin{figure}[tb]
\subfigure[]{
  \includegraphics*[width=0.45\textwidth]{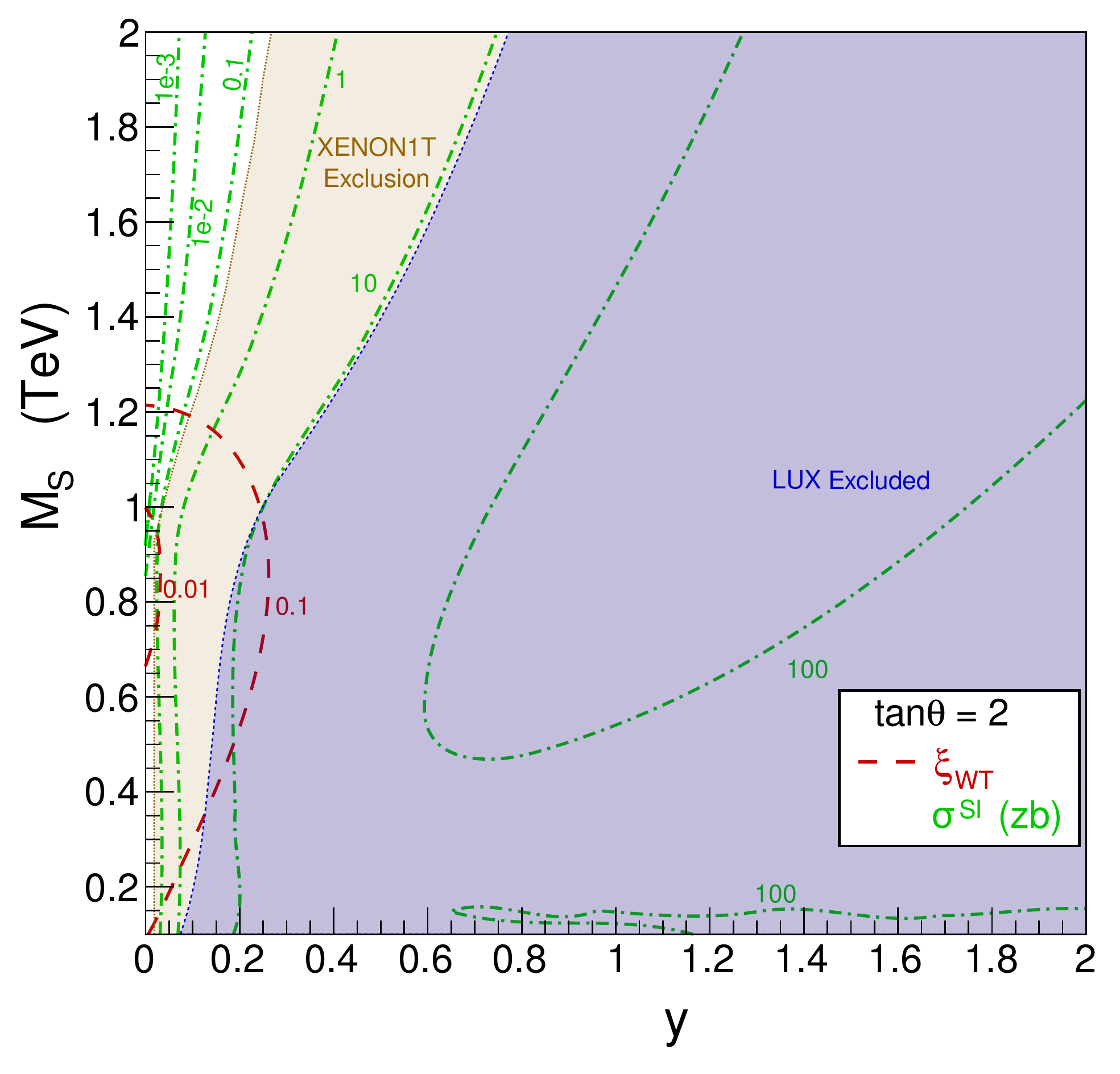}
  \label{fig:Plot2a_tantheta=2}}
\subfigure[]{
  \includegraphics*[width=0.45\textwidth]{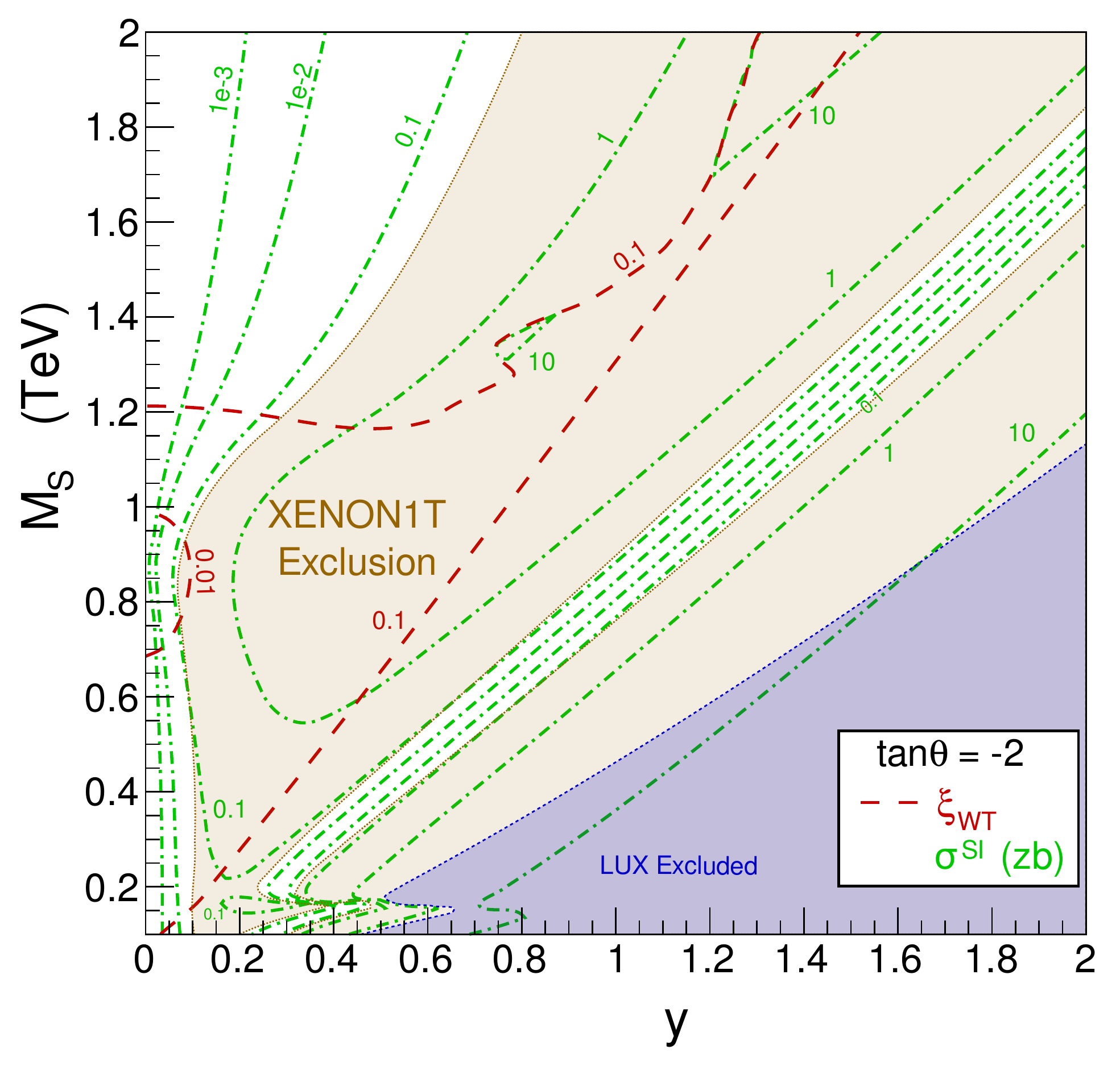}
  \label{fig:Plot2a_tantheta=-2}}
\subfigure[]{
  \includegraphics*[width=0.45\textwidth]{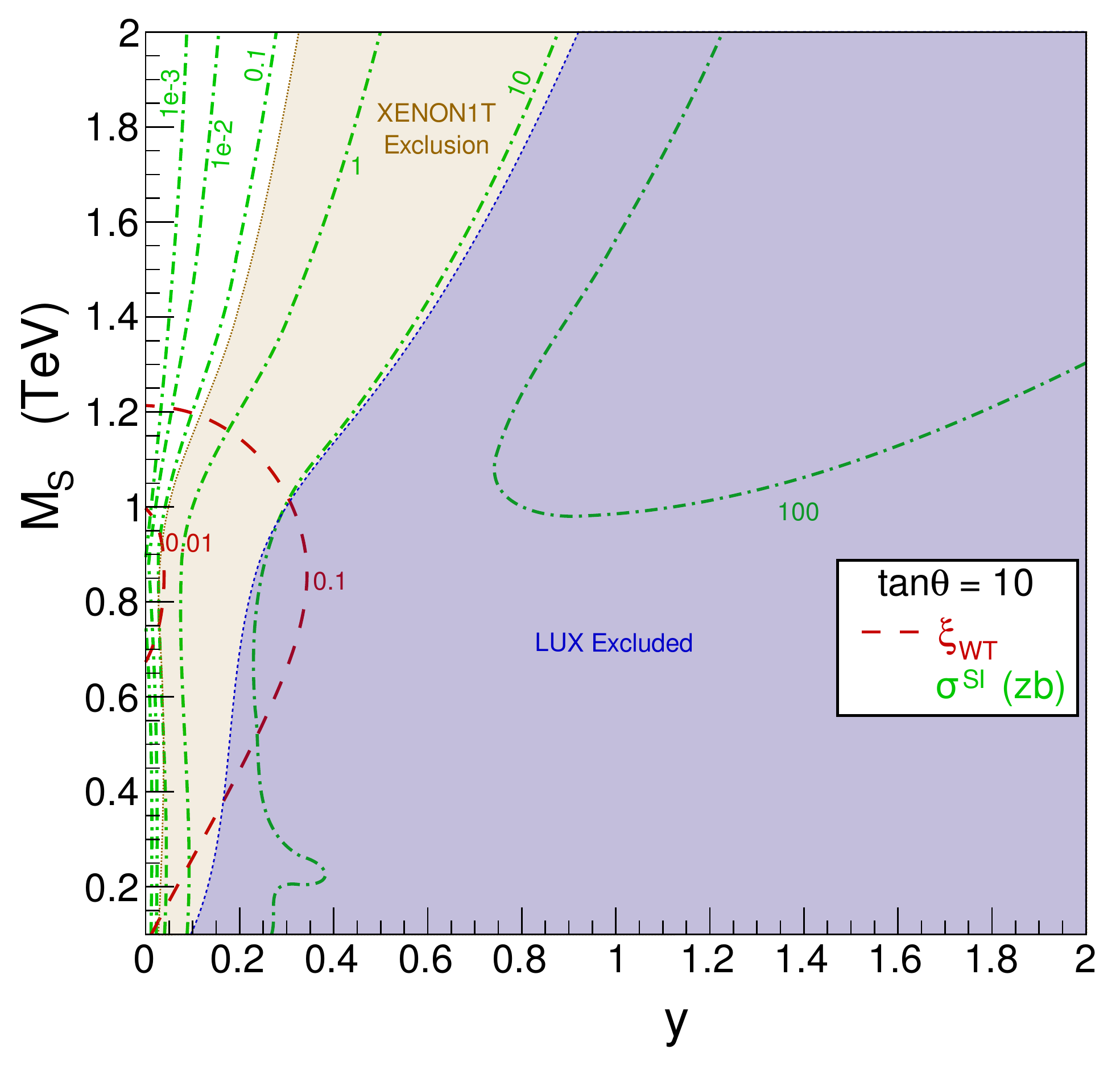}
  \label{fig:Plot2a_tantheta=10}}
\subfigure[]{
  \includegraphics*[width=0.45\textwidth]{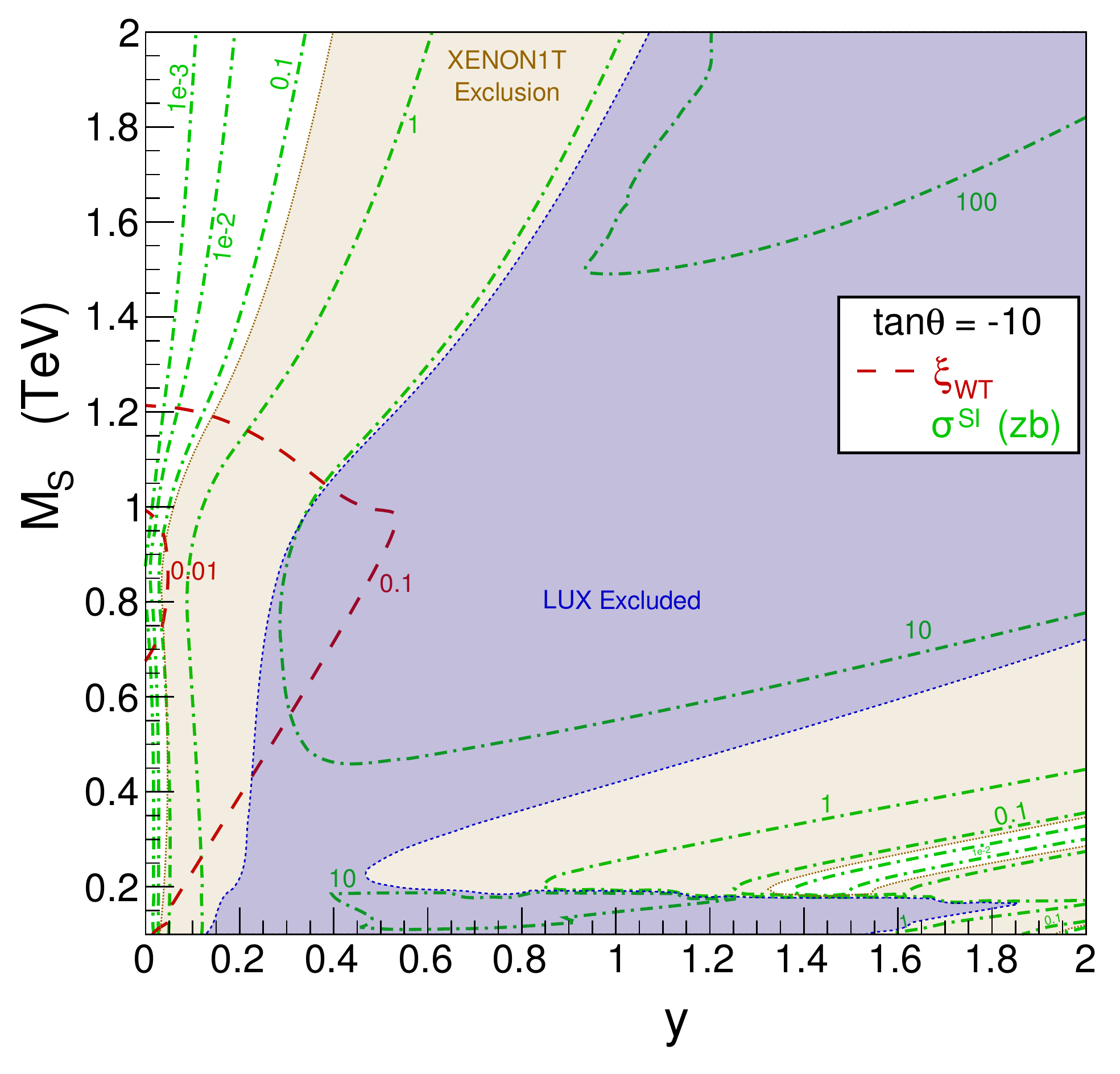}
  \label{fig:Plot2a_tantheta=-10}}
\vspace*{-.1in}
\caption{\label{fig:Plot2a} \textit{Well-tempering and direct
    detection prospects for fixed $\Omegachi=\Omega_{DM}$ as a
    function of $y$ and $M_S$.}  Shown are the well-tempering measure
  $\xiwt$ (red dashed) and direct detection cross-section $\sigmaSI$
  in zb (green, various styles).  Also shown are the regions currently
  excluded by LUX (blue shaded) and the projected reach of XENON1T
  (gold shaded).}
\end{figure}

\figref{Plot2a} depicts constraints in the $(y,M_S)$ plane, fixing
$\tan\theta$ to various values and setting $M_D$ so that $\eqomega$.
As before, the sign of $\tan\theta$ has a significant effect on DM
properties.  For $\tan\theta > 0$ there are no blind spots, and for $y
\lesssim 0.5$, the cross-section $\sigmaSI$ grows almost monotonically
with $y$ and decreases with $M_S \sim \mchi$.  In this region, DM
annihilation results primarily from gauge interactions typical to
nearly-pure doublet DM.  For $y \gtrsim 0.5$, however, Higgs-mediated
and $Z$-mediated diagrams become important, with their relative
contributions increasing with $y$.  As a result, $\sigmaSI$ actually
decreases for increasing $y$ for $y \gtrsim 1$, due to a reduction in
the mixing to limit the Higgs- and $Z$-mediated annihilation processes
that increase with $y$.  A small region of larger $\sigmaSI > 100~\zb$
is also present for $M_S \lesssim 200~\gev$ in the $\tan\theta = 2$
case, associated with large splitting which drives $\mchi < m_W$ and
thus requires a larger Higgs coupling to produce appropriate levels of
annihilation.  \figref{Plot2a} also shows that direct detection limits
from LUX bound $y \lesssim 0.3$ for $M_S \lesssim 1~\tev$ in the
$\tan\theta > 0$ case, while XENON1T reach covers all regions except
for extremely small $y$ for any $M_S \lesssim 1$ TeV for $\tan\theta$,
and even up to $M_S = 2~\tev$ XENON1T limits $y \lesssim 0.3$.

For $\tan\theta < 0$, both the general suppression in the $\chxx$ and
the existence of blind spots significantly affect the results.  The
shape of the $\sigmaSI$ contours for small $y$ is similar to the
$\tan\theta > 0$ case, but the values are significantly reduced.  For
$\tan\theta = -2$ in particular this suppression is sufficient that no
LUX exclusion exists for $y \lesssim 0.5$ even for small $M_S$, and
only bounds $M_S \gtrsim 1~\tev$ even for $y=2$.  Even with this
suppression, however, XENON1T still has strong constraining power down
to small values of $y$ for $M_S \lesssim 1~\tev$.  The position of
blind spots which also avoid XENON1T projected bounds is set by $M_D$,
and they occur primarily at larger values of $y$ since such values are
required to produce sufficient splitting between $M_S$ and $M_D$ for
the values of $\tan\theta$ shown.  These blind spots are more
difficult to accommodate at large $\tan \theta$, where they require a
larger hierarchy between $\mchi$ and $M_D$.

These plots also depict the well-tempering parameter, $\xiwt$, which
characterizes the level of degeneracy required among the mixed neutral
states in order to yield the correct relic abundance.  The existing
LUX bound still allows for thermal relics with moderate
well-tempering, $\xiwt \sim 0.1$, especially for $\tan \theta < 0$.
It also places moderate limits on $y \lesssim 0.3$ for $\tan \theta >
0$, and allows larger values of $y$ for $\tan \theta < 0$.  However,
XENON1T will strongly alter the available parameter space of this
simplified model.  Projected limits from XENON1T typically constrain
$\xiwt \lesssim 0.1$ throughout for $\mchi \lesssim 1$ TeV, with
$\xiwt \lesssim 10^{-2}$ for $700~\gev \lesssim \mchi \lesssim
1~\tev$, which is a substantial degree of tuning .  A notable
exception is exactly on blind spots, where there is no direct
detection limit; however, residing on these cancellation regions
constitutes an additional tuning.

\begin{figure}[tb]
\subfigure[]{
  \includegraphics*[width=0.48\textwidth]{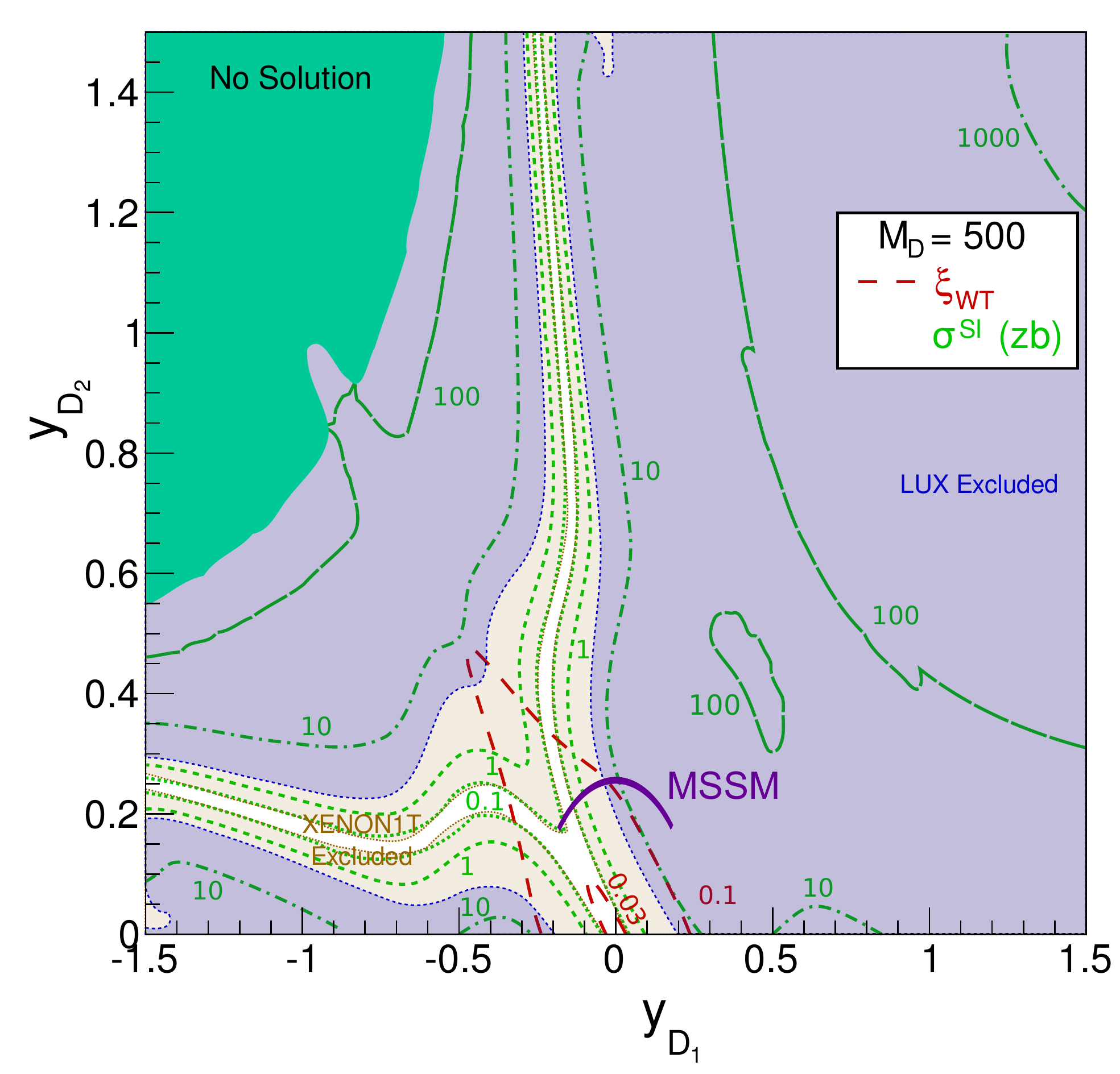}
  \label{fig:Plot2b_MD=500}}
\subfigure[]{
  \includegraphics*[width=0.48\textwidth]{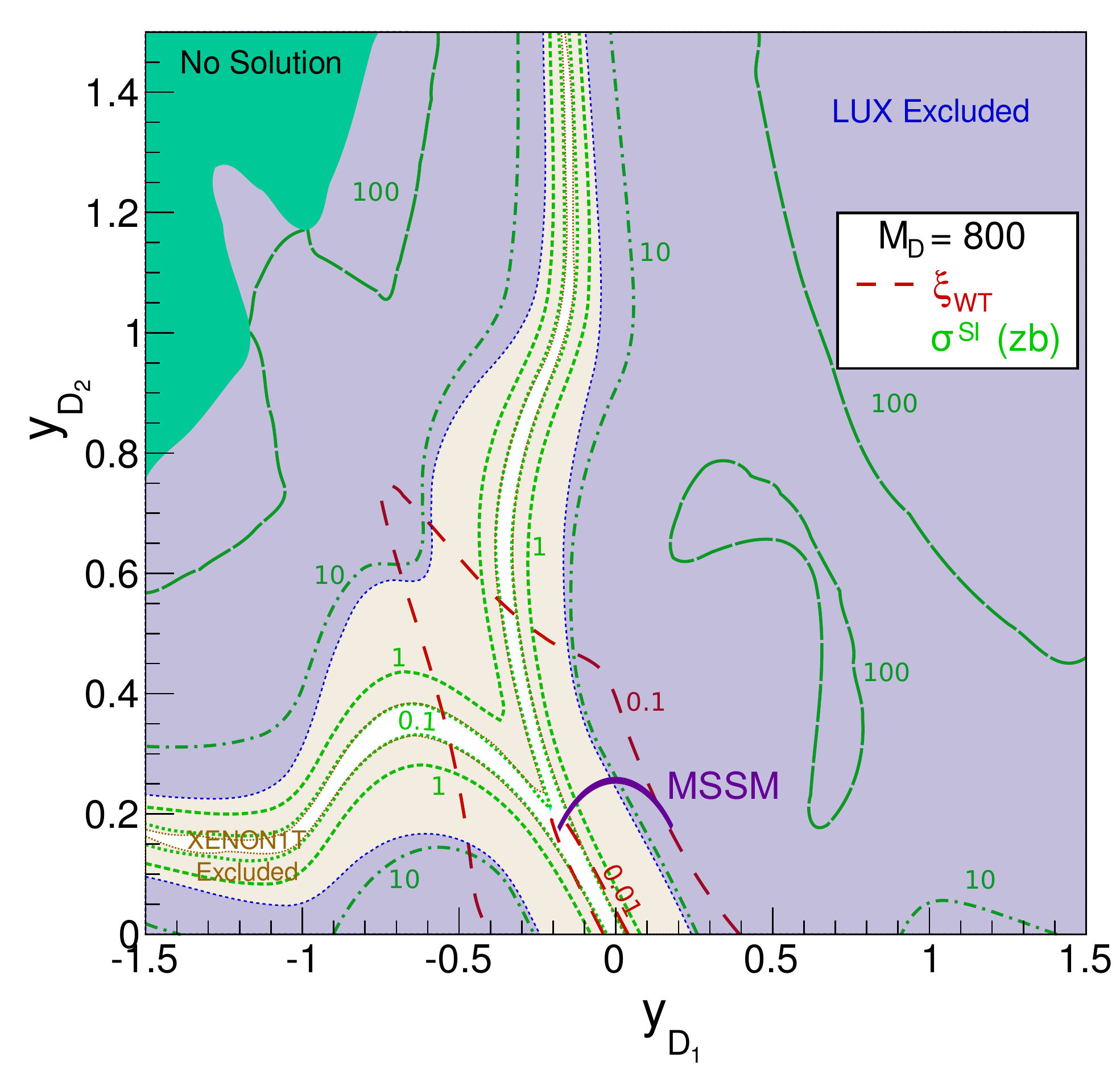}
  \label{fig:Plot2b_MD=800}}
\vspace*{-.1in}
\caption{\label{fig:Plot2b} \textit{Well-tempering and direct
    detection prospects for fixed $\Omegachi=\Omega_{\rm DM}$ as a function of $y_{D_1}$ and $y_{D_2}$.}
  Contours shown are the same as in \figref{Plot2a}.}
\end{figure}

\figref{Plot2b} depicts direct detection constraints in the thermal
relic parameter space, only shown in the $(y_{D_1}, y_{D_2})$ plane at
fixed $M_D$, and fixing $M_S$ to satisfy the relic constraint.  For
certain regions within this plane there are multiple solutions for
$\Omegachi = \OmegaDM$; in such cases the solution with the largest
value of $\mchi$ is displayed\footnote{Secondary solutions for
  $\tan\theta > 0$ tend to be at a lower mass and thus more strongly
  constrained by direct detection; secondary solutions for $\tan\theta
  < 0$ have similar masses and $\sigmaSI$ to the primary solutions.}.
Here $\tan\theta = y_{D_2} /y_{D_1}$ is positive (negative) in the
right (left) quadrant.  As in \figref{Plot2a}, $\xiwt$ is small for
small $y$ and grows as $y$ becomes large.  However, in contrast with
\figref{Plot2a} where $M_D$ was allowed to grow to offset the
increasing value of $y v$, in \figref{Plot2b} $M_D < 1~\tev$
throughout.  As such, some mixing is still required to dilute the
annihilation strength, so $M_S$ is never even approximately decoupled.
For $\tan\theta > 0$ ($y_{D_1} > 0$) the large $y$ region induces a
significant mass splitting, and is thus dominated by small values of
$\mchi \lesssim m_t$ and correspondingly large $\sigmaSI$, closing
annihilation channels and reducing coannihilation to compensate for
enhanced annihilation strength.  For $\tan\theta < 0$ ($y_{D_1} < 0$)
the induced mass splitting is significantly smaller, particularly for
$\tan\theta \approx -1$, and so no such low mass solutions exist.
This produces a region for which $\Omegachi=\OmegaDM$ is unachievable
by varying $M_S$ alone.  The size of both the small mass and no
solution regions decreases with increasing $M_D$.

At present, LUX strongly constrains this simplified model for $y_{D_1}
> 0$, only allowing a small region at small $y$; however, once again a
blind spot exists for $y_{D_1} < 0$ which leaves open large swaths of
available parameter space.  This blind spot extends to arbitrarily
large values of $y$ (limited, of course, by perturbativity), though
the trajectory varies with mass at large $y$.  However, after XENON1T,
only models very close to this blind spot will still be viable.
Regarding well-tempering, for $y_{D_1} > 0$, only a very small value
of $y$ will be allowed after XENON1T, corresponding to a significant
degree of well-tempering, $\xiwt \lesssim 0.03$.  For $y_{D_1} < 0$,
well-tempering is still substantial, but is alleviated when residing
precisely on the blind spot.

\figref{Plot2b} also contains a line corresponding to bino-Higgsino DM
in the MSSM.  In this case, $y =g'/\sqrt{2} $, and $\theta = \beta$ is
restricted to be in the range $-\pi/4$ to $+\pi/4$ required for a
perturbative top Yukawa coupling.  Here, negative values of $\beta$
correspond to a negative $\mu$ parameter in the MSSM.

\begin{figure}[tb]
\subfigure[]{
  \includegraphics*[width=0.48\textwidth]{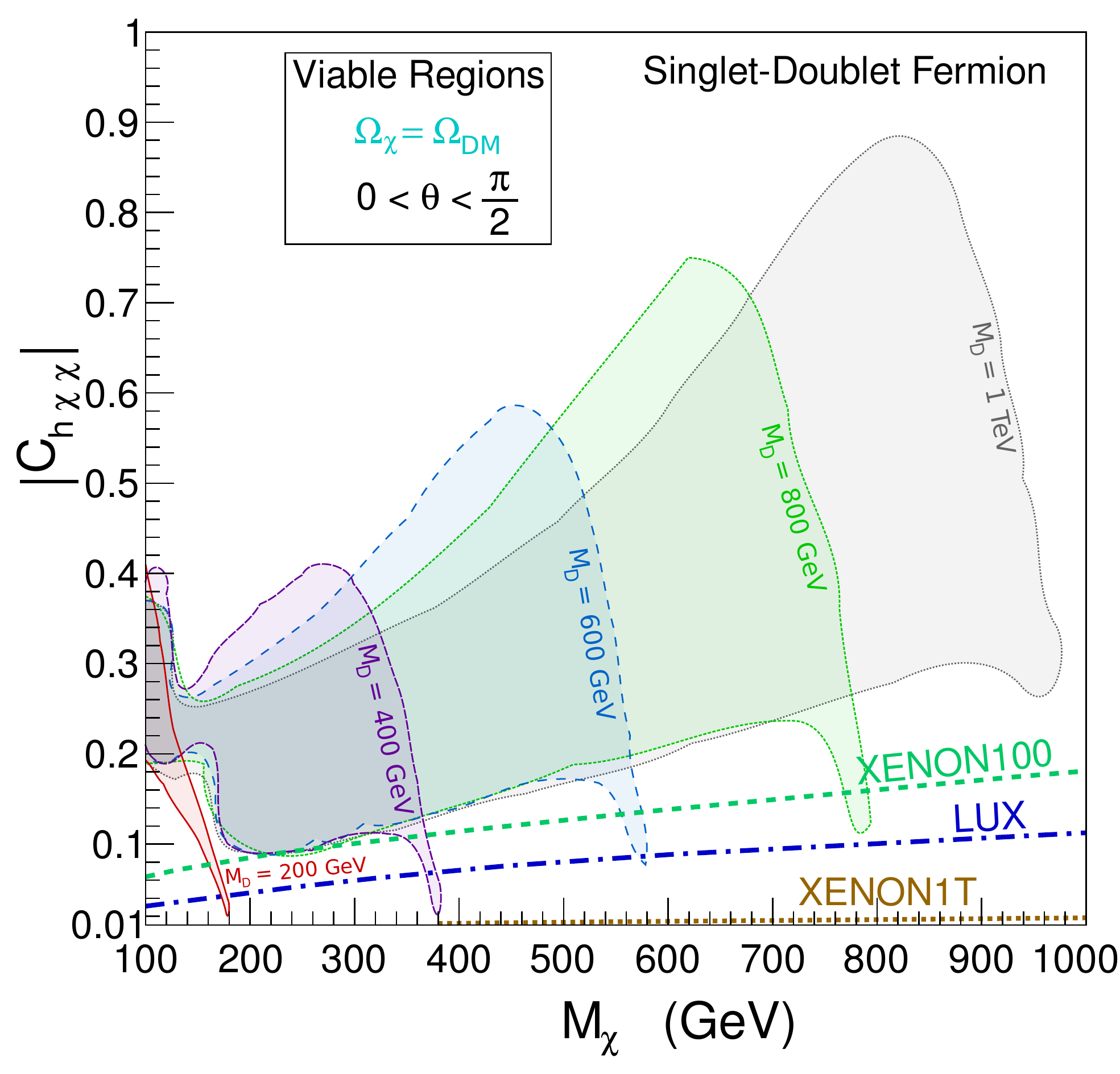}
  \label{fig:Plot2c_thetaPos}}
\subfigure[]{
  \includegraphics*[width=0.48\textwidth]{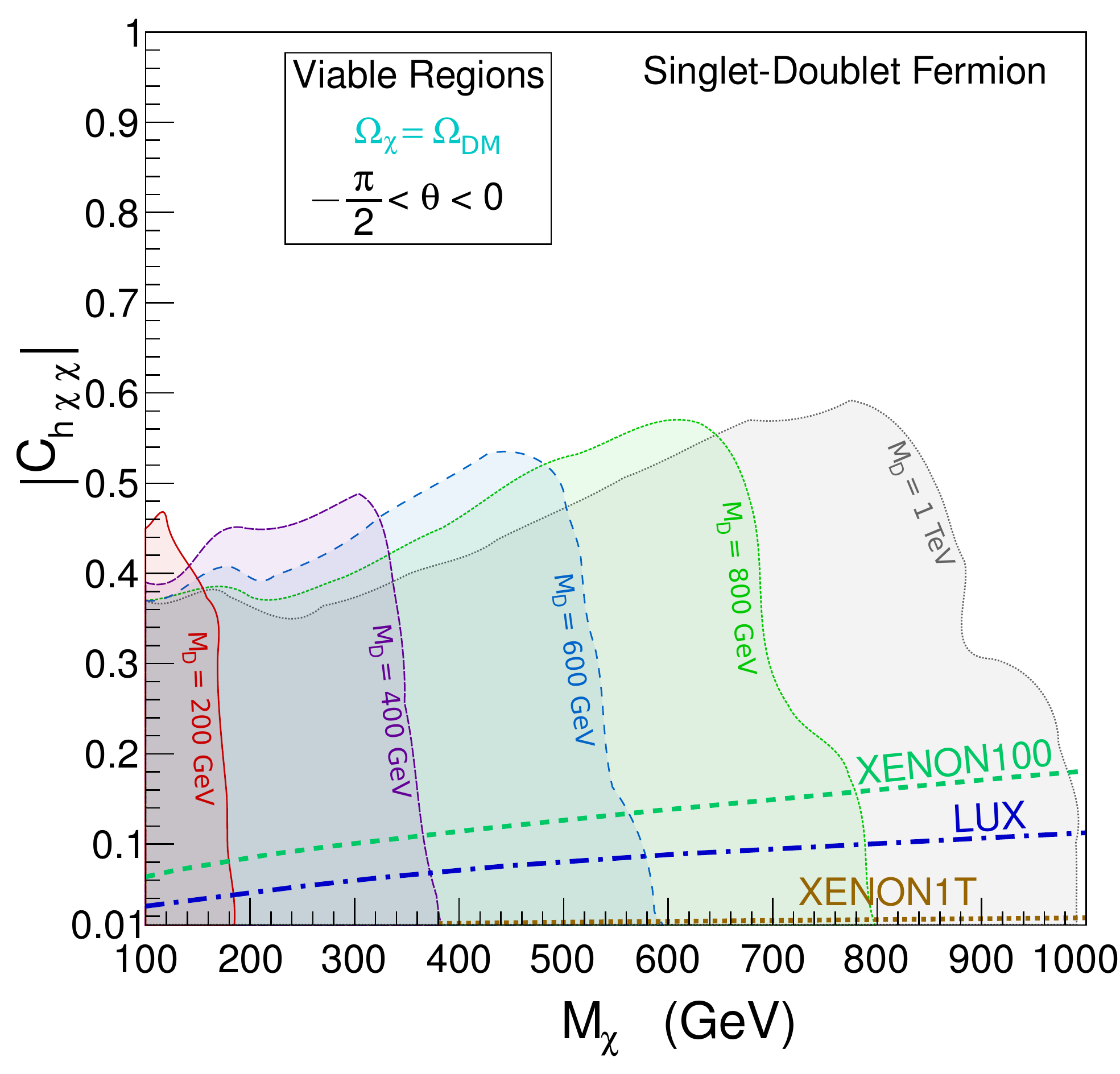}
  \label{fig:Plot2c_thetaNeg}}
\vspace*{-.1in}
\caption{\label{fig:Plot2c} \textit{Viable regions with
    $\Omegachi=\Omega_{\rm DM}$ as a function of $\mchi$ and
    $c_{h\chi\chi}$.}  Regions are shown for various values of $M_D$
  and for the ranges $0 < \theta <\pi/2$ (left) and $-\pi/2 < \theta <
  0$ (right).  Also shown are upper limits on $|\chxx|$ allowed by
  XENON100 (teal dashed), LUX (blue dot-dashed) and projected upper
  limits from XENON1T (gold dashed).  }
\end{figure}

Lastly, \figref{Plot2c} shows the theoretically available parameter
space with $\eqomega$ in terms of more physical quantities: the DM
mass, $\mchi$, and the DM-Higgs coupling, $\partial \mchi /\partial v
= c_{h \chi\chi}$. By definition, the blind spot is defined at the
bottom of the plot where $|c_{h\chi\chi}|=0$.  We can substitute the
model inputs $M_S$ and $y$ for these more physical parameters using
the relations
\begin{eqnarray}
M_S & = & \mchi - \frac{1}{2} \chxx v \frac{ \Delta M^2 B }{
  \Delta M^2 B + \chxx v \left( \Delta M^2 + 2 \mchi B
  \right)/2} \\
y^2 & = & - \frac{\chxx}{v B^2} \left( \Delta M^2 (M_S-M_\chi) + B
\left( \Delta M^2 + 2 \mchi (M_S-M_\chi) \right) \right) \\
\Delta M^2 & = & M_D^2 - \mchi^2 \\
B & = & \mchi + M_D \sin 2\theta \,.
\label{eq:replace}
\end{eqnarray}
We have marginalized over all values of $\tan\theta$ positive (left)
and negative (right) to fix $\Omegachi=\OmegaDM$ for various values of
$M_{\chi^\pm} = M_D$.  In keeping with comparison to direct detection
the sign of $\chxx$ is also left undetermined.  This issue will be
discussed further in \secref{sdfermionthermaldd}.

For $\tan\theta > 0$ both the upper and lower edges of the allowed
range increase with $\mchi$.  The position of the upper edge depends
strongly on $M_D$, and increases more quickly for small $M_D$, while
the position of the lower edge is almost independent of $M_D$.  This
behavior is modified as $\mchi$ and $M_D$ become degenerate, with the
upper edge peaking at $\mchi \approx M_D - 100~\gev$ and dropping
steadily thereafter.  The lower edge also drops below the general
trend line in the same mass range.  Both effects are due to
coannihilation, which requires a spectrum with relatively suppressed
Higgs couplings.  The presence of a general lower bound on $\chxx$ is
due to the lack of a blind spot for $\tan\theta > 0$.  However, a
small region with $\chxx \rightarrow 0$ exists for $\tan\theta > 0$ in
the $\mchi \rightarrow M_D$ limit as small cross-terms are required
for mixing.  Note that the absence of this region in
\figref{Plot2c_thetaPos} is due to a binning artifact.

For $\tan\theta < 0$, the behavior of the upper edge is similar to
that for $\tan\theta > 0$, though found at somewhat different values
of $|\chxx|$.  There is no lower bound, however, due to the
possibility of blind spot cancellations for $\tan\theta < 0$.  In
terms of constraints, LUX currently excludes most of the $\tan\theta >
0$ regions, including everything above the generic lower limit, and
excludes a large portion of the $\tan\theta < 0$ regions as well.
XENON1T will further exclude most of the parameter space of this
simplified model, though for the cases of extreme coannihilation and
blind spots for $\tan\theta < 0$, XENON1T will not be able to
eliminate the model.

\subsection{Exclusion Plots (Thermal Relic and Marginal Exclusion)}
\label{sec:sdfermionthermaldd}

So far, we have reduced the dimensionality of the parameter space by
imposing $\eqomega$.  However, we can further reduce the parameter
space by fixing $\sigmaSI$, which is controlled by $\chxx$ coupling.
In particular, we will now focus on the parameter space of thermal
relic DM that exactly saturates present LUX limits or projected
XENON1T reach (in addition to fixing $\Omegachi=\OmegaDM$).  This
defines a space of marginally excluded, thermal relic DM models which
can accommodate the observed abundance today.  This space of
marginally excluded thermal relics represents the class of models at
the edge of direct detection limits.  Thus, comparing the space fixed
to XENON1T projected constraints against that fixed to current LUX
limits unambiguously demonstrates the effect of improved direct
detection sensitivity.  This is aided by the additional benefit of
reducing the parameter space to only two dimensions; hence the entire
surviving parameter space can be described in a single plane.
Moreover, because the direct detection limits bound $\sigmaSI$ from
above, the values $y$ shown can be identified as the maximal DM-Higgs
coupling allowed by a given experiment.  The corresponding $\mchi$
should be interpreted as the minimal DM mass allowed by direct
detection, at least for $\mchi > 200$ GeV, when experimental limits on
the cross-section begin to scale linearly as $\propto
\mchi$.

\begin{figure}[tb]
\subfigure[]{
  \includegraphics*[width=0.45\textwidth]{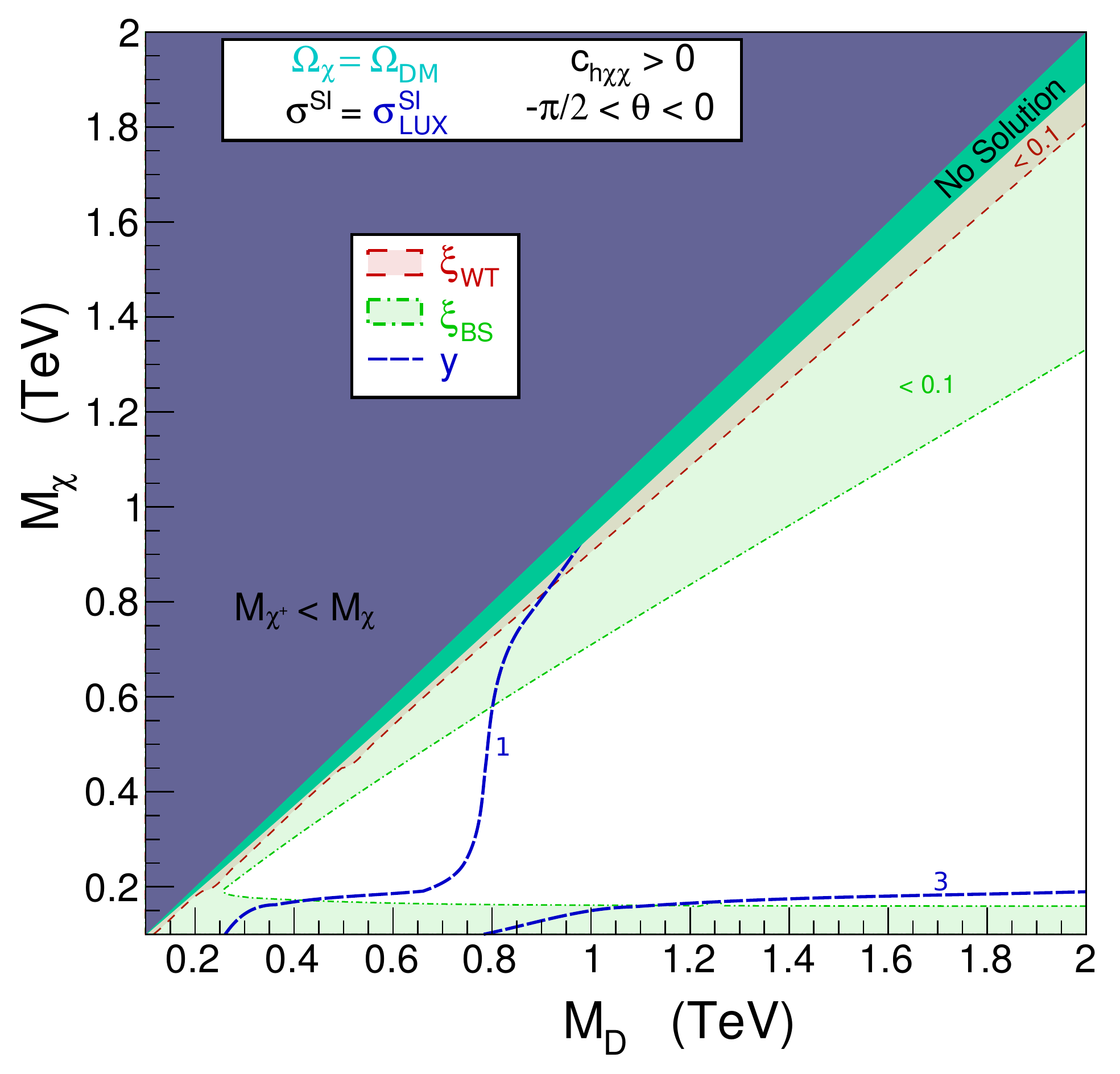}
  \label{fig:Plot3a_LUX_chxxPos_thetaNeg}}
\subfigure[]{
  \includegraphics*[width=0.45\textwidth]{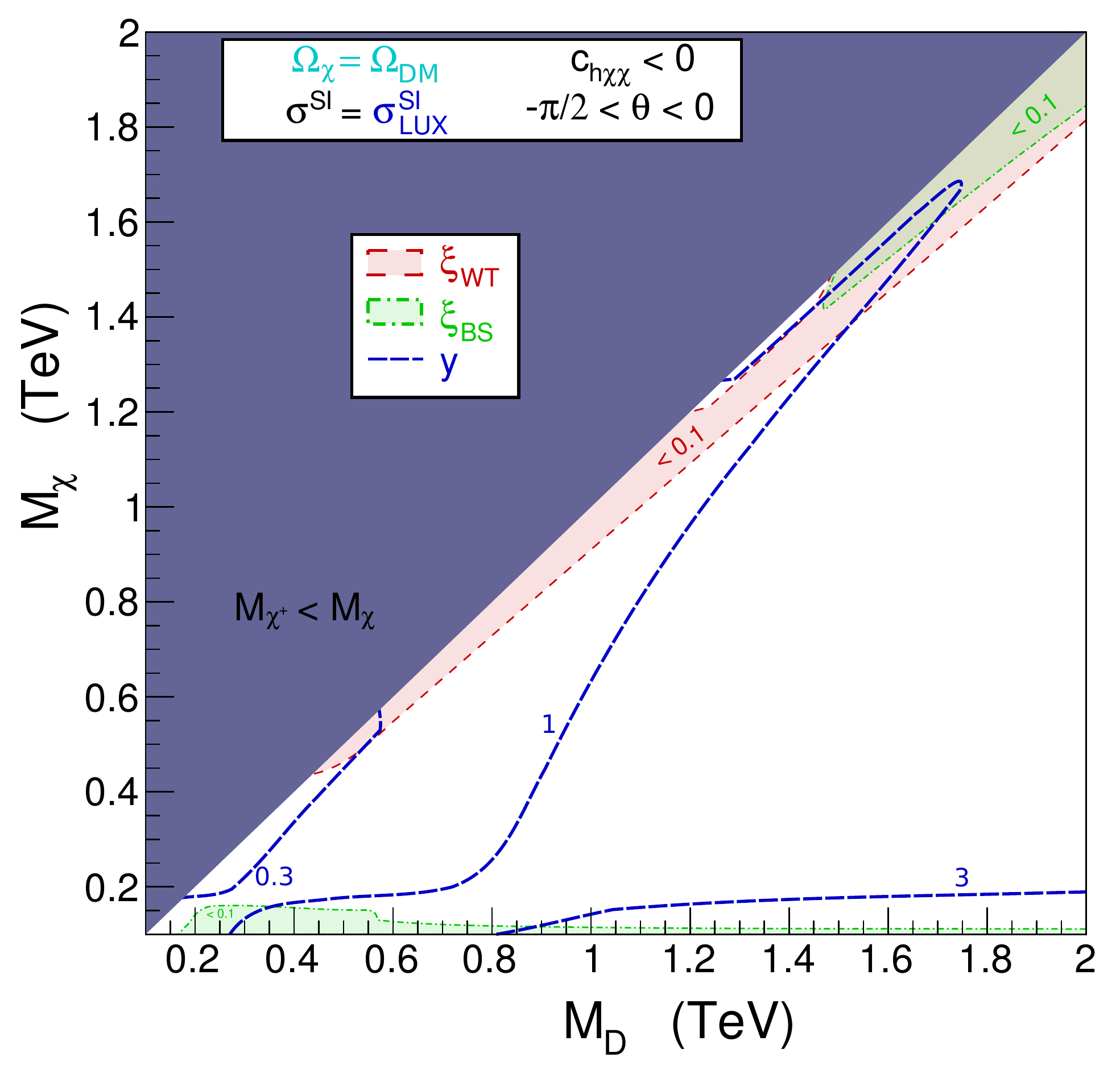}
  \label{fig:Plot3a_LUX_chxxNeg_thetaNeg}}
\subfigure[]{
  \includegraphics*[width=0.45\textwidth]{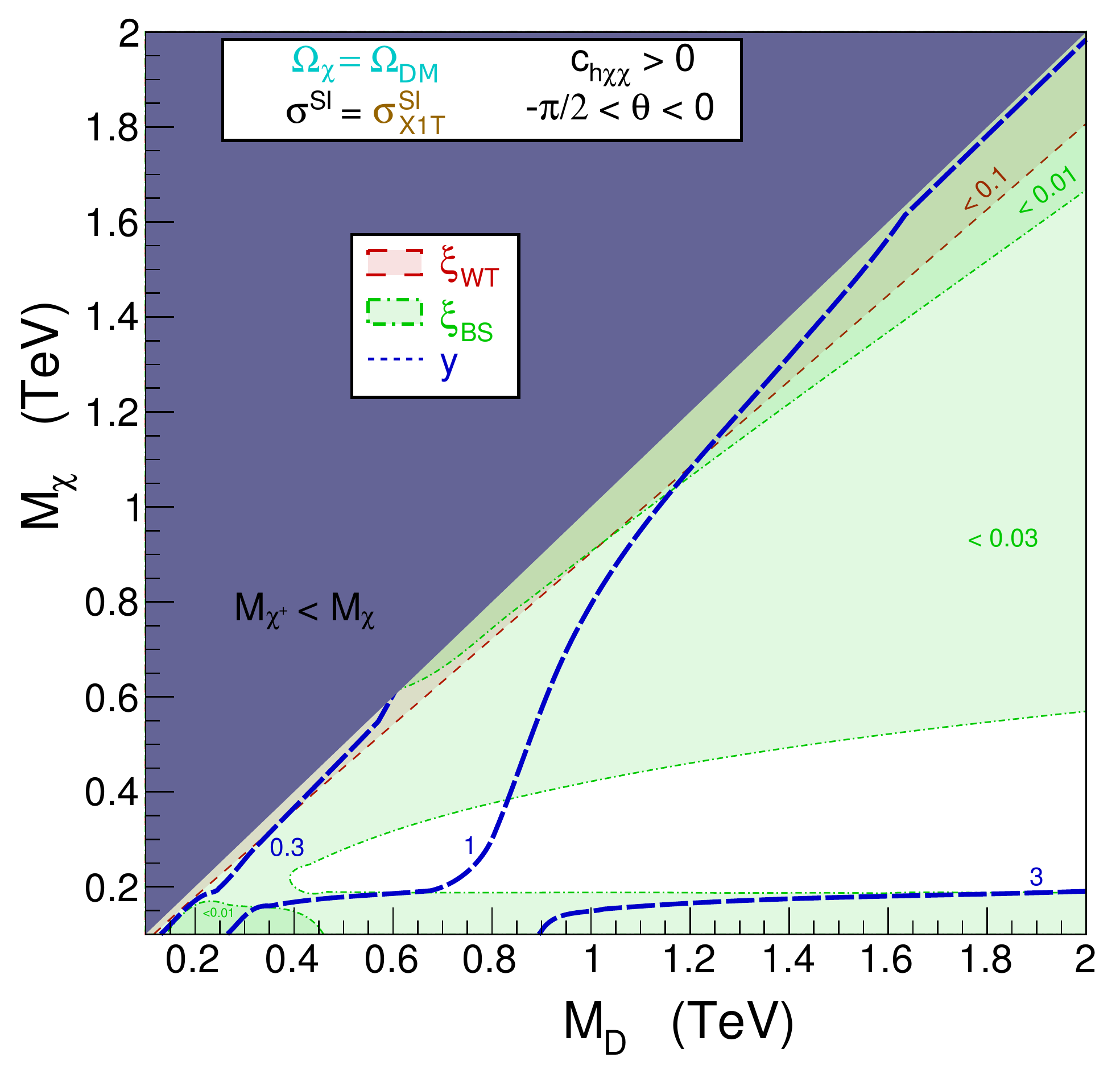}
  \label{fig:Plot3a_XENON1ton_chxxPos_thetaNeg}}
\subfigure[]{
  \includegraphics*[width=0.45\textwidth]{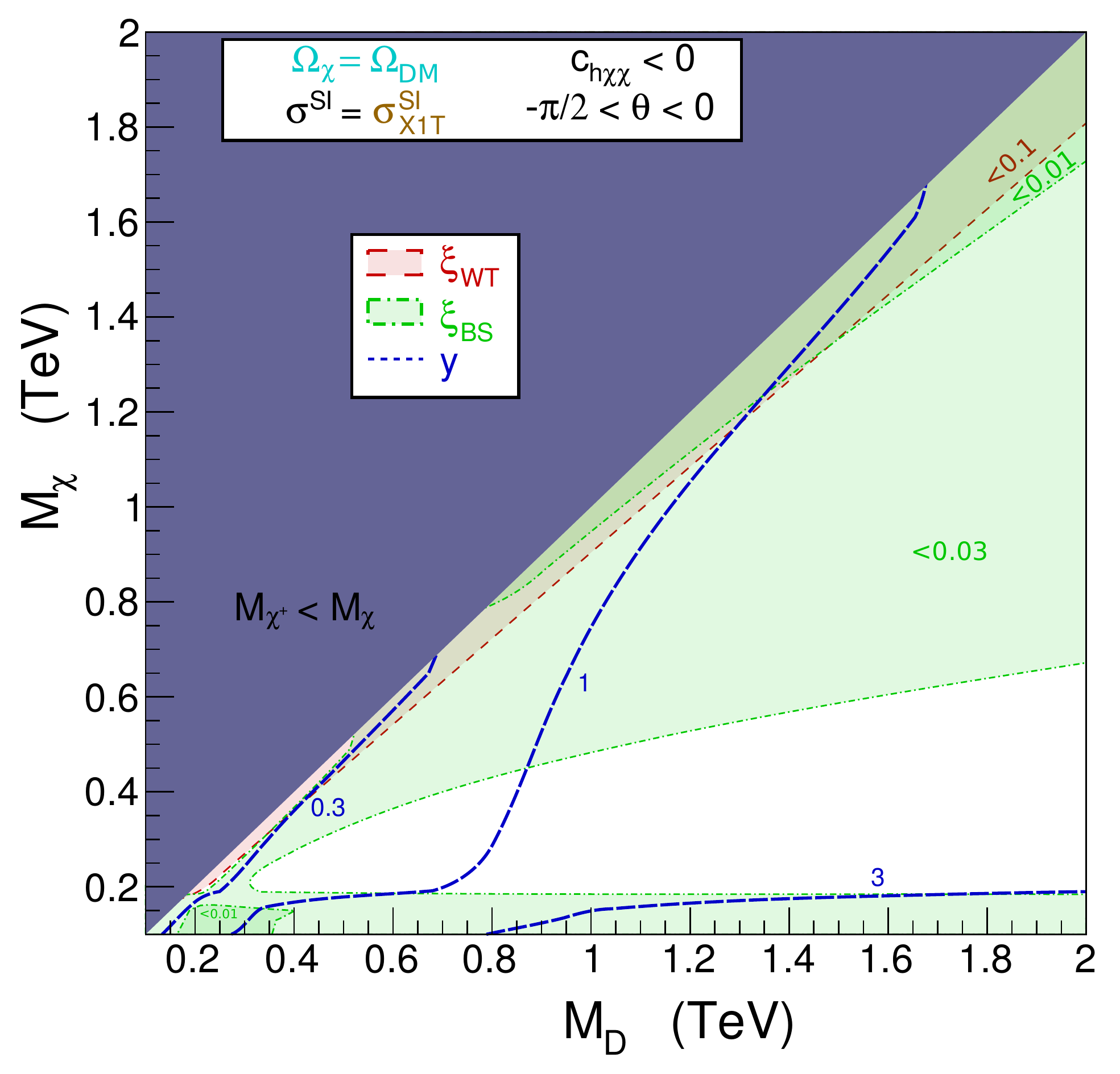}
    \label{fig:Plot3a_XENON1ton_chxxNeg_thetaNeg}}
\vspace*{-.1in}
\caption{\label{fig:Plot3a} \textit{Tuning for models with fixed
    $\Omegachi$ and $\sigmaSI$ as a function of $M_D$ and $\mchi$.}
  Shown are points which satisfy the relic density constraint and
  exactly saturate $\sigmaSI$ limits from LUX (top) or projected reach
  at XENON1T (bottom).  Regions with significant tuning are shown for
  $\xiwt$ (red shaded) and $\xibs$ (green shaded).  The white regions
  correspond to parameters with lesser tuning than the shaded regions:
  $\xiwt, \xibs \gtrsim 0.1$ for the upper panels, and $\xiwt \gtrsim
  0.1$ and $0.1 \gtrsim \xibs \gtrsim 0.03$ in the lower panels.  The
  left panels correspond to $c_{h\chi\chi} > 0$ and the right panels
  to $c_{h\chi\chi} < 0$; in both cases $\tan\theta < 0$. }
\end{figure}

First we consider the parameter space of marginally excluded thermal
relics in the $\left(M_D, \mchi \right)$ plane for $\tan\theta < 0$.
For every pair of $\left(\mchi, M_D \right)$, there are a maximum of
four viable solutions -- each with $c_{h \chi\chi}$ either positive or
negative and $\tan\theta$ either positive or negative.  From
\eqref{eq:sdfermionhiggs},
\begin{equation}
\chxx \propto - \left(\mchi + M_D \sin 2\theta \right)\,,
\end{equation}
so it is negative definite for $\tan\theta$ positive, leaving a
maximum of three viable solutions.  Furthermore, from the discussion
in \secref{sdfermionthermal} it is clear that the viable region for
$\tan\theta > 0$ occurs only for small values of $y$ and significant
well-tempering, and thus would occupy only a small sliver along the
$\mchi = M_D$ line in this plane; an explicit scan confirms this
assumption.

The two remaining interesting cases are shown in \figref{Plot3a} for
$\tan\theta < 0$ and $\chxx > 0$ (left) or $\chxx < 0$ (right).  For
each point the required value of $y$ is shown, along with regions
showing various degrees of well-tempering and blind spot tuning.
Models marginally excluded by LUX (top) have minimal tuning, with most
regions subject to $\xiwt, \xibs \gtrsim 0.1$ (shown in white) for
both positive and negative $\chxx$.  For models within the marginal
sensitivity of XENON1T, however, $\xibs < 0.1$ throughout the entire
plane for both signs of $\chxx$, and $\xibs < 0.03$ throughout most of
the plane.  While the parameter space remains viable for DM, it is
clear that some degree of fine-tuning is necessary for consistency
with XENON1T projected limits.

\begin{figure}[tb]
\subfigure[]{
  \includegraphics*[width=0.45\textwidth]{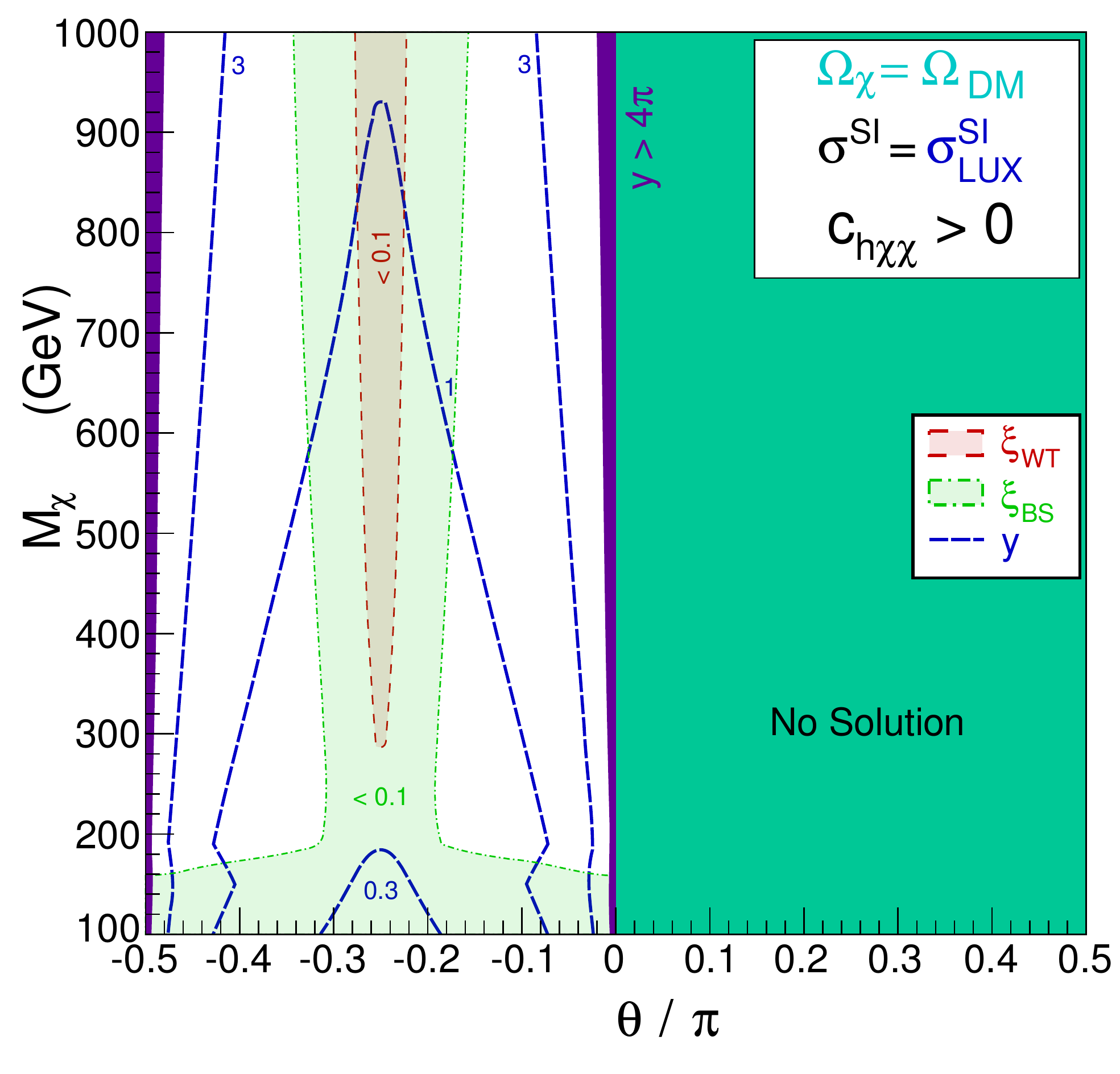}
  \label{fig:Plot3b_XENON100_chxxPos}}
\subfigure[]{
  \includegraphics*[width=0.45\textwidth]{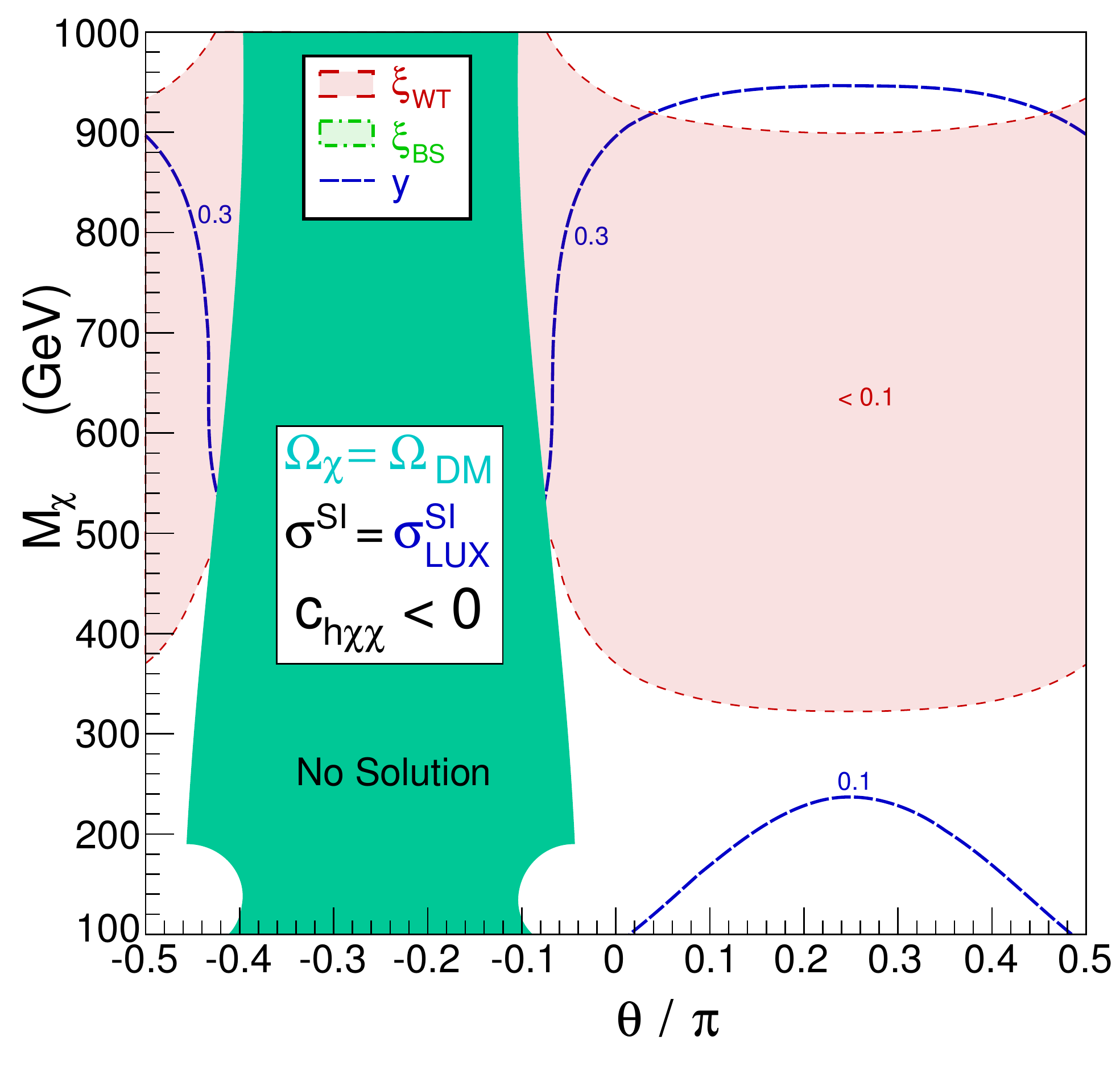}
  \label{fig:Plot3b_XENON100_chxxNeg}}
\subfigure[]{
  \includegraphics*[width=0.45\textwidth]{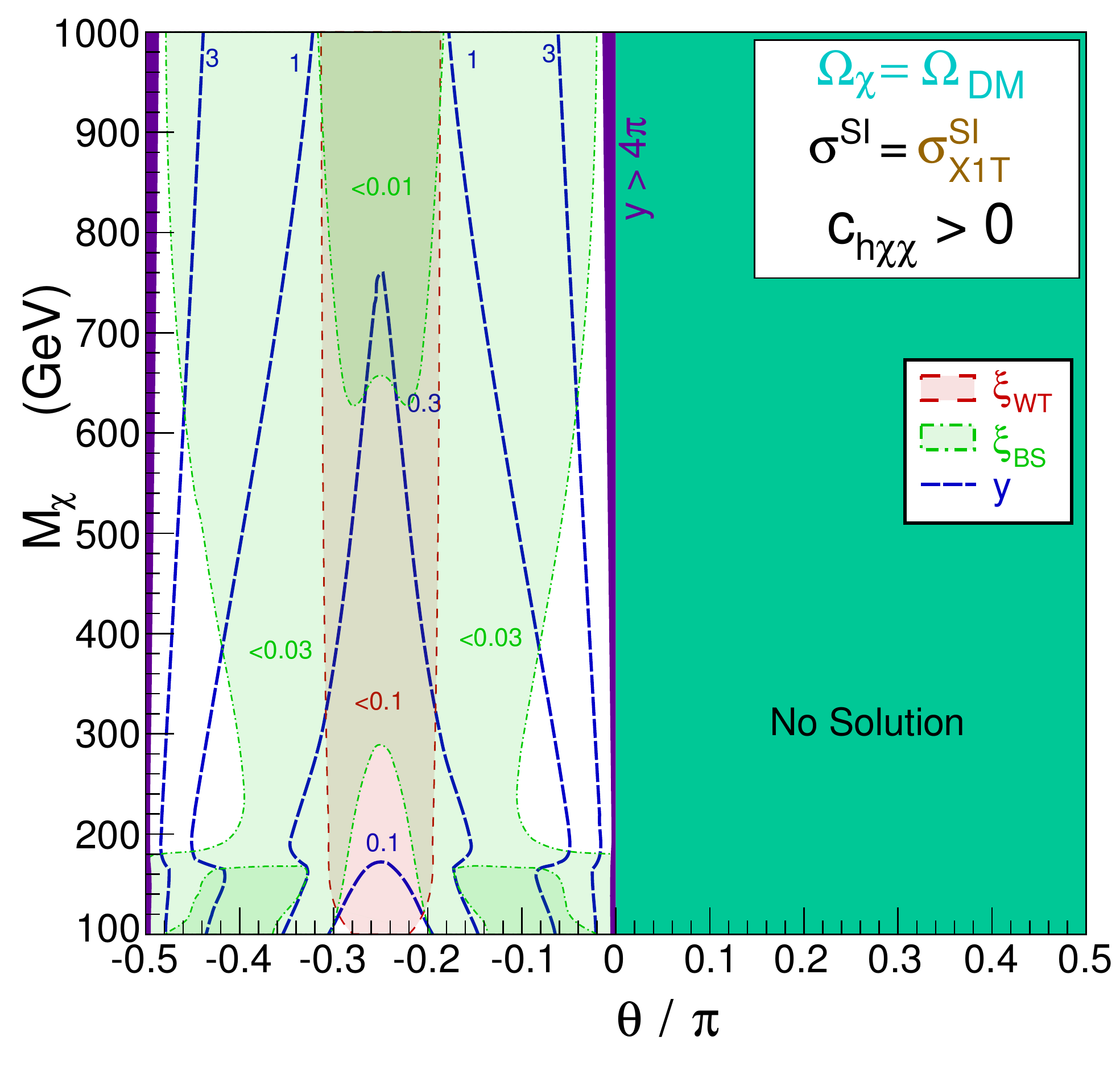}
  \label{fig:Plot3b_XENON1ton_chxxPos}}
\subfigure[]{
  \includegraphics*[width=0.45\textwidth]{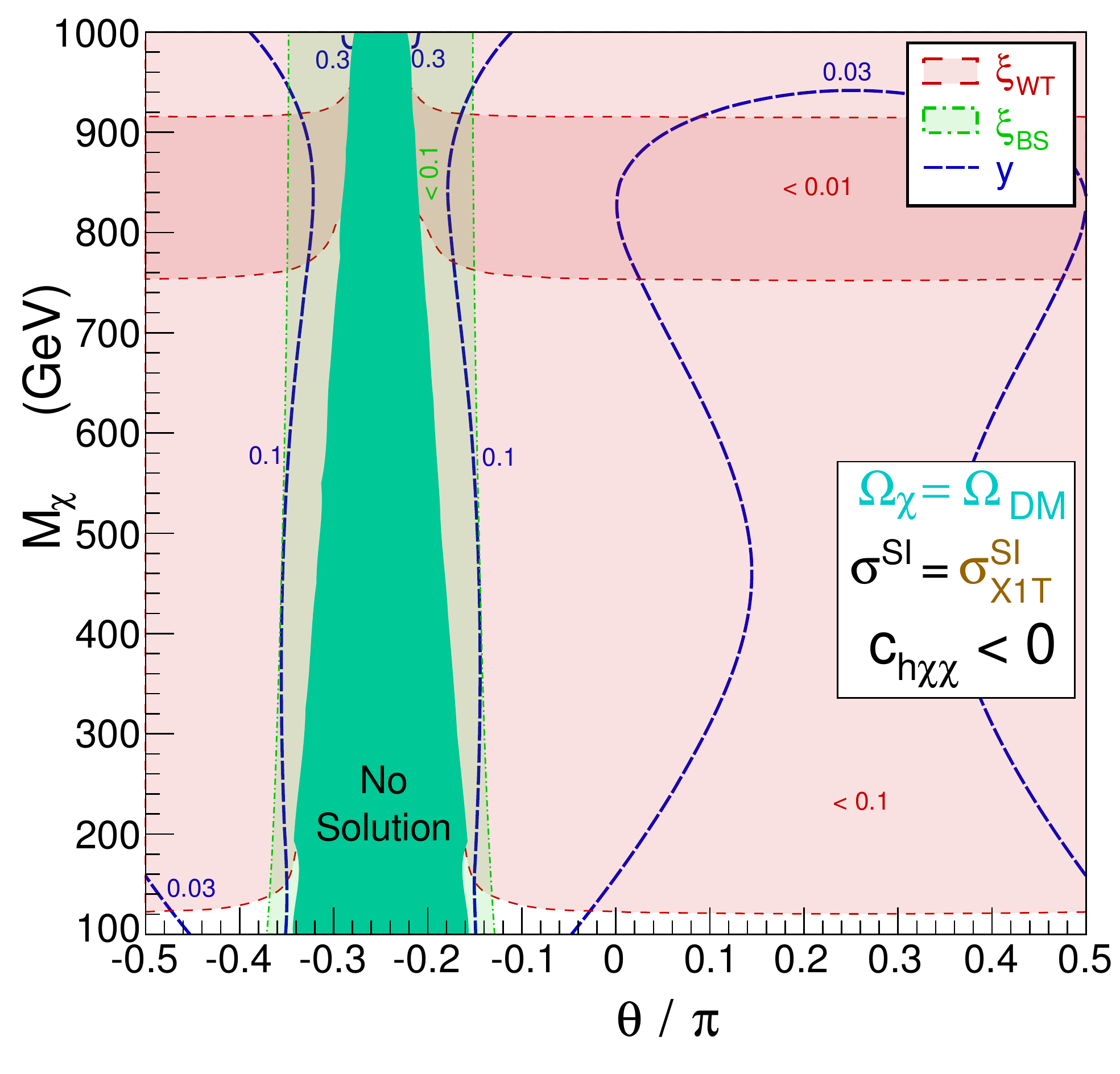}
  \label{fig:Plot3b_XENON1ton_chxxNeg}}
\vspace*{-.1in}
\caption{\label{fig:Plot3b} \textit{Tuning for models with fixed
    $\Omegachi$ and $\sigmaSI$ as a function of $\theta$ and $\mchi$.}
  Contours and regions shown are the same as in \figref{Plot3a}.  }
\end{figure}

\figref{Plot3b} show the parameter space of marginally excluded
thermal relic DM in the $\left(\mchi, \theta \right)$ plane, with
$M_D$ set to fix $\Omegachi = \OmegaDM$.  As discussed above, no
solution exists for $\chxx > 0$ and $\tan\theta > 0$.  In principle,
solutions exist throughout the plane for both signs of $\chxx$ and
$\tan\theta < 0$.  However, at sufficiently large values of $y$, the
potential solutions for $\chxx < 0$ do not exist due to over-efficient
annihilation, resulting in no solution near $\theta = - \pi/4$.  For
$\chxx > 0$ the results mirror those shown in \figref{Plot3a} -- there
are significant regions with $\xiwt, \xibs \gtrsim 0.1$ for LUX, while
$\xibs < 0.1$ everywhere and $\xibs < 0.03$ throughout most of the
region for XENON1T.  In contrast, for $\chxx < 0$ the dominant
fine-tuning lies in well-tempering.  For $\tan\theta > 0$, we find
that $\xibs=1$ by definition because no cancellations are possible to
eliminate the DM-Higgs coupling.  However, to evade direct detection,
we still need tiny values of $y$, requiring coannihilation to produce
$\eqomega$.  This produces $\xiwt < 0.1$ for $400~\gev \lesssim \mchi
\lesssim 900~\gev$ for LUX, while for XENON1T, $\xiwt < 0.1$
throughout nearly the entire region shown and $\xiwt < 0.01$ for
$750~\gev \lesssim \mchi \lesssim 900~\gev$.

\section{Model B: Singlet-Doublet Scalar DM}
\label{sec:sdscalar_analysis}

Next, we will analyze the case of singlet-doublet scalar DM.  This
simplified model has qualitative similarities to the fermionic version
discussed in \secref{sdfermion_analysis}.  However, there is an
important physical difference: while a pure fermion singlet is inert,
a pure scalar singlet can have renormalizable interactions to the SM
through the Higgs boson.  In particular, for scalar DM the quartic
couplings between the DM and the Higgs can have a substantial effect
on the thermal relic abundance and direct detection constraints.  The
addition of a quartic Higgs coupling to the singlet produces a line in
the $\mchi$ vs. $\sigmaSI$ plane consistent with the thermal relic
density for pure singlet~\cite{Cline:2013gha} DM, while a quartic
Higgs coupling to the doublet shift the value of $M_D$ which gives
$\eqomega$ independent of $M_S$ for pure doublet~\cite{Hambye:2009pw}
DM.  However, we are specifically interested in the case of mixed DM,
whereby the dominant annihilation channels relevant to freeze-out
derive from mixing of the singlet and non-singlet states.  Thus, we
will focus on the parameter space where the cubic term $A \left[ S H
  D^* + h.c. \right]$ is large enough to induce significant mixing.

From the practical standpoint of analyzing constraints, scalar DM
models tend to have more parameters than fermionic DM -- in the case
of singlet-doublet DM, the scalar model has seven while the fermionic
has only four.  The large number of parameters renders a comprehensive
parameter scan like that used for fermions in
\secref{sdfermion_analysis} impractical.  Fortunately, four of the
free parameters for scalar DM are quartic couplings of the Higgs
directly to the singlet or doublet states, and thus do not induce
singlet-doublet mixing.  We will therefore focus on the case of
$\lambda_S=\lambda_D=\lambda_D'=\lambda_D''=0$ as the ``minimal case''
for mixed singlet-doublet scalar DM.  As we move away from this
simplifying limit, mixing becomes less important for the relic
abundance and direct detection constraints, and the DM properties
approach that of a pure singlet or doublet.

In principle, there are three distinct Higgs couplings to the doublet
component of DM.  However, only one combination of these enters into
the couplings of the neutral mixed doublet state.  Thus, many
combinations of the three couplings will result in the same dynamics
for the DM state alone.  On the other hand, these different
combinations will result in modified dynamics for processes involving
DM and another doublet state, which here is limited to coannihilation
for nearly pure doublet states.  While these effects may be important
in certain regions of parameter space, they are not the primary focus
here.

The dominant effect of including Higgs couplings to pure states is a
modification of $\ahxx$ and the coupling associated with the $\chi\chi
h h$ operator.  If re-expressed in terms of the singlet-doublet mixing
angles,
$\ahxx$ becomes~\cite{Cohen:2011ec}
\begin{eqnarray}
\ahxx & = & v \left( \lambda_S N_S^2 + \left[ \lambda_D + \lambda_D' +
  \lambda_D'' \right] N_D^2 \right) - 2 A N_S N_D \,.
\end{eqnarray}
As shown in \figref{Plot0b}, for the majority of the parameter space
either $N_S^2 \ll 1$ or $N_D^2 \ll 1$, so either the singlet or
doublet Higgs coupling contribution to $\ahxx$ and the associated
$\chi\chi h h$ operator will be sub-dominant.  Using this feature, we
further simplify the parameter space by taking
\begin{eqnarray}
\label{eq:SDscalarBS}
\lambda_S &=& \lambda_D  =  \lambda \\
\lambda_D' &=& \lambda_D'' =  0
\end{eqnarray}
This simplified parameter space carries most of the qualitative
features of the full theory, diverging primarily for $M_S \approx
M_D$.

\subsection{Exclusion Plots (General)}

\begin{figure}[tb]
\subfigure[]{
  \includegraphics*[width=0.31\textwidth]{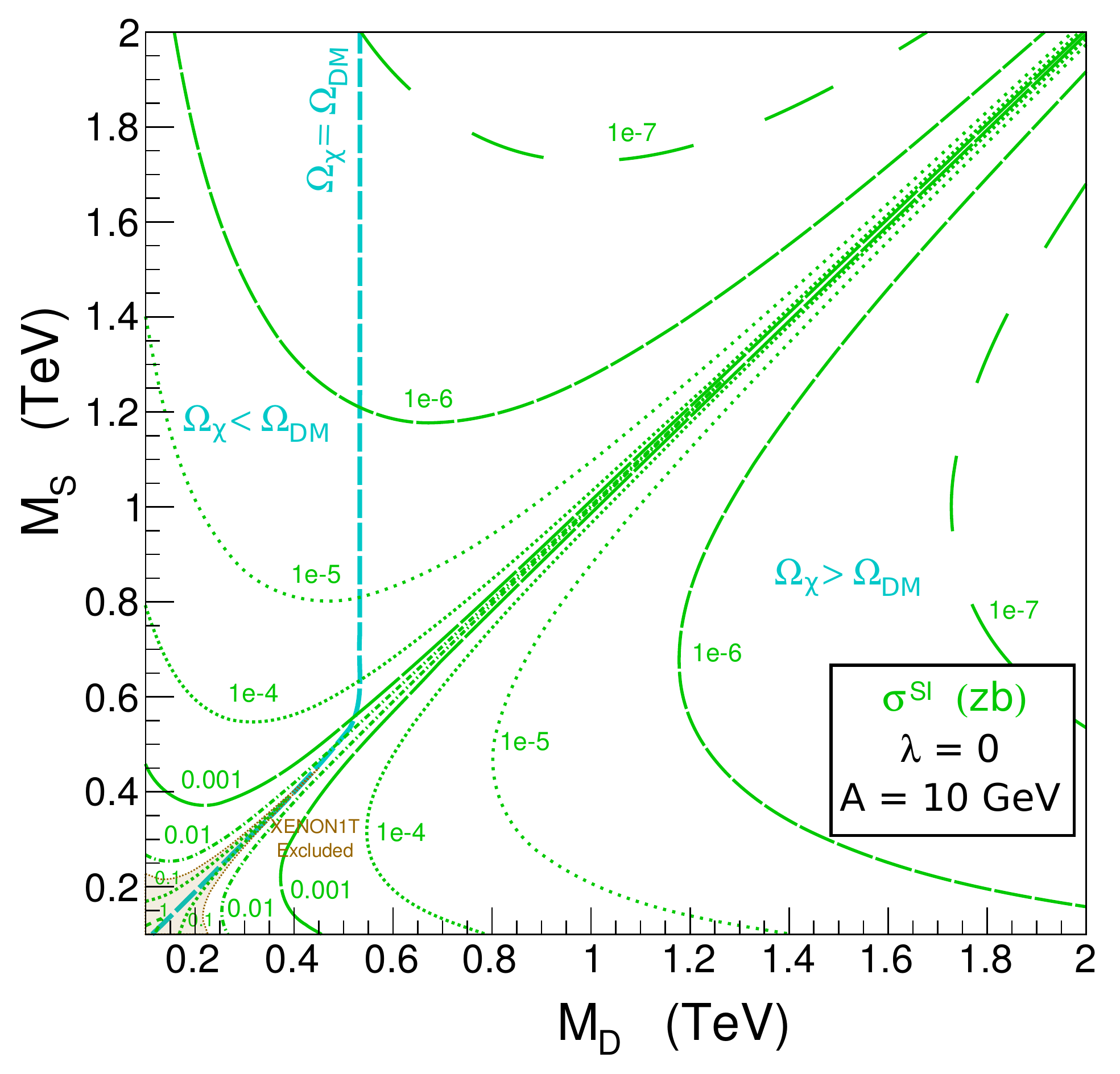}
  \label{fig:Plot4a_MSMD_A=10_y=0}}
\subfigure[]{
  \includegraphics*[width=0.31\textwidth]{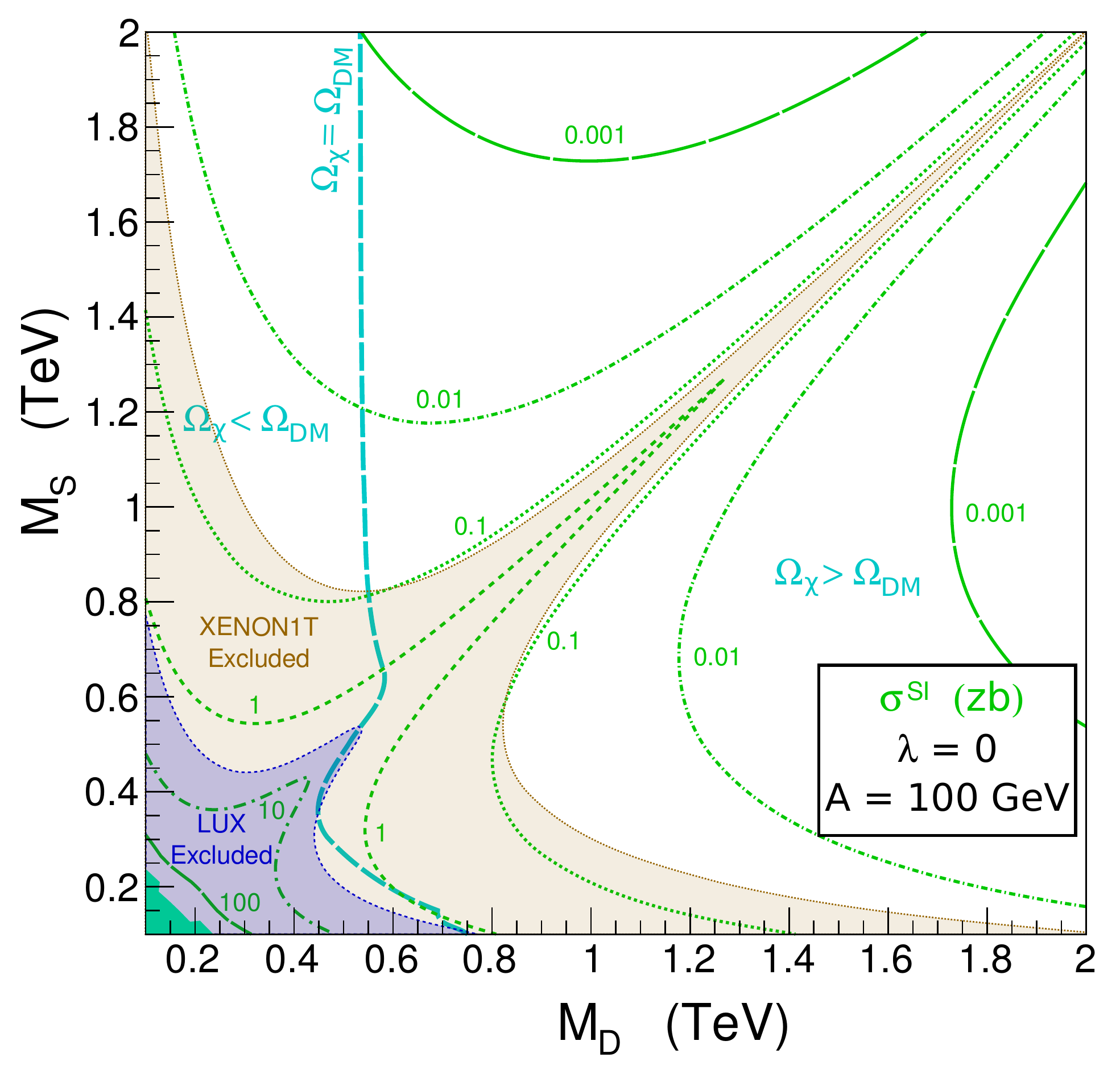}
  \label{fig:Plot4a_MSMD_A=100_y=0}}
\subfigure[]{
  \includegraphics*[width=0.31\textwidth]{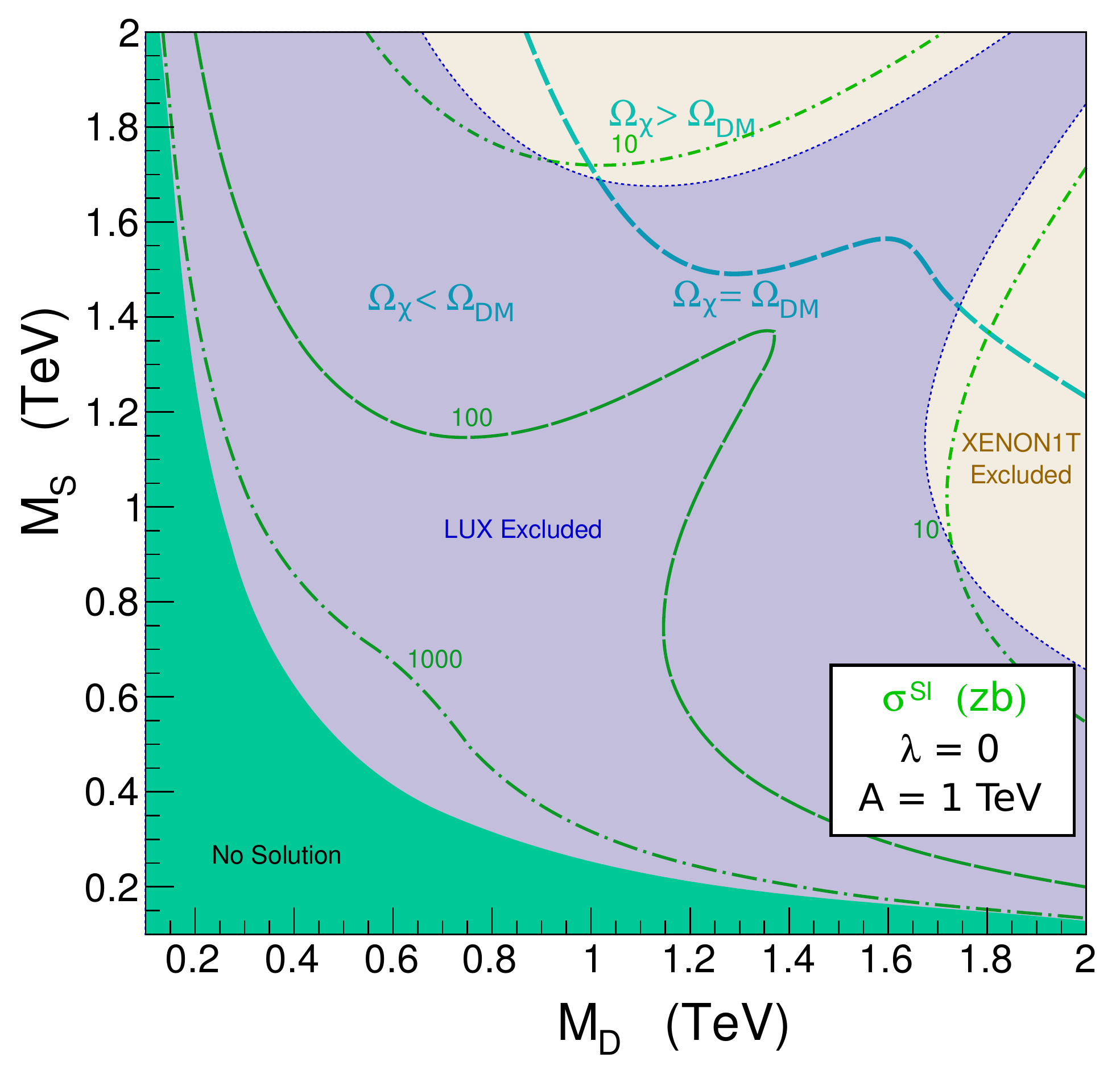}
  \label{fig:Plot4a_MSMD_A=1000_y=0}}
\vspace*{-.1in}
\caption{\label{fig:Plot4} \textit{Direct detection prospects for
    $\lambda = 0$ and $A = 10, 100, 1000~\gev$.}  Contours shown
  are the same as in \figref{Plot1_y=0.3}.}
\end{figure}

To begin, we analyze the unconstrained parameter space of scalar
singlet-doublet DM.  In \figref{Plot4} we restrict to the case
$\lambda=0$, fixing $A = 10, 100, 1000~\gev$ from left to right.  For
$A=10~\gev$, the associated cubic coupling is too weak to produce
significant splitting or contribute to the relic density directly,
thus requiring large mixing for $M_S \lesssim 500-550~\gev$.  For $M_S
\gtrsim 600~\gev$, DM with $\eqomega$ becomes a nearly pure doublet
with $M_D \approx 550~\gev$~\cite{Cirelli:2005uq,Hambye:2009pw}.  In
this case direct detection prospects are minimal, with XENON1T
sensitivity only for $M_{S,D} \lesssim 200~\gev$.

For $A=100~\gev$, the position of the $\eqomega$ line changes
substantially for $M_S \lesssim 800~\gev$, shifting to larger values
of $M_D$ as large as $\approx 750~\gev$.  However, it eventually
asymptotes to pure doublet behavior for larger $M_S$.  This shift is
qualitatively similar to the enhanced Higgs-mediated annihilation for
fermions in the case of large couplings, but the quantitative results
diverge substantially.  As discussed in \secref{sdfermion_analysis},
while Majorana fermion DM can annihilate through the Higgs, this
process appears at p-wave and only becomes important for large
couplings.  For scalars, however, annihilation via the Higgs is s-wave
and interferes strongly with t-channel annihilation diagrams involving
the other charged and neutral scalars.  Qualitatively, the relative
strength of Higgs annihilation can be seen in the direct detection
coverage -- the sensitivity of LUX and XENON1T shown in
\figref{Plot4a_MSMD_A=100_y=0} is much weaker for equivalent $\mchi$
than the full-plane coverage present for fermions in
\figsref{Plot1_y=0.3_tantheta=2}{Plot1_y=0.3_tantheta=10}, but despite
this achieving $\eqomega$ requires less well-tempering, at least for
$M_S \lesssim 800~\gev$.

The relevance of annihilation through the Higgs is even more
pronounced for $A=1~\tev$.  In this case, $\eqomega$ cannot be
achieved for small DM masses due to Higgs-mediated annihilation.  The
$\eqomega$ line occurs at $M_S \gtrsim 1.2~\tev$, corresponding to DM
masses well above the value of $\sim 550~\gev$ required for a pure
doublet thermal relic.  The $\eqomega$ line occurs at somewhat smaller
DM mass for $M_D < M_S$ than for $M_D > M_S$ due to the more efficient
annihilation from the typical doublet annihilation processes, but in
both cases the relatively small mixing angle still produces large
direct detection cross-sections.  As a result, a portion of the
$\eqomega$ line remains outside of LUX bounds despite the relatively
large values of $\sigmaSI$, though it can be probed at XENON1T.

\begin{figure}[tb]
\subfigure[]{
  \includegraphics*[width=0.31\textwidth]{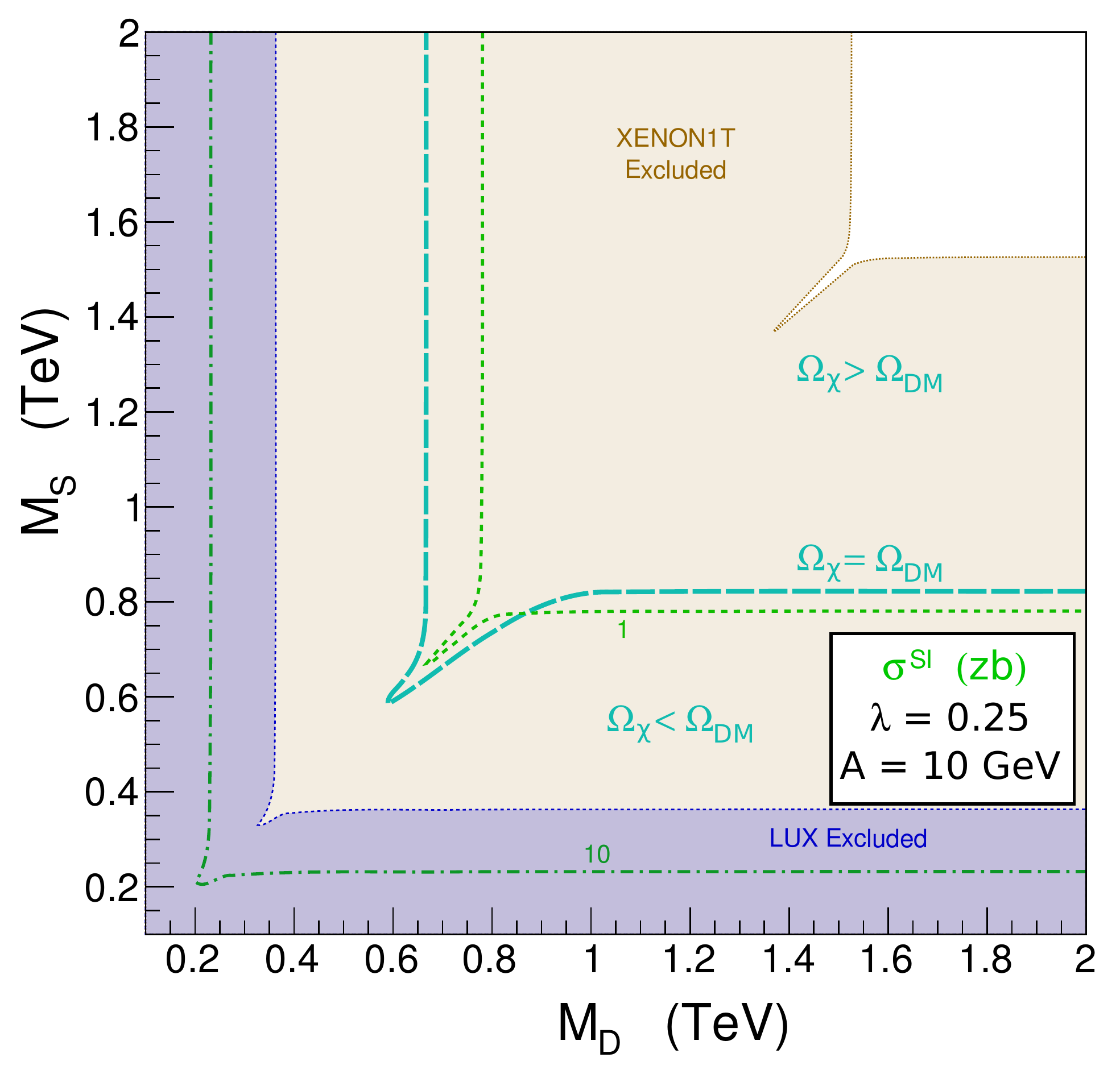}
  \label{fig:Plot4b_MSMD_A=10_yPos}}
\subfigure[]{
  \includegraphics*[width=0.31\textwidth]{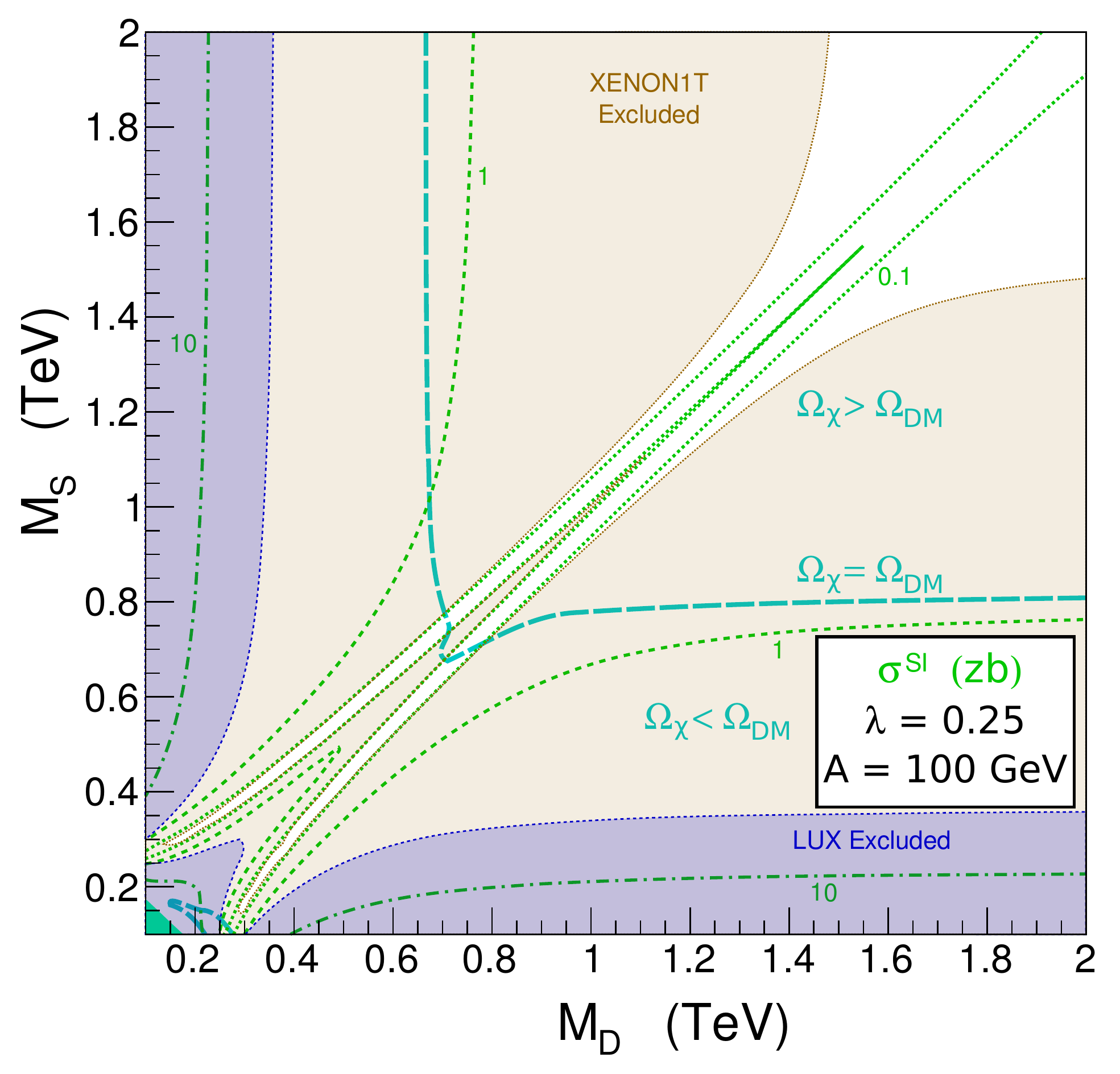}
  \label{fig:Plot4b_MSMD_A=100_yPos}}
\subfigure[]{
  \includegraphics*[width=0.31\textwidth]{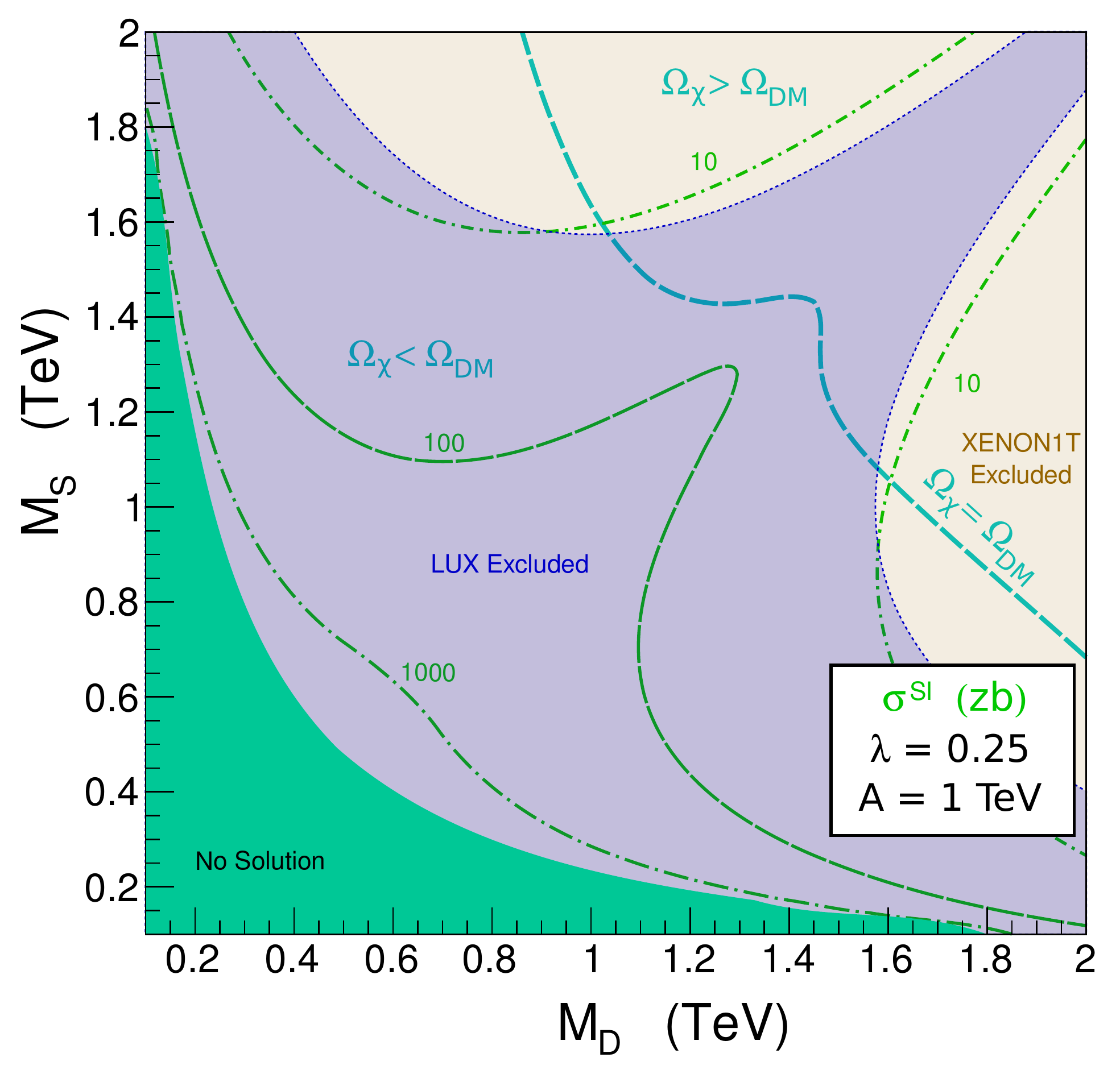}
  \label{fig:Plot4b_MSMD_A=1000_yPos}}
\subfigure[]{
  \includegraphics*[width=0.31\textwidth]{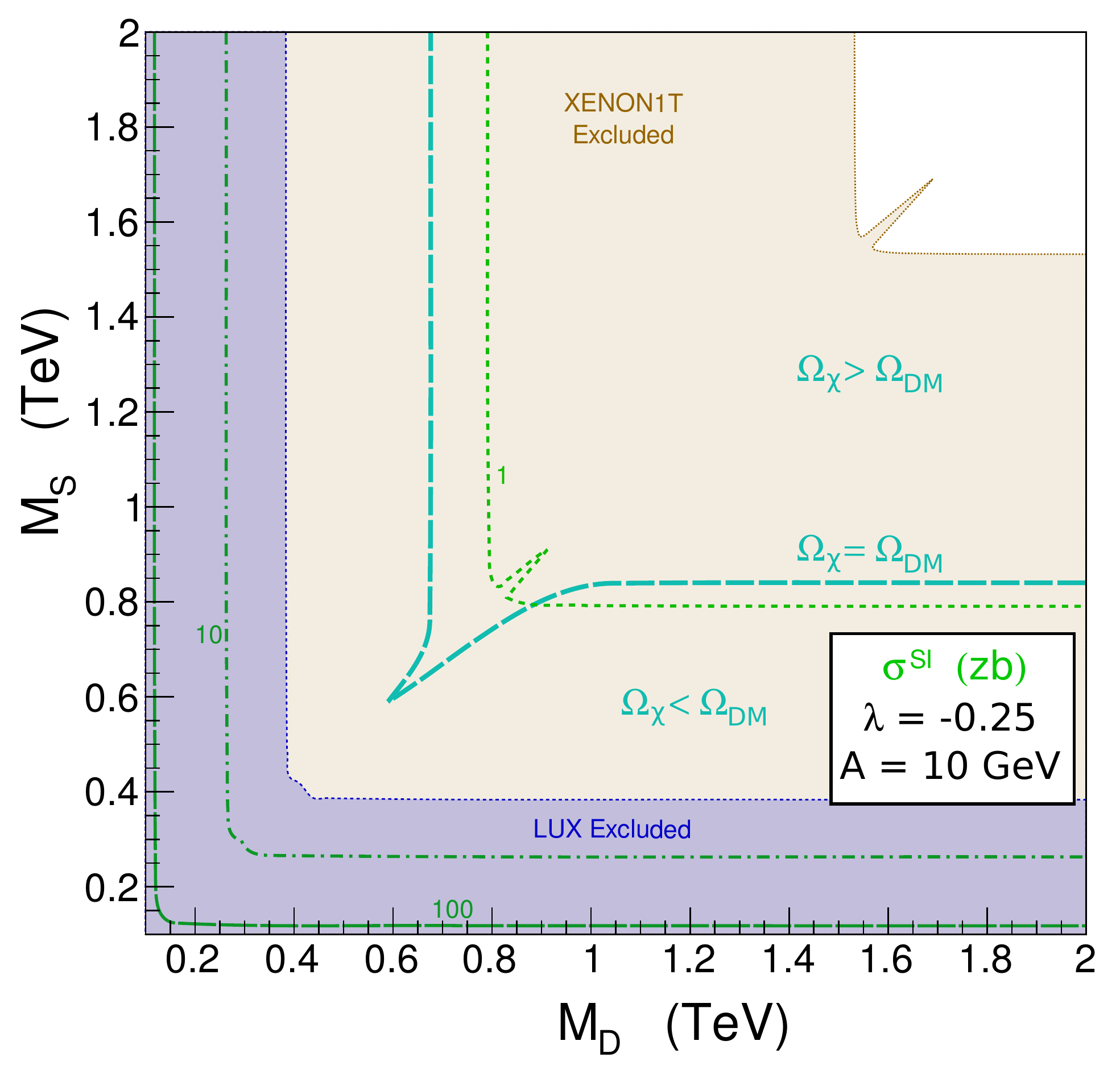}
  \label{fig:Plot4b_MSMD_A=10_yNeg}}
\subfigure[]{
  \includegraphics*[width=0.31\textwidth]{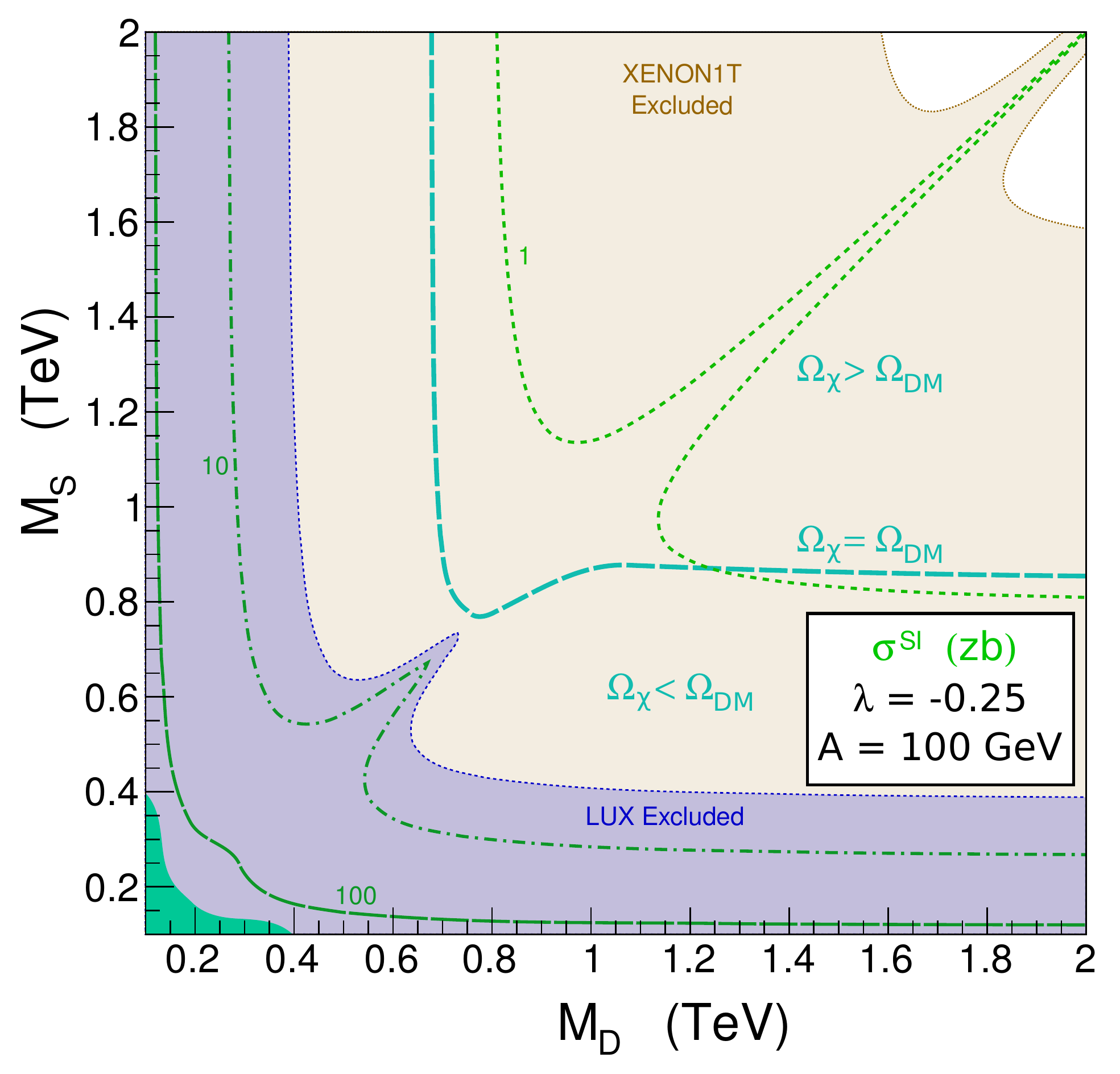}
  \label{fig:Plot4b_MSMD_A=100_yNeg}}
\subfigure[]{
  \includegraphics*[width=0.31\textwidth]{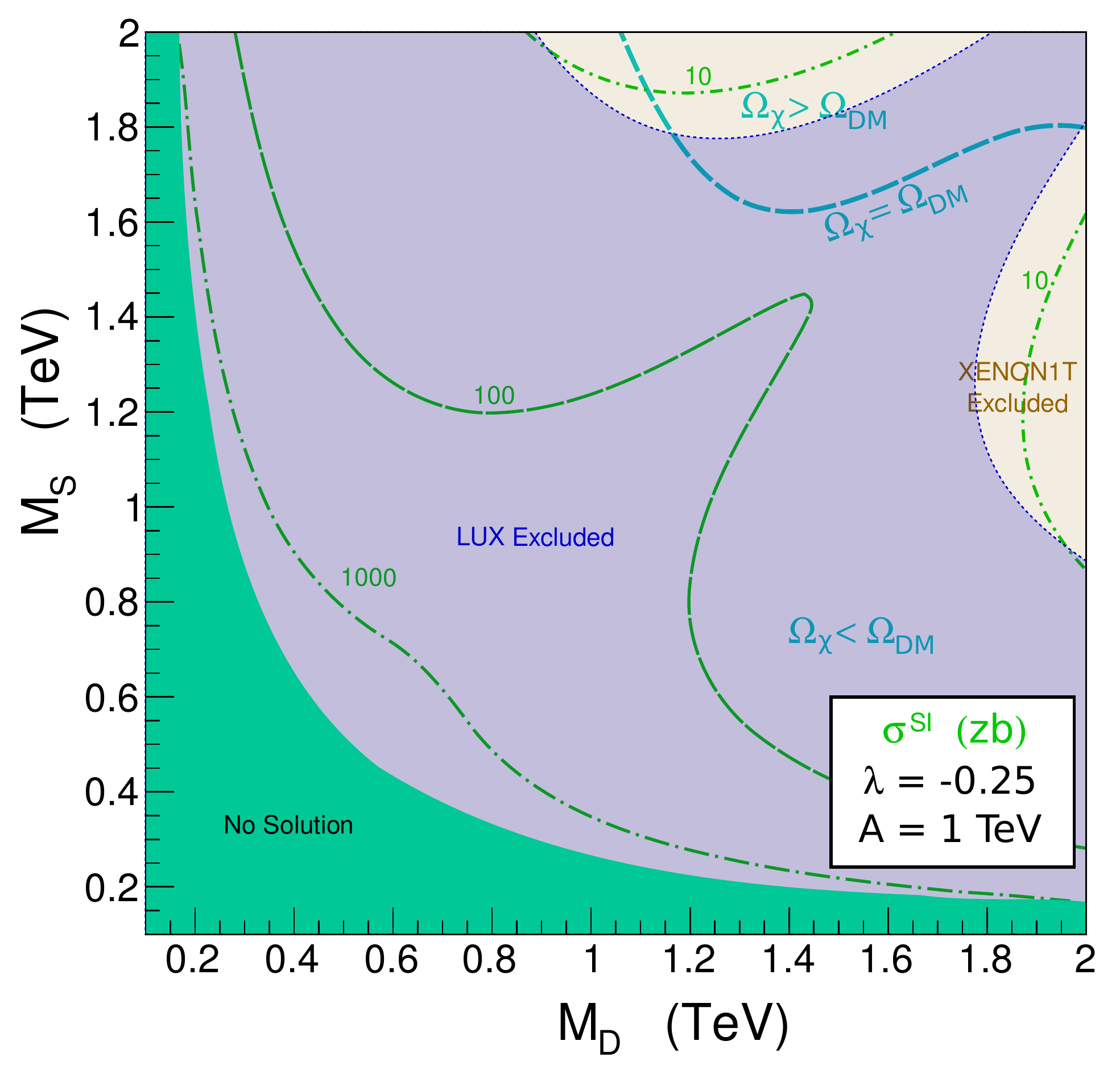}
  \label{fig:Plot4b_MSMD_A=1000_yNeg}}
\vspace*{-.1in}
\caption{\label{fig:Plot4b} \textit{Direct detection prospects for
    $\lambda = 0$ and $A = 10, 100, 1000~\gev$.}  Contours shown
  are the same as in \figref{Plot1_y=0.3}.}
\end{figure}

The situation changes drastically if there are quartic couplings
between the Higgs and the singlet and doublet components of the DM,
shown in \figref{Plot4b} for $\lambda=\pm0.25$.  The most distinctive
feature is the possibility of $\eqomega$ for pure singlet DM,
occurring at $M_S \approx 800~\gev$ for a small mixing term $A =
10~\gev$.  Furthermore, a Higgs coupling to the doublet also shifts
the mass of the $\eqomega$ for pure doublet DM to $M_D \approx
650~\gev$.  For $A \ll \lambda v$ the Higgs couplings to pure states
dominates, and the sign of $\lambda$ is irrelevant except for $M_S
\approx M_D$.  Direct detection sensitivity is strong even for the
modest value of $\lambda = \pm 0.25$, with LUX sensitivity reaching
$M_S$ or $M_D$ of just under $400~\gev$, and XENON1T sensitivity
reaching out to $M_S, M_D \lesssim 1.5~\tev$, covering the entire
$\eqomega$ line.

The sign of $\lambda$ becomes important for $A \sim \lambda v$.  As
can be seen from \eqref{eq:ahxx_sdscalar_simplified}, the contribution
of $A \neq 0$ to the DM-Higgs coupling is always negative, leading to
an enhancement of the couplings for $\lambda < 0$ and a suppression
for $\lambda > 0$.  The corresponding blind spot cancellation regions
are located near but slightly offset above and below the $M_S = M_D$
line in \figref{Plot4b_MSMD_A=100_yPos} for $\lambda = 0.25$, with no
corresponding behavior present in \figref{Plot4b_MSMD_A=100_yNeg} for
$\lambda = -0.25$.  For $\lambda = 0.25, A = 100~\gev$, certain points
in the blind spot region are actually consistent with $\eqomega$,
while for $\lambda = -0.25, A = 100~\gev$ the entire $\eqomega$
contour is within the sensitivity range of XENON1T.  A small region
also exists for $\lambda = 0.25, A = 100~\gev$ with $\eqomega$ for
$M_S, M_D \lesssim 200~\gev$, resulting from the cancellation in the
Higgs coupling; by varying values of $A$ for appropriate suppression
this low mass contour can be shifted anywhere with $M_S < M_D$.

For $A = 1~\tev \gg \lambda v$, the mixing term dominates for both
signs of $\lambda$, so no blind spot is present; however, $\sigmaSI$
is enhanced for $\lambda = -0.25$ relative to $\lambda = 0.25$
throughout the plane and constraints are stronger, as shown in
\figsref{Plot4b_MSMD_A=1000_yPos}{Plot4b_MSMD_A=1000_yNeg}.  As in
\figref{Plot4a_MSMD_A=1000_y=0}, when $A=1~\tev$, the Higgs coupling
is so large that the majority of parameter space is already excluded
by LUX.  The relative sign of $\lambda$ also has a significant effect
on the location of the $\eqomega$ line, with the enhanced coupling for
$\lambda = -0.25$ pushing the contour to larger mass, while the
suppressed coupling for $\lambda = 0.25$ shifts the line to lower
mass.  As for $\lambda = 0$, the portions of the $\eqomega$ lines in
the region shown are not excluded by LUX, but are within XENON1T
sensitivity for $\lambda = \pm 0.25$.

\subsection{Exclusion Plots (Thermal Relic)}

Next, we consider the singlet-doublet scalar model for $\eqomega$.  In
keeping with our interest in mixed DM, we fix to various values of
$\lambda$, leaving only three variables in the remaining parameter
space, $(M_S, M_D, A)$.  Within this sub-space, for fixed $(M_S, A)$
or $( M_D, A)$ a varying number of solutions exist for $\eqomega$,as
evidenced by \figsref{Plot4}{Plot4b}.  In particular, for $\lambda
\neq 0$ sufficiently large, a solution always exists for which
$\eqomega$ is independent of $M_D$ and $M_S$, respectively, above a
certain critical value.  This limits the usefulness of an analog to
\figref{Plot2a} or \figref{Plot2b} where either $M_S$ or $M_D$ is used
to fix the relic density.

\begin{figure}[tb]
\subfigure[]{
  \includegraphics*[width=0.31\textwidth]{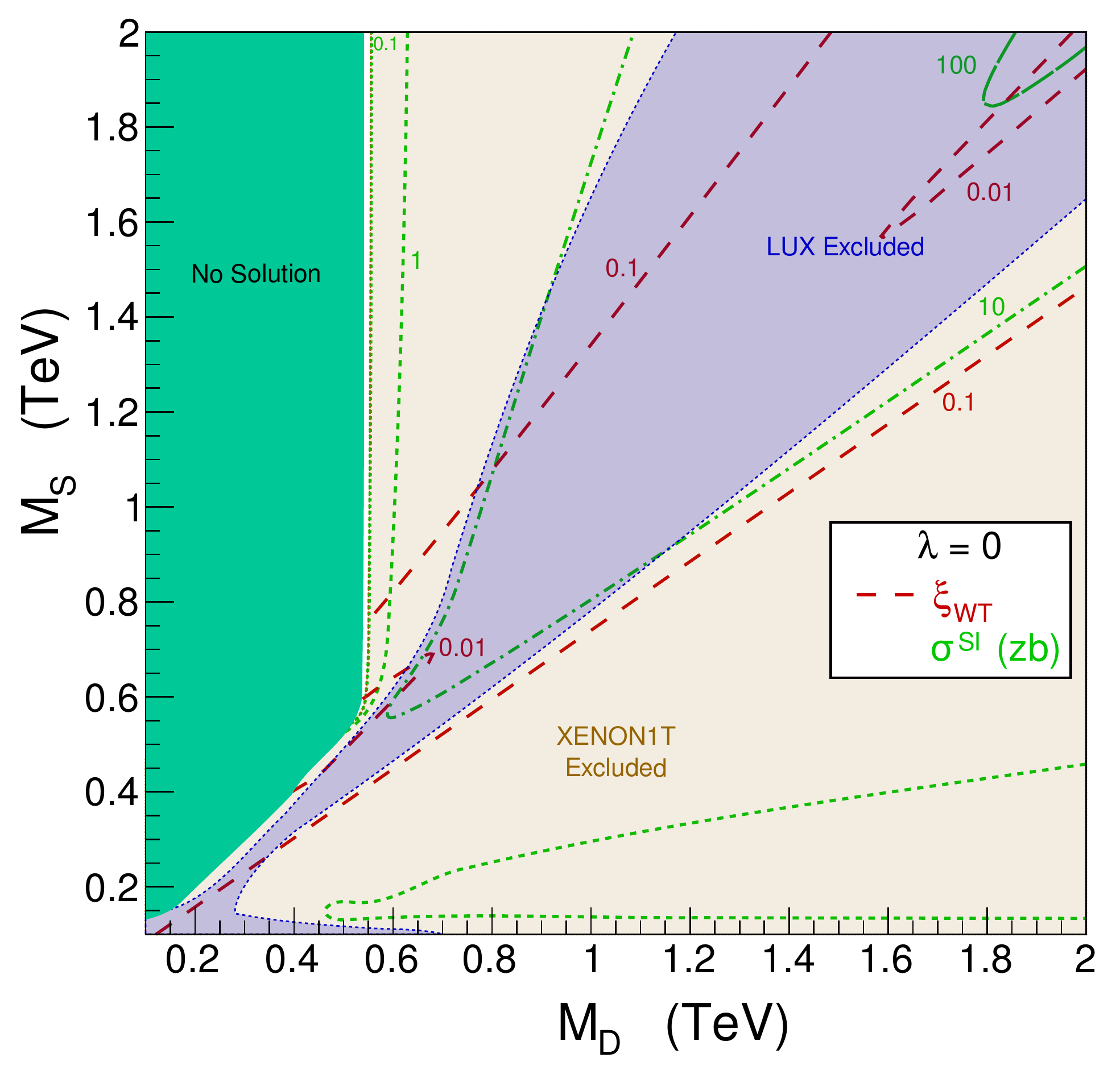}
  \label{fig:Plot5a_MSMD_y=0}}
\subfigure[]{
  \includegraphics*[width=0.31\textwidth]{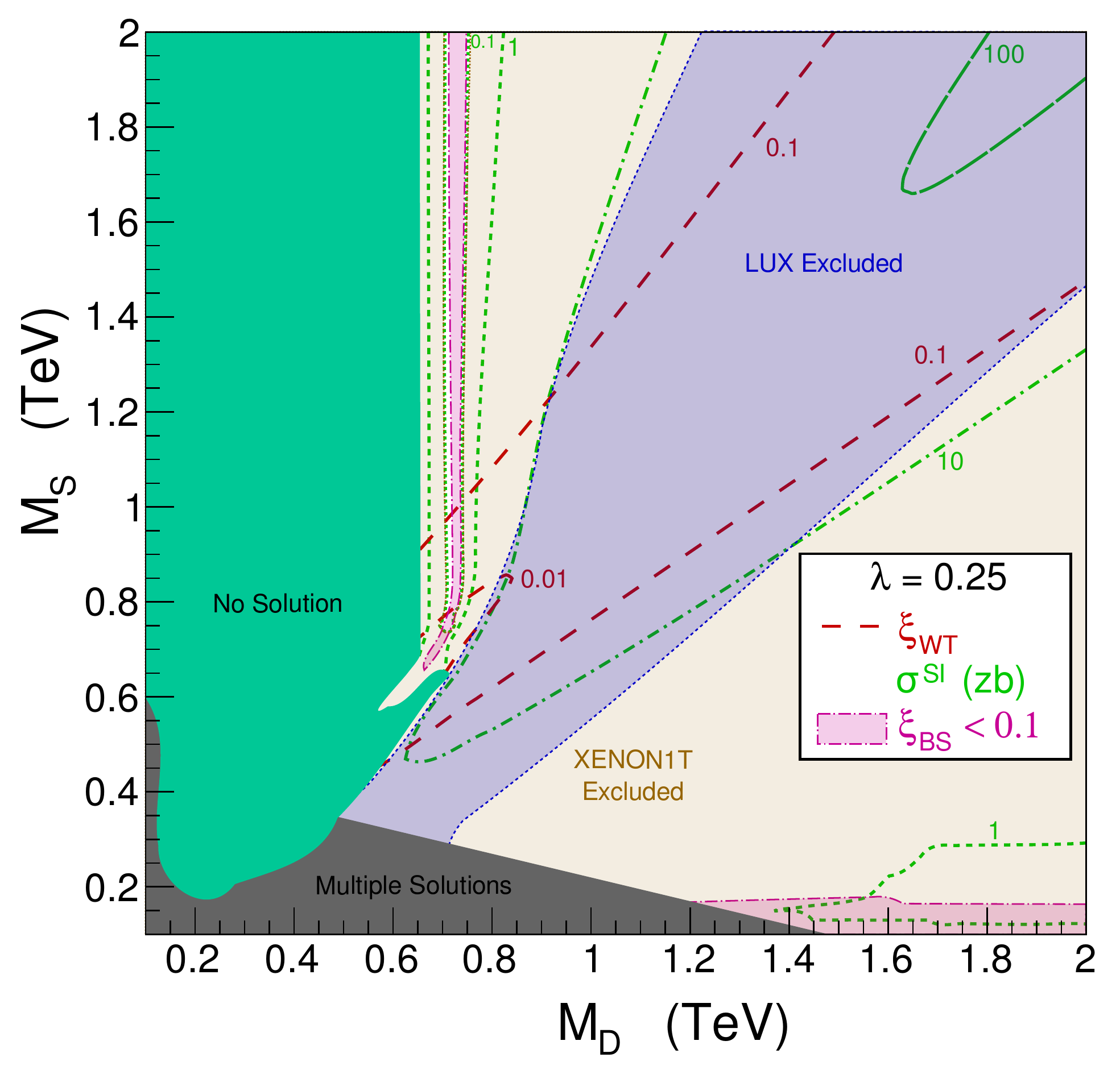}
  \label{fig:Plot5b_MSMD_yPos}}
\subfigure[]{
  \includegraphics*[width=0.31\textwidth]{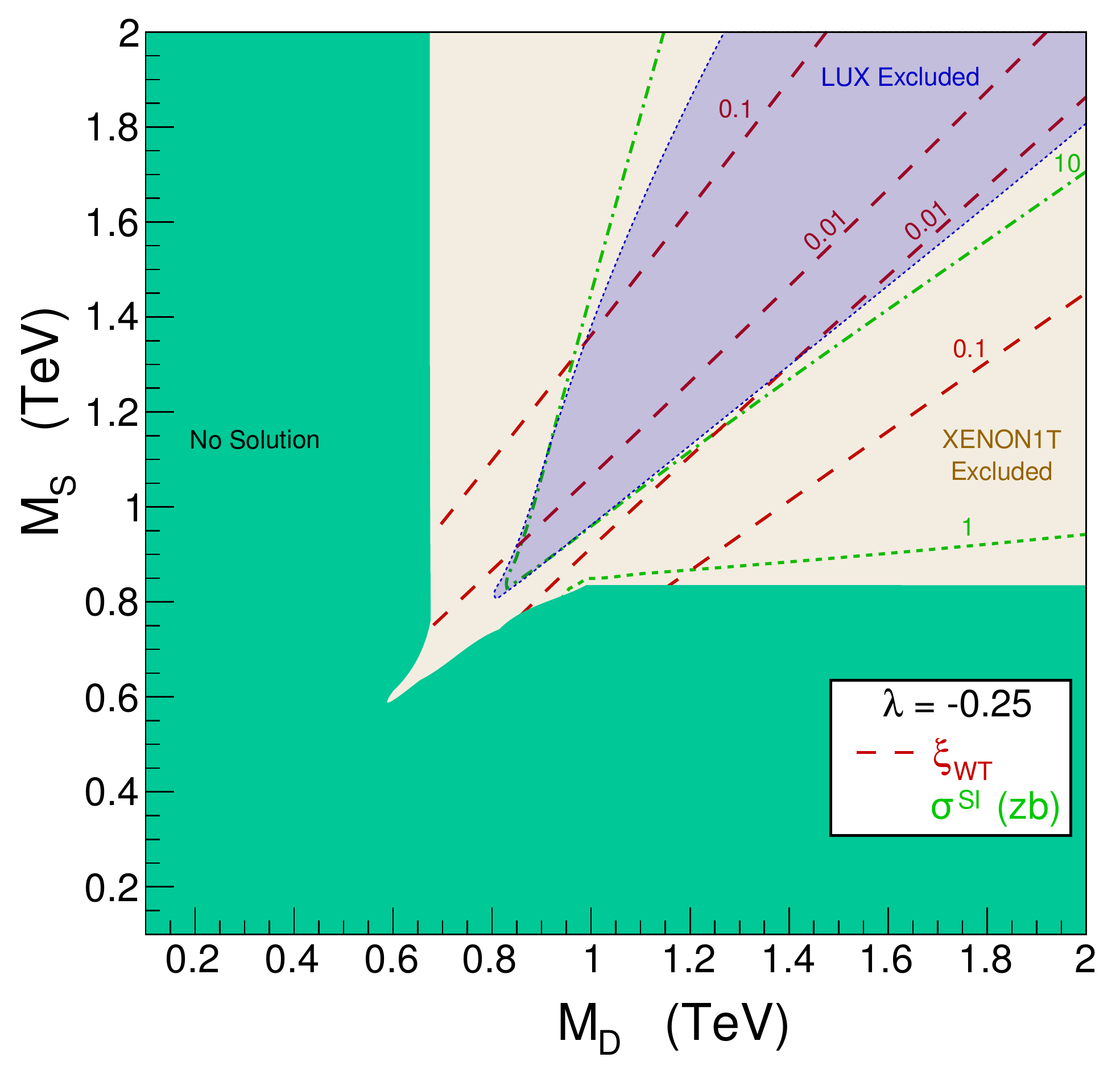}
  \label{fig:Plot5b_MSMD_yNeg}}
\vspace*{-.1in}
\caption{\label{fig:Plot5} \textit{Well-tempering and direct detection
    prospects for fixed $\Omegachi=\Omega_{DM}$ as a function of $M_D$
    and $M_S$.} Contours shown are the same as in \figref{Plot2a}, with
  the addition of a region corresponding to $\xibs < 0.1$ (pink
  shaded) in \figref{Plot5b_MSMD_yPos}.}
\end{figure}

As such, we consider the $(M_D, M_S)$ plane in \figref{Plot5} with
$\lambda = 0, \pm 0.25 $ and $A$ fixed to produce $\eqomega$.  For
$\lambda = 0$, no solution is present for $M_D < M_S$ and $M_D
\lesssim 550~\gev$, as a nearly pure doublet is under-dense in this
region.  The same effect is present for $\lambda = \pm 0.25$ for $M_D
\lesssim 650~\gev$.  Throughout the rest of the plane in
\figref{Plot5a_MSMD_y=0}, direct detection sensitivity to the
$\eqomega$ scenario is strong, with the best sensitivity present for
$M_S \approx M_D$ and dropping off as the masses become less
degenerate.  In fact, direct detection sensitivity grows at large DM mass
because the mixing required for the thermal relic abundance requires
$(vA)^2 \sim \mchi^4$ at large mass, resulting in $\chxx \sim \mchi /
v$.  The DM-nucleon cross-section thus scales as $\sigmaSI \propto \chxx^2 \propto (\mchi/v)^2$, which grows faster than 
the direct detection limits weaken:
$\sigmaSI_{\mathrm{LUX}},\sigmaSI_{\mathrm{X1T}} \propto \mchi$.  For
$\lambda = 0$, the blind spot is present for $M_S > M_D$ and $M_D$ near
the ``no solution'' region.

For $\lambda \neq 0$ the interplay of mixing and non-mixing Higgs
couplings modifies the location of the ``no solution'' and blind spot
regions.  For $\lambda = -0.25$, no solution exists for $M_S \lesssim
800~\gev$ except for $M_S \approx M_D$, as the Higgs-mediated
annihilation cross-section is too large in this region.  However,
despite an enhanced Higgs coupling, the LUX exclusion is marginally
\textit{weaker} for $\lambda = -0.25$ -- the direct coupling alone is
insufficient to saturate the LUX bound for $M_S \not\approx M_D$, but
it does reduce the required degree of mixing necessary to produce
$\eqomega$.  However, the entire viable region for $\lambda = -0.25$
falls within project XENON1T sensitivity.  For $\lambda = 0.25$, the
LUX exclusion is correspondingly stronger than for $\lambda = 0$;
however, the blind spot is shifted away from the ``no solution''
region to $M_D \approx 725~\gev$ and $M_S \gtrsim 750~\gev$.  A region
with multiple solutions for $\eqomega$ also exists for small $M_S$ and
$\lambda = 0.25$, resulting from the interplay of Higgs coupling
cancellation with varying $A$.  In this region XENON1T constrains most
solutions, but a set exists for which the singlet and doublet states
are highly mixed but $\ahxx$ vanishes due to a cancellation between
the contributions, producing a blind spot.

The blind spot tuning measure, $\xibs$, only has meaningful
implications for the $\lambda > 0$ parameter space.  As defined in
\eqref{eq:sdscalar_tuning}, $\xibs = 1$ identically for $\lambda \le
0$, with partial cancellation in the top term possible only for
$\lambda > 0$.  This gives $\xibs \lesssim 0.1$ throughout the region
allowed by XENON1T in the $\lambda > 0$ plane.  The region has a
widths of $50~\gev$, which na\"{i}vely is somewhat thinner than $10\%$
turning would indicate; however, the size of a fine-tuned region
should be compared to the characteristic mass, making a 10\% tuning
reasonable for $\mchi \approx 700~\gev$.

\begin{figure}[tb]
  \includegraphics*[width=0.48\textwidth]{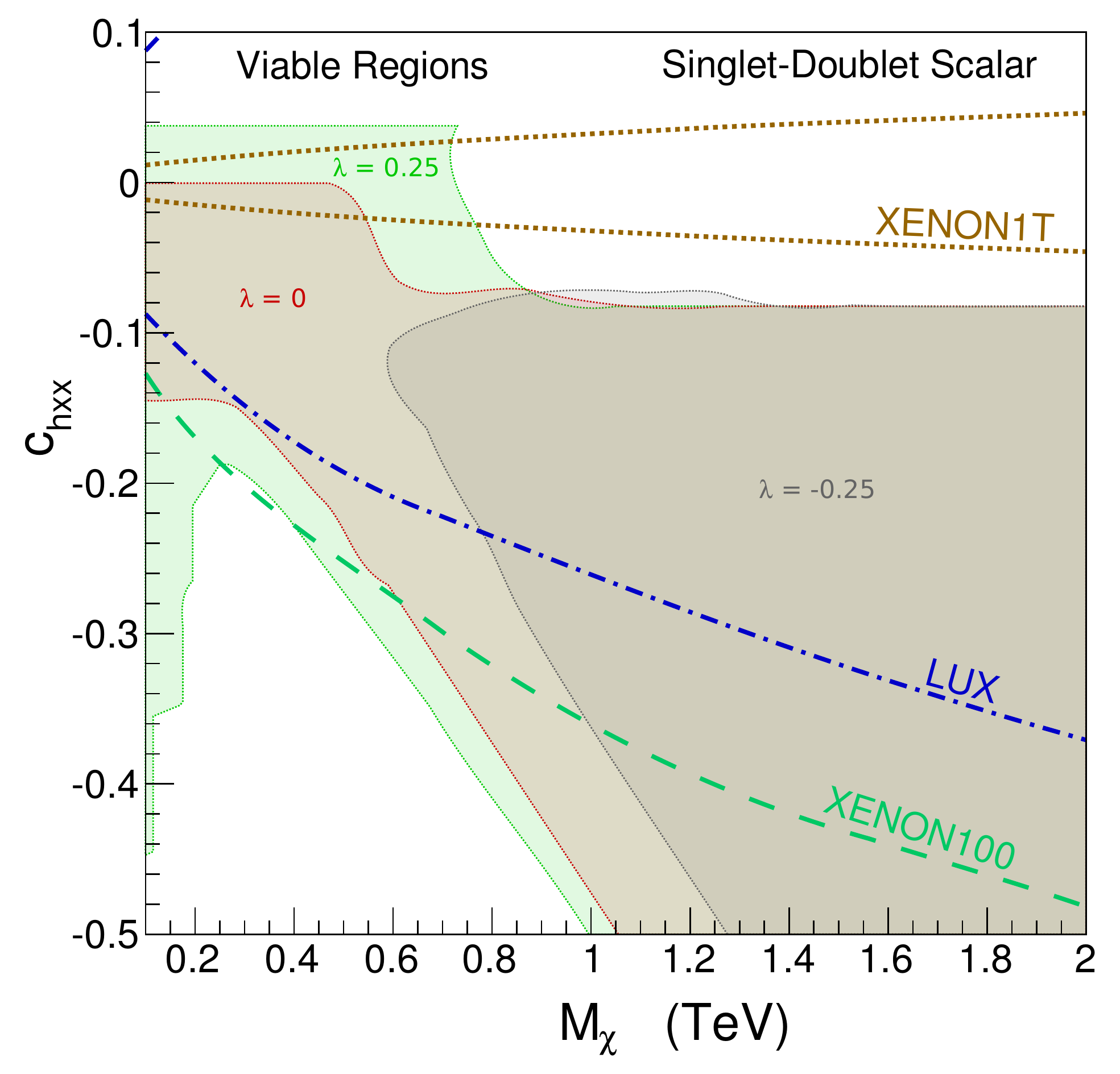}
\vspace*{-.1in}
\caption{\label{fig:Plot6} \textit{Viable regions with
    $\Omegachi=\Omega_{\rm DM}$ as a function of $\mchi$ and
    $c_{h\chi\chi}$.} Regions are shown for $\lambda = 0$ (red
  shaded), $\lambda = 0.25$ (green shaded) and $\lambda = -0.25$ (gray
  shaded) with $M_D$ fixed to produce the correct relic density.  Also
  shown are upper limits on $|\chxx|$ from XENON100 (teal dashed), LUX (blue
  dot-dashed), and projected upper limits from XENON1T (gold dashed).}
\end{figure}

The experimentally viable region for singlet-doublet scalar DM is
depicted in \figref{Plot6} in the plane of physical variables,
$(\mchi, \chxx)$.  To produce this figure we have replaced the model
parameters $(M_S, A)$ with $ (\mchi, \chxx)$, and set $M_D$ to a value
in order to accommodate $\eqomega$.  As noted before, this produces
multiple solutions at most points in each region, particularly for
$|\chxx| \gtrsim 0.1$.  For $|\chxx| \gtrsim 0.1$ we find that
$\Omega_\chi < \Omega_{\rm DM}$ both for $M_D \approx \mchi$ due to
coannihilation and for $M_D \rightarrow \infty$ due to pure Higgs
coupling.  However, $\Omega_\chi > \Omega_{\rm DM}$ for an
intermediate range of $M_D$ due to destructive interference between
pure gauge and Higgs-mediated annihilation diagrams, producing at
least two solutions for $\eqomega$.

According to \eqref{eq:ahxx_sdscalar_simplified}, $\chxx \le 0$ for
$\lambda = 0$.  For $\mchi \lesssim 550~\gev$, arbitrarily small
values of $\chxx$ are viable, since a sufficiently small splitting
$|M_D - M_S|$ can always be chosen to produce the correct degree of
mixing through pure gauge diagrams.  For $\mchi \gtrsim 550~\gev$,
some degree of Higgs-mediated annihilation is required to produce
$\eqomega$, and the region boundaries are set by the relative size of
Higgs-mediated versus pure gauge annihilation diagrams.  The upper
boundary along $\chxx \approx 0.8$ for $\mchi \gtrsim 550~\gev$ has
the relic density set entirely by Higgs-mediated diagrams, with gauge
diagrams suppressed by $M_D \gg \mchi$.  In this case $A^2 / M_D^2
\sim \mchi / v$, so the underlying Higgs coupling must increase
dramatically with $\mchi$ and will eventually become non-perturbative.
For larger values of $|\chxx|$ the degree of mixing is increased, with
the lower boundary set by near-maximal mixing.  The largest values of
$\chxx$ are achieved when significant interference is present between
pure gauge and Higgs-mediated diagrams, resulting in ${\rm BR} \left(
\chi \chi \rightarrow W^+ W^-, Z Z \right) \rightarrow 0$ and ${\rm
  BR} \left( \chi \chi \rightarrow t \bar{t} \right) \rightarrow 1$.
The nearly pure doublet case for $\mchi \gtrsim 550~\gev$ is found at
intermediate points with $M_S \gg \mchi$ and Higgs-mediated diagrams
compensating for insufficient pure gauge annihilation.

For $\lambda \neq 0$, the viable region is modified significantly at
low mass but remains similar in shape at high mass.  This is a result
of the scaling of the terms in \eqref{eq:ahxx_sdscalar_simplified},
with the unmixed coupling term $\lambda v$ becoming sub-dominant for
increasing $\mchi$, while the term proportional to $A^2$ remains
constant or increases with increasing $\mchi$.  For $\lambda = 0.25$,
positive $\chxx$ is possible for $\mchi \lesssim 700~\gev$, and
significantly larger values of $|\chxx|$ are viable for $\mchi \lesssim
200~\gev$ due to interference effects in annihilation to gauge bosons
for masses insufficiently large to allow for annihilation to
$t\bar{t}$.  For large $\mchi$, however, the upper boundary asymptotes
to $\chxx \approx 0.8$ as in the $\lambda = 0$ case, while the lower
boundary is offset to more negative $\chxx$ by a small value.  For
$\lambda = -0.25$, no solution is present for $\mchi \lesssim
600~\gev$ -- both contributions to $\chxx$ are negative, and the
resulting value always produces $\Omegachi < \OmegaDM$.  While
relatively large values of $|\chxx|$ were feasible at low mass for
$\lambda = 0.25$, in such cases $A$ and $|M_D - M_S|$ could be
relatively large due to cancellation between the contributions,
resulting in a suppression of coannihilation contributions, while for
$\lambda = -0.25$ the contributions are always additive.  For large
$\mchi$ the upper boundary of the $\lambda = -0.25$ region also
asymptotes to $\chxx \approx 0.8$, while the lower boundary is offset
to less negative $\chxx$ by an amount somewhat larger than the offset
for $\lambda = 0.25$.

As previously shown in \figref{Plot5}, the constraints from direct
detection experiments become stronger for larger masses in
\figref{Plot6}.  As noted in \secref{sdscalar}, direct detection
limits place a constraint on the quantity $c_{h\chi\chi}^2/ \mchi$,
while the region with a viable thermal relic abundance is bounded by
constant $\chxx$ and constant $\chxx / \mchi$.  The blind spot for
small $\mchi$ remains constrained to low $\mchi$, though larger values
of $\mchi$ would become viable for increased $\lambda$.  A significant
region of parameter space remains unconstrained by LUX limits,
particularly at large mass, but XENON1T projected sensitivity covers
all viable regions except for the low mass blind spot.  If XENON1T
yields null results, then theories with negative $\lambda$ will be
strongly constrained.

\section{Model C: Singlet-Triplet Scalar DM}
\label{sec:stscalar_analysis}

Finally, we examine the experimental limits on mixed singlet-triplet
scalar DM.  The phenomenology of the singlet-triplet scalar model is
similar to that of the singlet-doublet scalar model -- the relic
density is set primarily by DM-gauge interactions and Higgs-mediated
diagrams.  Moreover, these contributions can interference
substantially in a way that strongly affects the final relic density
determination.  However, several features significantly alter the
detailed phenomenology.  First, the cubic DM-gauge interaction
vanishes because of the $SU(2)_L$ symmetry.  The leading DM-gauge
interaction is quartic, and has a coupling which is effectively four
times stronger than in the singlet-doublet scalar case simply due to
group theory factors.  Thus, the model can accommodate $\eqomega$
primarily through gauge interactions with sufficient well-tempering up
to larger values of $\mchi$.  Secondly, for a given mixing angle, the
trilinear DM-Higgs coupling is larger by a factor of 2 for
singlet-triplet DM as compared to singlet-doublet DM.  This modifies
the relative strength of contributions to annihilation and scattering.
More significantly, however, is the DM-Higgs quartic interaction
induced by the $\kappa$ mixing term.  While a similar interaction is
induced at tree level by the mixing term in the singlet-doublet scalar
case, the contribution from the direct coupling for the
singlet-triplet is larger for a spectrum with similar mass splittings.

Singlet-triplet scalar DM involves the same basic processes as
singlet-doublet scalar DM, so in the following analysis we focus
primarily on the differences between these theories.  Once again
either $N_S^2 \ll 1$ or $N_T^2 \ll 1$ for most of the parameter space,
so the region where $\lambda_S$ and $\lambda_T$ both play an important
role in the dynamics is limited to $M_S \approx M_T$; thus for
simplicity we set
\begin{eqnarray}
\lambda_S &=& \lambda_T  =  \lambda
\end{eqnarray}
In the singlet-triplet case this is a somewhat better approximation,
as the absence of additional quartic couplings reduces the range of
possible divergent results for $M_T \not\approx M_S$.

\subsection{Exclusion Plots (General)}

\begin{figure}[tb]
\subfigure[]{
  \includegraphics*[width=0.45\textwidth]{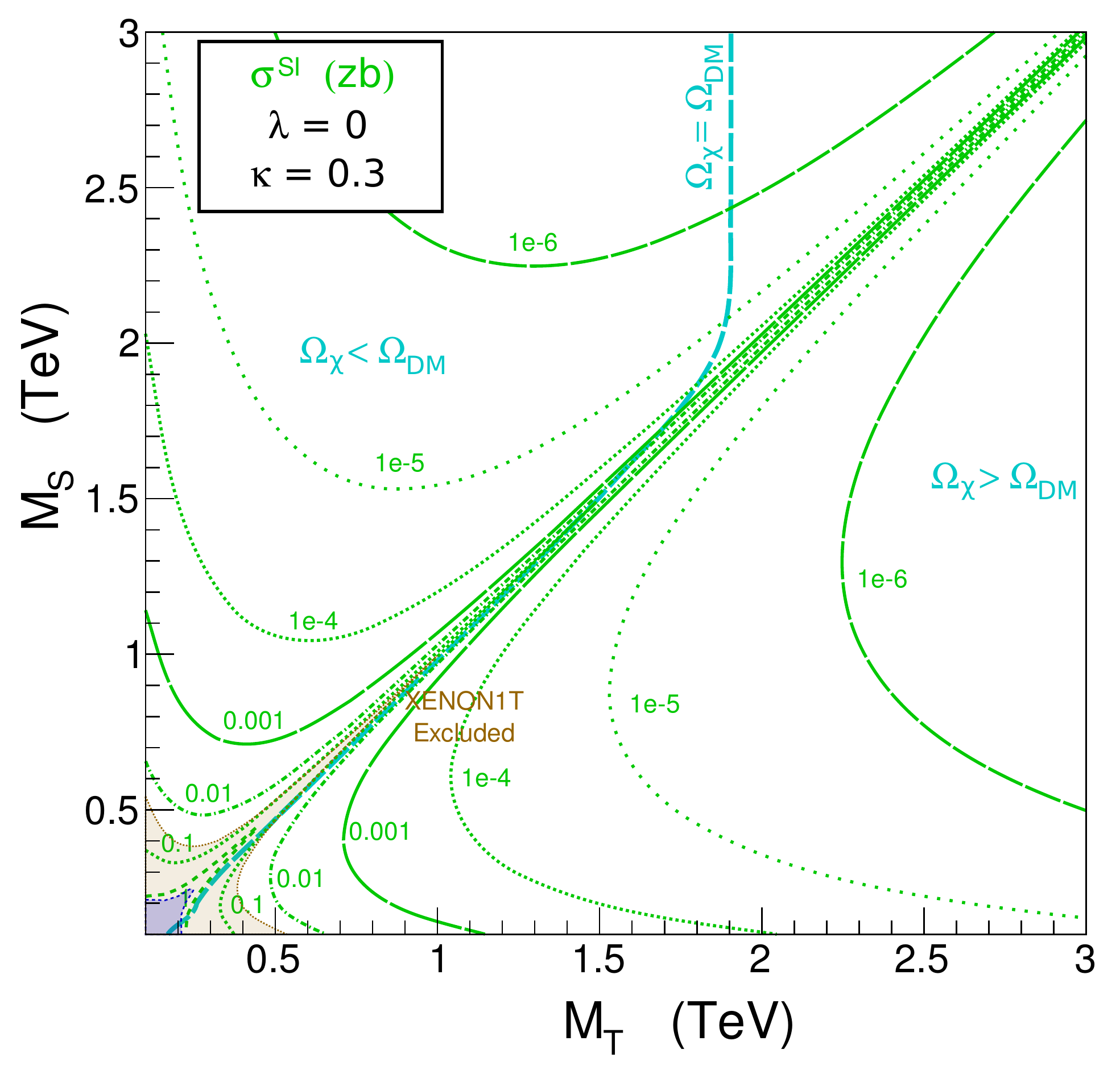}
  \label{fig:Plot7_MSMT_yST=0.3}}
\subfigure[]{
  \includegraphics*[width=0.45\textwidth]{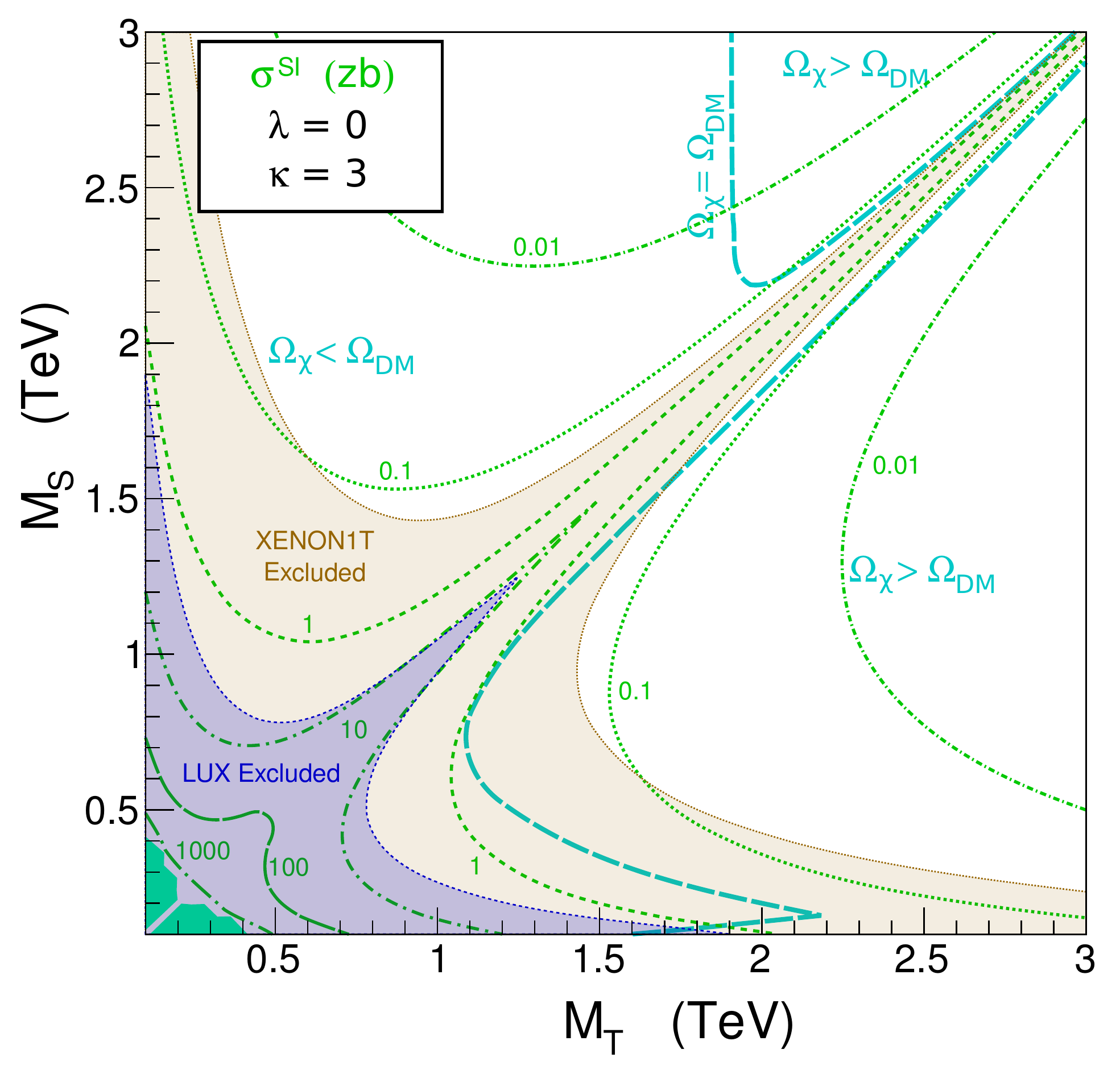}
  \label{fig:Plot7_MSMT_yST=3}}
\subfigure[]{
  \includegraphics*[width=0.45\textwidth]{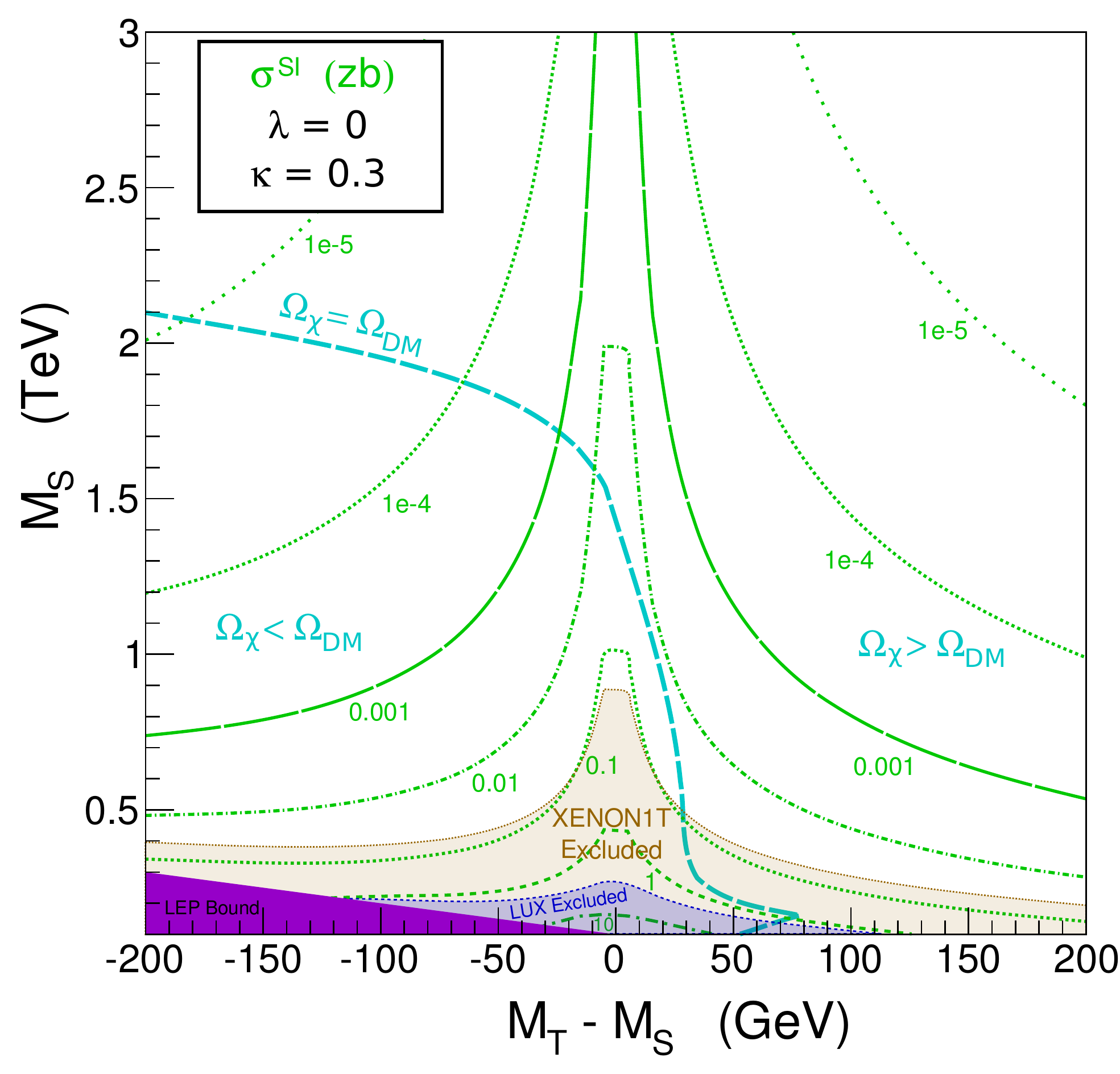}
  \label{fig:Plot7_MSMT_Zoomed_yST=0.3}}
\subfigure[]{
  \includegraphics*[width=0.45\textwidth]{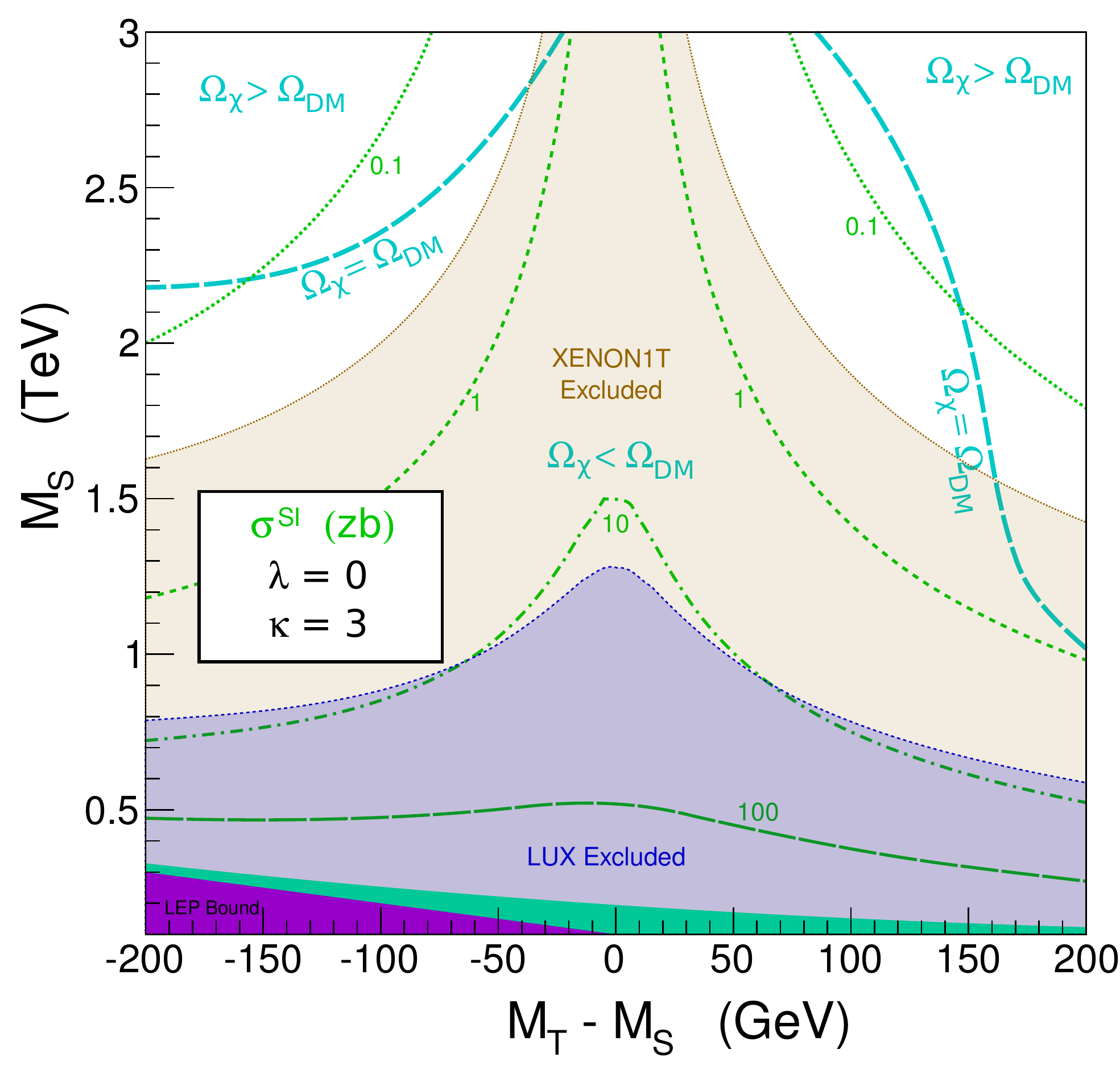}
  \label{fig:Plot7_MSMT_Zoomed_yST=3}}
\vspace*{-.1in}
\caption{\label{fig:Plot7} \textit{Direct detection prospects for
    $\lambda = 0$ and $\kappa = 0.3, 3$.}  Contours shown are the same
  as in \figref{Plot1_y=0.3}.}
\end{figure}

To analyze the unconstrained singlet-triplet DM parameter space, we
fix $\lambda=0$ to eliminate the effects of the Higgs-DM quartic
couplings.  In the case of singlet-triplet DM, mixing is controlled by
the dimensionless coefficient $\kappa$, which we set to $\kappa =
0.3 ,3$ in \figref{Plot7}.  The top two panels of \figref{Plot7}
depict the thermal relic abundance and direct detection sensitivity in
the $(M_T, M_S)$ plane.  For the case of $\kappa =0.3$ in
\figref{Plot7_MSMT_yST=0.3}, $M_S$ and $M_T$ must be very degenerate
to accommodate a thermal relic consistent with observations, and
$\sigmaSI$ is small throughout the plane.  For $M_S \not\approx M_T$,
LUX only constrains $M_S, M_T \lesssim 200~\gev$, and even XENON1T
only has sensitivity for $M_S, M_T \lesssim 400~\gev$.  In
\figref{Plot7_MSMT_yST=3}, the relatively large value of $\kappa=3$
shifts the $\eqomega$ line substantially, primarily due to an increase
in the annihilation process $\chi \chi \rightarrow hh$ with $\kappa$.
This depletes the DM abundance and implies that very large DM masses
-- upwards of multi-TeV -- may be necessary to accommodate the
observed DM relic abundance.  While the both LUX and XENON1T strongly
constrain the plane, with generic sensitivity to $M_S, M_T \lesssim
800~\gev$ and $M_S, M_T \lesssim 1.5~\tev$ respectively, the shift in
the $\eqomega$ line places some regions with a viable thermal relic
beyond XENON1T reach.

The lower panels of \figref{Plot7} show the same information as the
top panels, only plotted in the $(M_T - M_S, M_S)$ plane so as to
focus on the diagonal region in which the singlet and triplet are well
mixed.  Sensitivity at direct detection experiments to the $M_S
\approx M_T$ region is significantly improved relative to other parts
of the plane.  For $\kappa =0.3$ LUX bounds remain weak, but the
projected limits from XENON1T for $\kappa =0.3$ are covering the
$\eqomega$ contour up to $M_S \lesssim 500~\gev$ and bound $M_S
\approx M_T \lesssim 900~\gev$.  While experimental sensitivity
improves for $\kappa =3$, with reach for $M_S \approx M_T > 3~\tev$,
even in this case the $\eqomega$ contour is only within XENON1T reach
for $M_S \lesssim 1.5~\tev$.

\begin{figure}[tb]
\subfigure[]{
  \includegraphics*[width=0.45\textwidth]{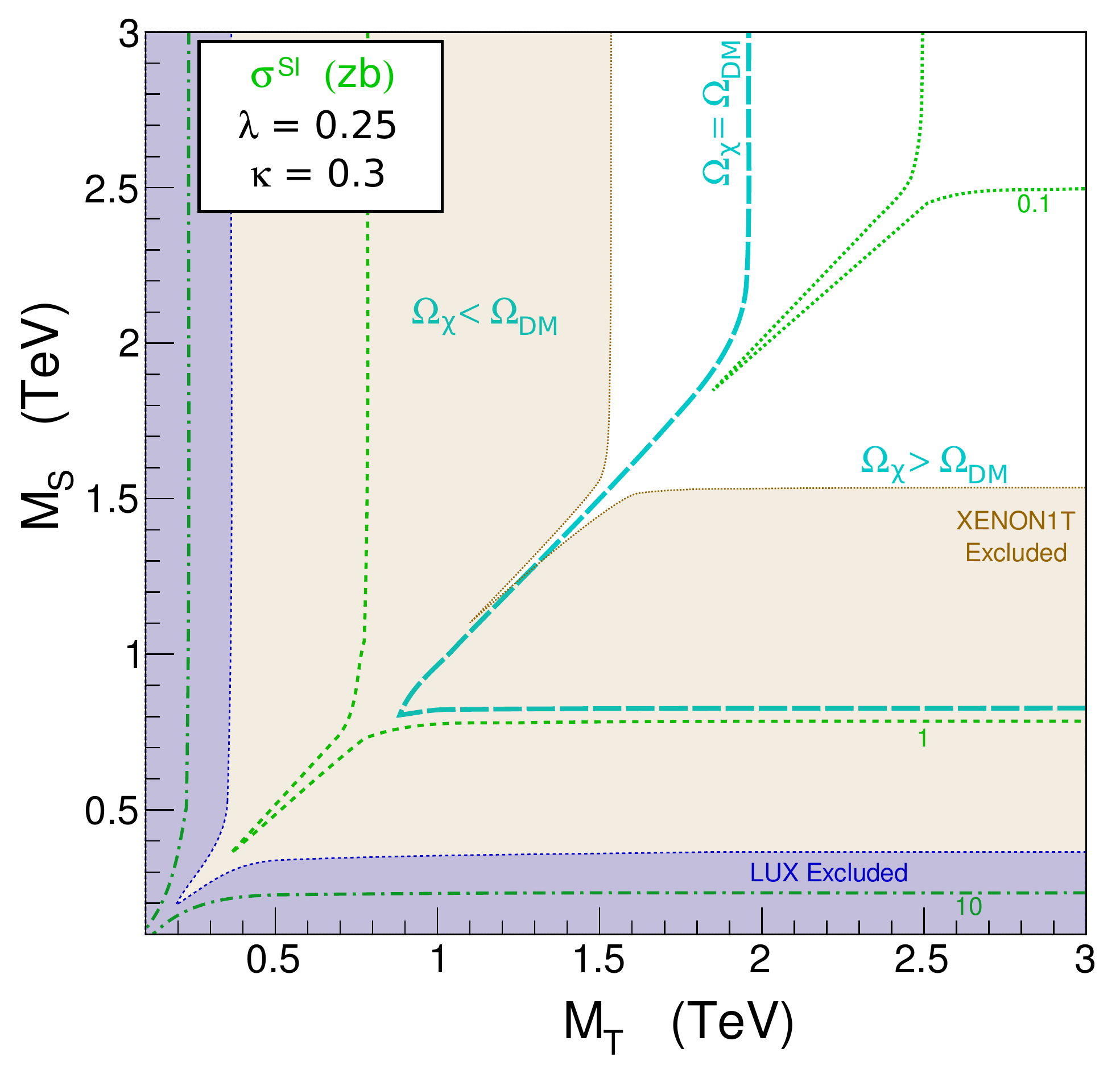}
  \label{fig:Plot7b_MSMT_yST=0.3_yPos}}
\subfigure[]{
  \includegraphics*[width=0.45\textwidth]{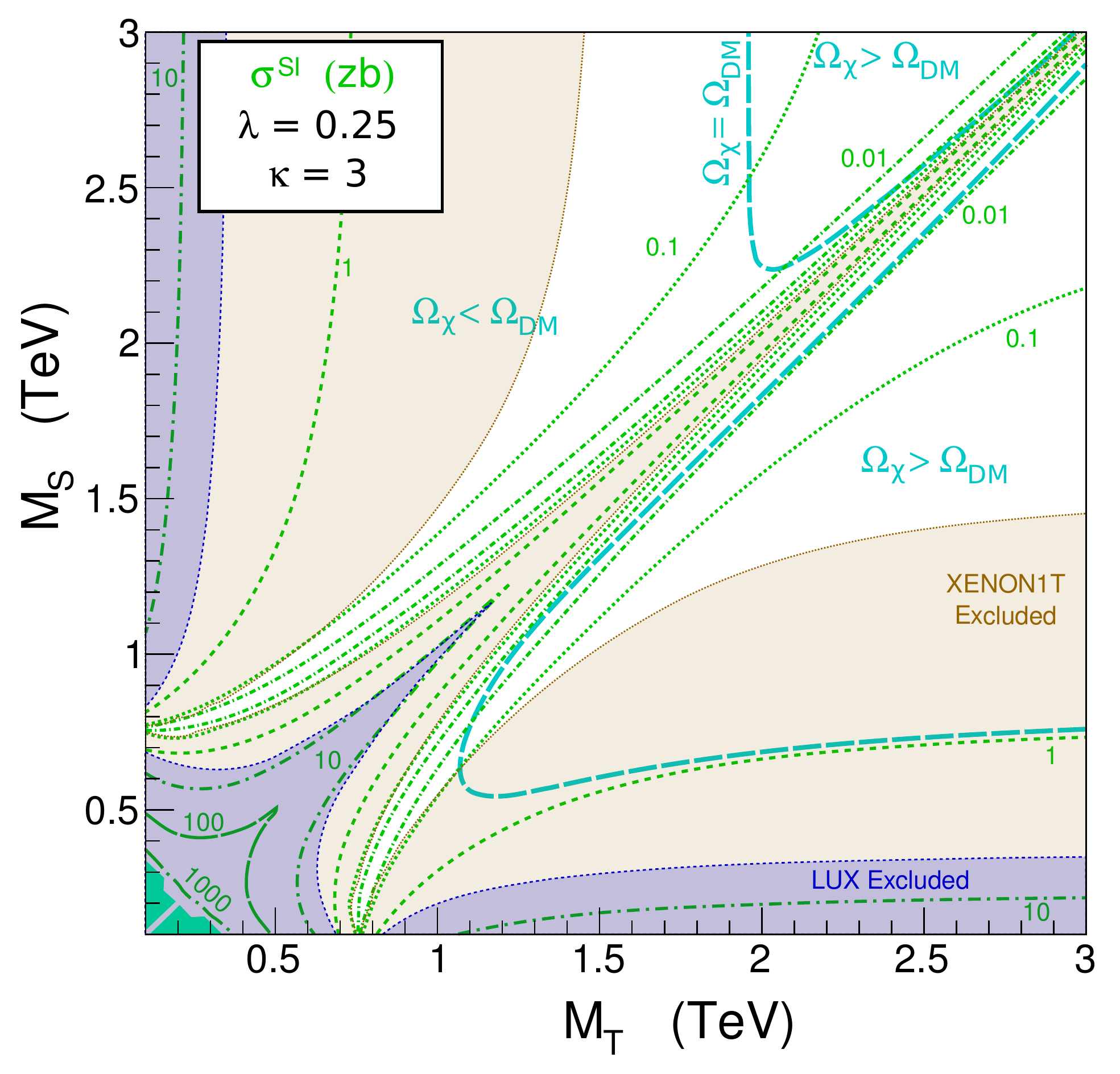}
  \label{fig:Plot7b_MSMT_yST=3_yPos}}
\subfigure[]{
  \includegraphics*[width=0.45\textwidth]{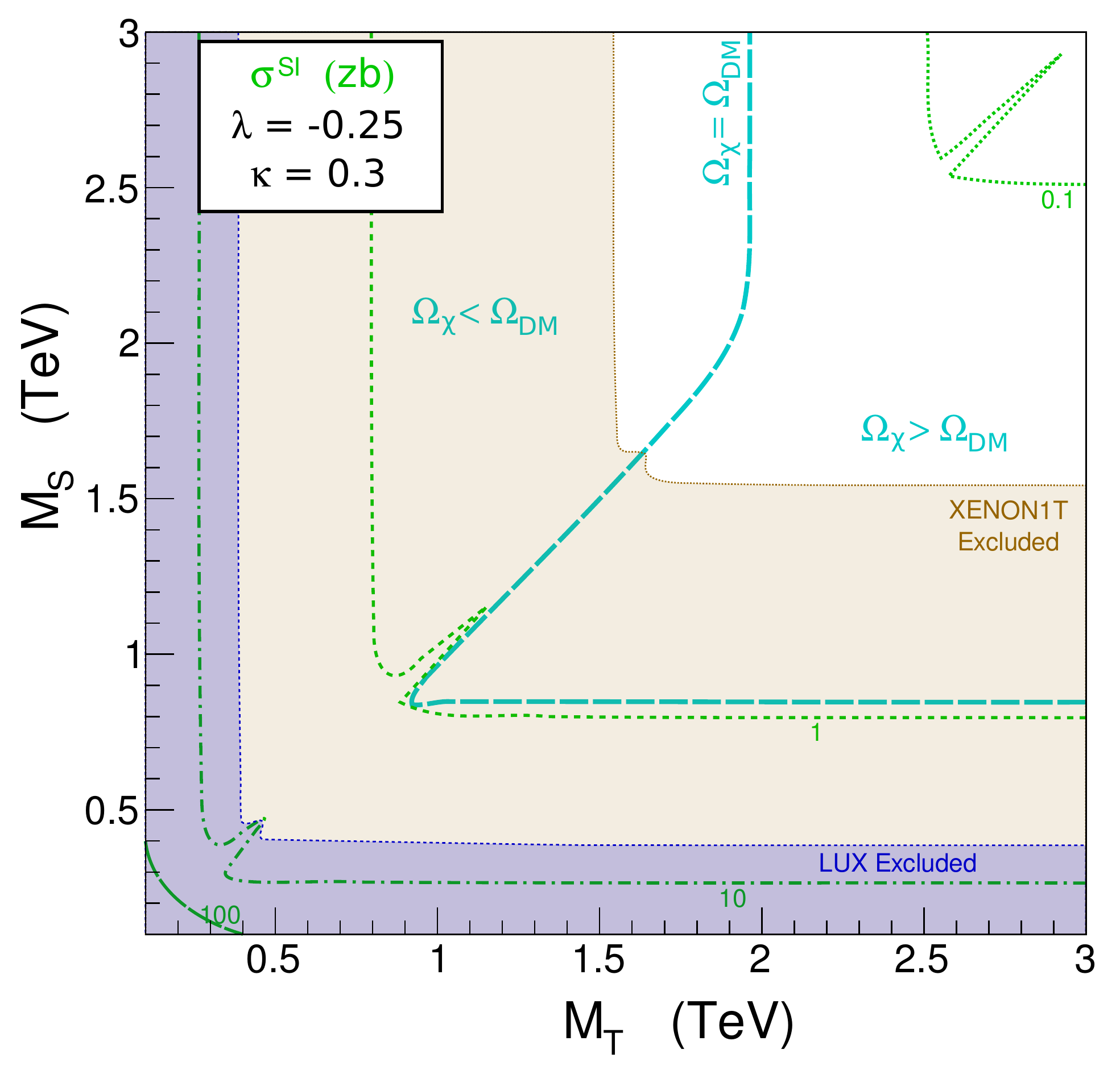}
  \label{fig:Plot7b_MSMT_yST=0.3_yNeg}}
\subfigure[]{
  \includegraphics*[width=0.45\textwidth]{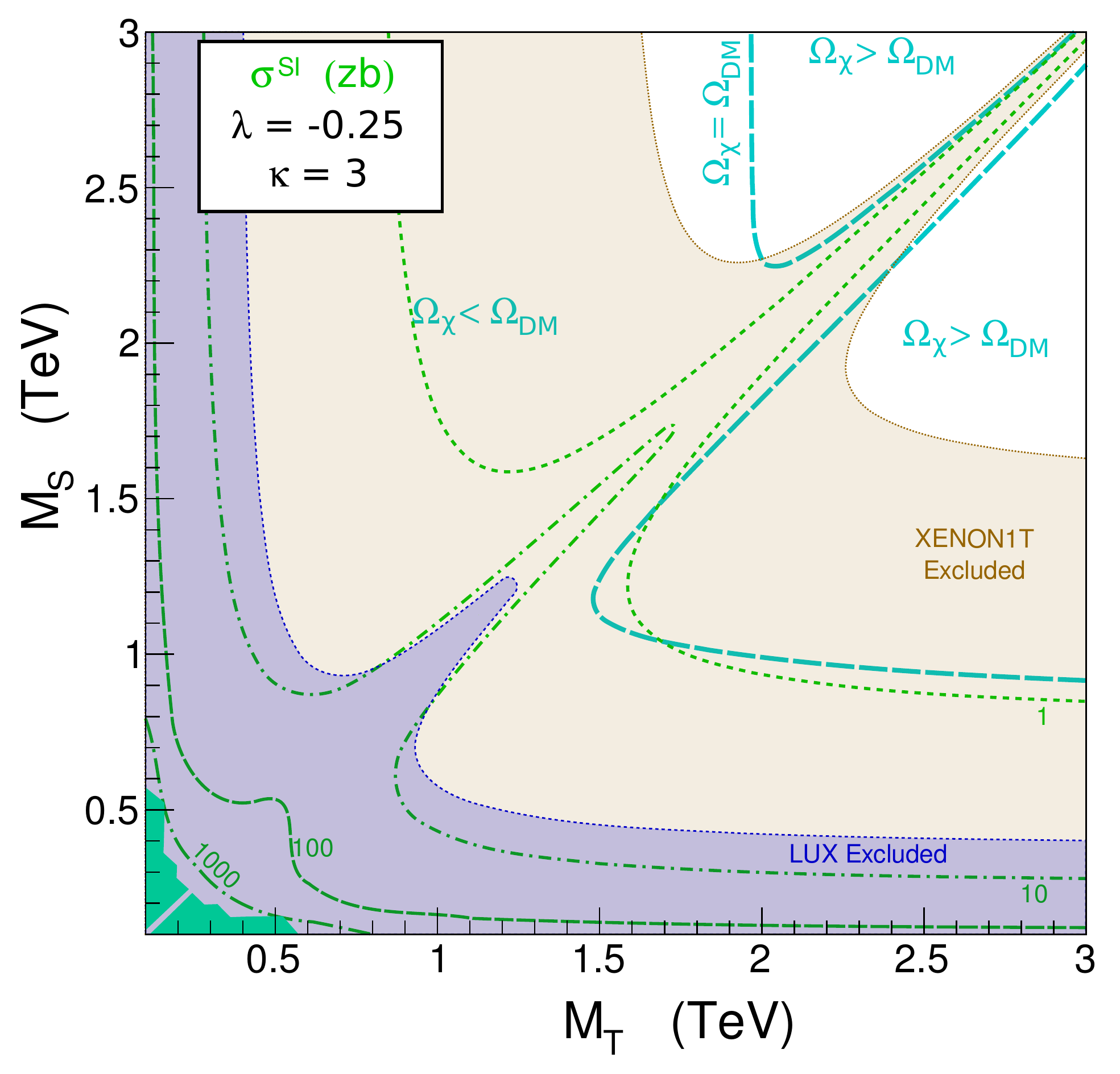}
  \label{fig:Plot7b_MSMT_yST=3_yNeg}}
\vspace*{-.1in}
\caption{\label{fig:Plot7b} \textit{Direct detection prospects for
    $\lambda \neq 0$ and $\kappa = 0.3, 3$.}  Contours shown are the
  same as in \figref{Plot1_y=0.3}.}
\end{figure}

The cases of $\lambda = \pm 0.25$ are shown in \figref{Plot7b}, which
is otherwise identical to the upper panels of \figref{Plot7}.  As can
be seen from \eqref{eq:ahchichiST}, a blind spot cancellation can
occur if $\lambda$ is positive, in which case there will be
destructive interference against Higgs exchange arising from the
$\kappa$ induced mixing.  This blind spot is visible in
\figref{Plot7_MSMT_yST=3} for points near (but not directly on) the
$M_S \approx M_T$ line.  Similarly to the behavior in
\figref{Plot4b_MSMD_A=100_yPos} for singlet-doublet scalars, the
DM-Higgs coupling is controlled by mixing induced by $\kappa$ near
$M_S \approx M_T$.  Away from this line, the DM-Higgs coupling is
controlled by the direct quartic DM-Higgs couplings proportional to
$\lambda$.  Partial cancellation is also present in
\figref{Plot7_MSMT_yST=0.3}, though $\kappa = 0.3$ is too small to
produce a true blind spot for $\lambda = 0.25$.  For $\lambda = -0.25$
the direct detection cross-sections is increased relative to $\lambda
= 0$, with a corresponding LUX constraint of $M_S, M_T \gtrsim 1~\tev$
for even moderate mixing.  Meanwhile, XENON1T will exclude $M_S, M_T
\lesssim 1.6~\tev$, with even greater sensitivity for $M_S \approx
M_T$.

\subsection{Exclusion Plots (Thermal Relic)}

\begin{figure}[tb]
\subfigure[]{
  \includegraphics*[width=0.31\textwidth]{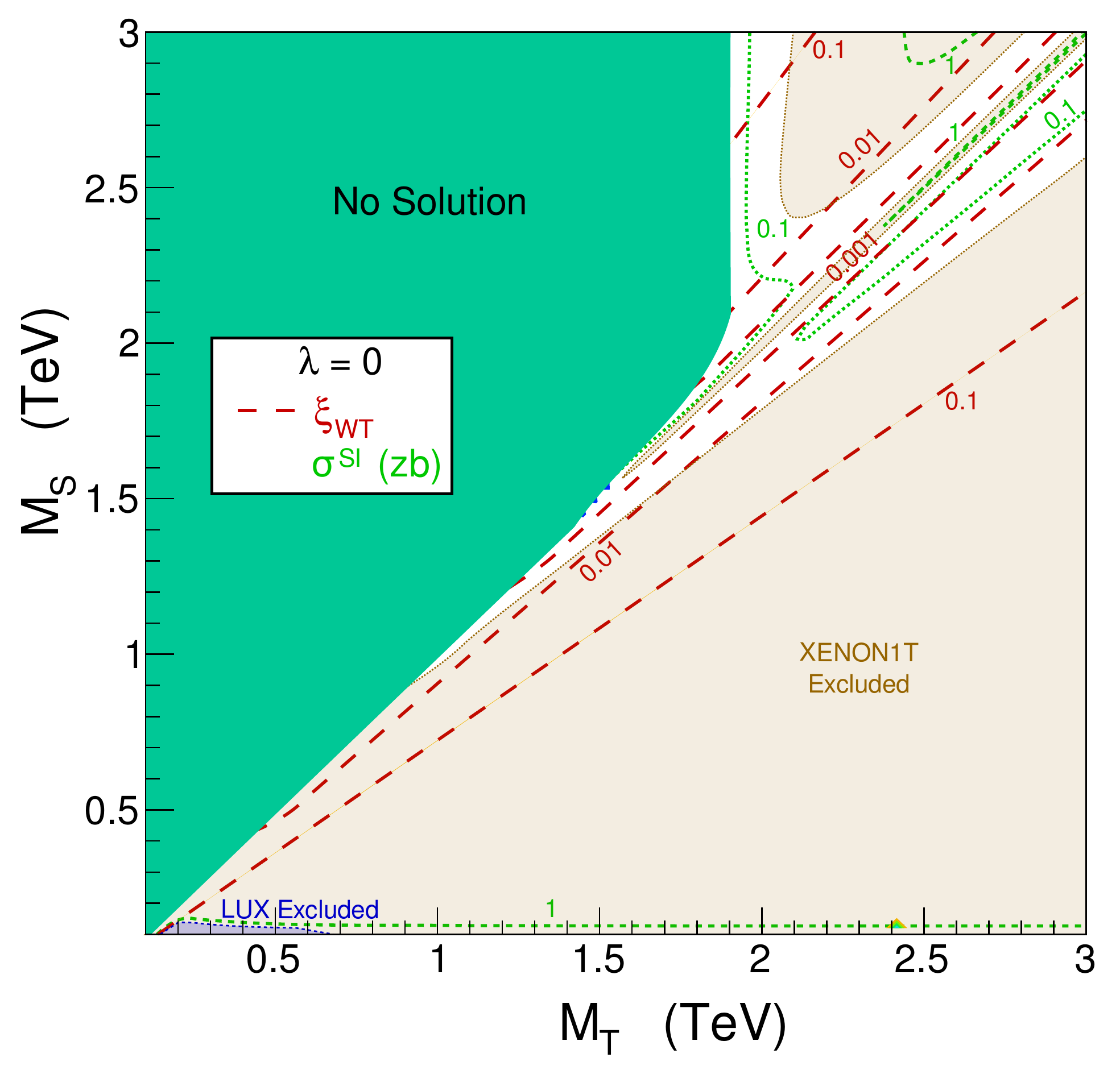}
  \label{fig:Plot8a_MSMD_ZeroCouplings}}
\subfigure[]{
  \includegraphics*[width=0.31\textwidth]{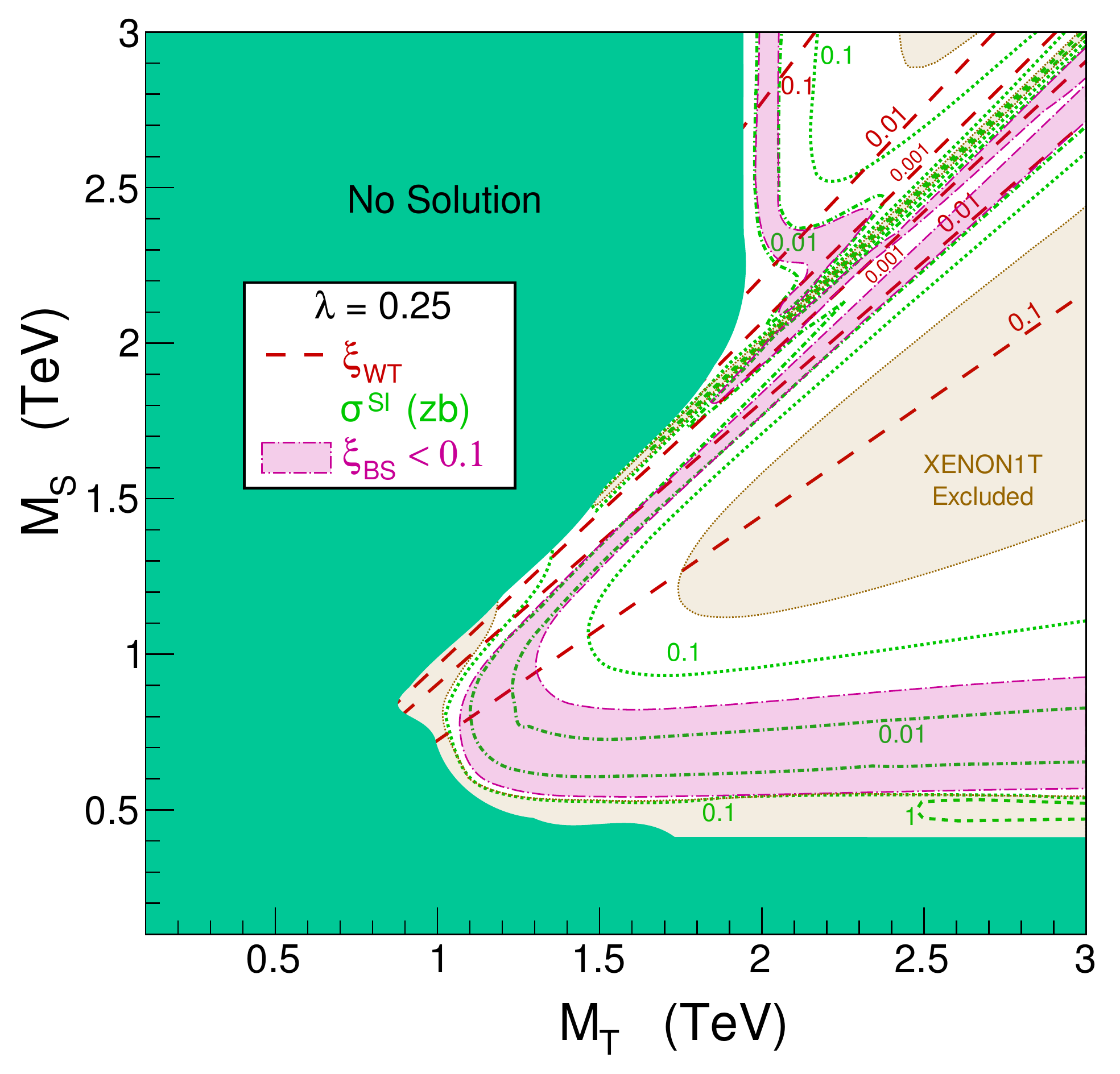}
  \label{fig:Plot8b_MSMD_yPos}}
\subfigure[]{
  \includegraphics*[width=0.31\textwidth]{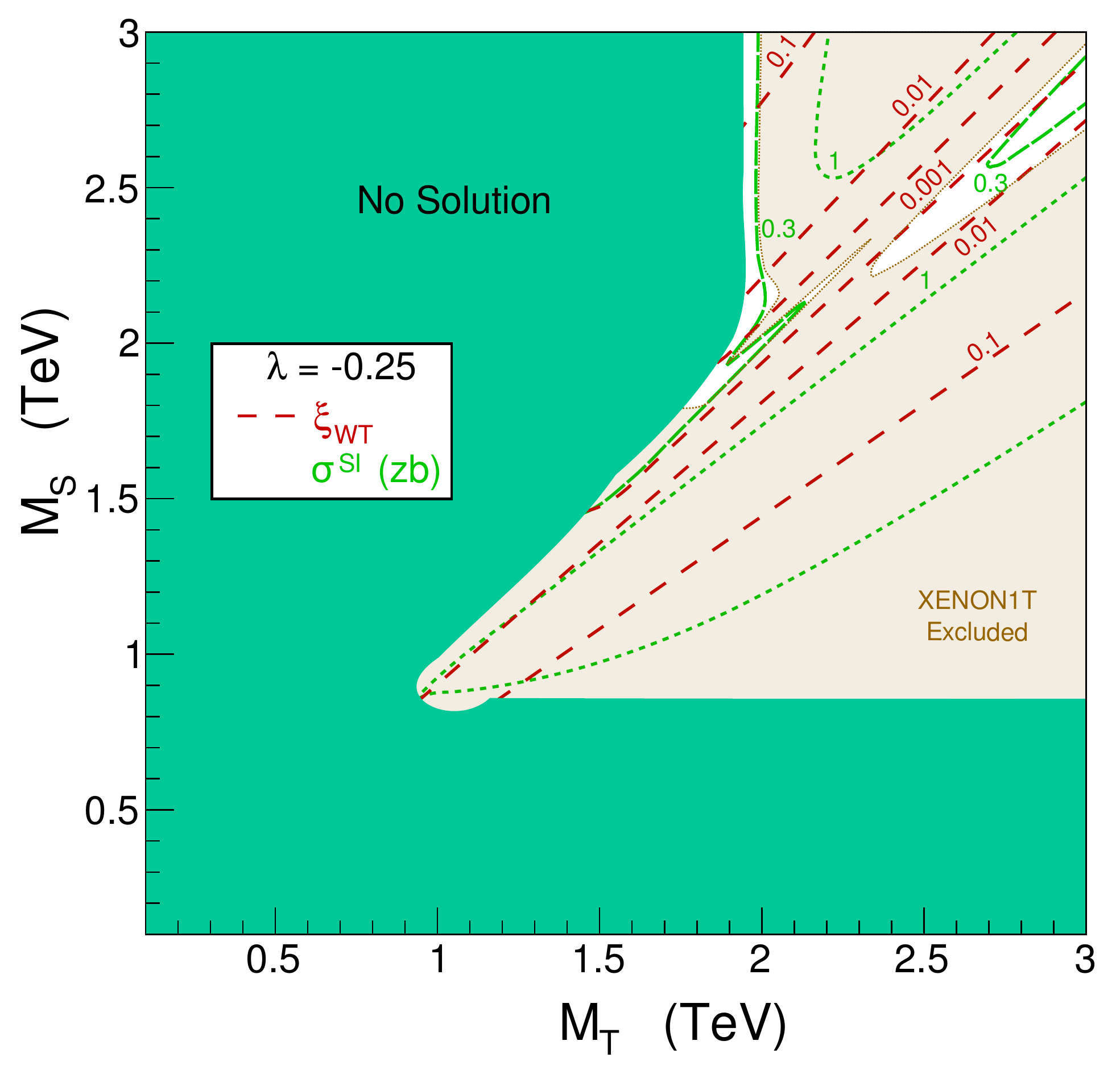}
  \label{fig:Plot8b_MSMD_yNeg}}
\vspace*{-.1in}
\caption{\label{fig:Plot8} \textit{Well-tempering and direct detection
    prospects for fixed $\Omegachi=\Omega_{\rm DM}$ as a function of $M_T$
    and $M_S$.} Contours shown are the same as in \figref{Plot2a}.}
\end{figure}

For the singlet-triplet scalar model, we set $\eqomega$ by varying
$\kappa$ in the $(M_T, M_S)$ plane in order to produce the observed
relic density.  \figref{Plot8} depicts the viable parameter space of
thermal relic singlet-triplet DM subject to current constraints from
LUX and projected reach from XENON1T.  For $\lambda = 0$, there is no
viable thermal relic for $M_T < M_S$ and $M_T \lesssim 1.9~\tev$,
since $\Omegachi < \OmegaDM$ for all such models.  For $M_T > M_S$ and
$1.9~\tev \lesssim M_T < M_S$, $\eqomega$ can be achieved throughout
for sufficiently large $\kappa$, with $M_T \approx 1.9~\tev, M_S \gg
M_S$ corresponding to pure triplet DM with the correct thermal relic
density~\cite{Cirelli:2005uq,Hambye:2009pw}.  Moreover,
$\sigmaSI$ varies by just over an order of magnitude over most of the
viable range shown.  This occurs because the dominant annihilation
channel, $\chi \chi \rightarrow h h$, scales with mixing terms in the
same way as the direct detection cross-section when $\kappa$ is
sufficiently large.  Hence, the direct detection and relic abundance
are correlated.  For $M_S \approx M_T$ annihilation is enhanced for
identical values of $\kappa$ reducing $\sigmaSI$ at large masses by
roughly an order of magnitude.  Because an equivalent value of $\ahxx$
implies stronger annihilation than in the singlet-doublet scalar case,
the current LUX bounds only constrain a few points at small mass.
However, because $\sigmaSI$ is only weakly dependent on DM mass,
XENON1T projected bounds constrain the majority of the parameter space
shown.  The only blind spot occurs along the edge of the ``no
solution'' region for $M_T \approx 1.9~\tev$, where the DM is nearly
pure triplet.

Unlike for the singlet-doublet scalar case, for $\lambda = 0.25$ in
\figref{Plot8b_MSMD_yPos} the ``no solution'' region extends into the
low mass $M_S < M_T$ region.  For $M_S \lesssim 400~\gev$, the direct
singlet-Higgs quartic coupling results in $\Omegachi < \OmegaDM$, and
increasing $\kappa$ sufficiently to cancel the direct coupling
contributions induces sufficient annihilation in other channels that
$\chi$ is remains under-abundant, resulting in no viable solution for
any value of $\kappa$. The allowed region still extends to lower $M_S$
than the $\eqomega$ contour in \figref{Plot7b_MSMT_yST=0.3_yPos},
however, and a blind spot occurs at $M_S \approx 650~\gev$ consistent
with this cancellation.  This blind spot extends to large masses below
and roughly parallel to the $M_S = M_T$ line.  A second blind spot
lies above this line and extends to large $M_S$ for $M_T\approx
2~\tev$.  The blind spot region is larger here in comparison to the
singlet-doublet scalar case because the value of $\ahxx$ required to
accommodate the observed relic density is significantly smaller.  In
the singlet-doublet case, gauge interactions are insufficient to set
$\eqomega$ for any mixing angle for $\mchi \gtrsim 550~\gev$, and the
induced DM-Higgs quartic coupling is relatively small.  For
singlet-triplet scalar DM, however, gauge interactions are strong
throughout the entire range shown, and the DM-Higgs quartic coupling
is sufficiently large that small values of $\chxx$ are viable and even
preferred.  LUX has no constraining power for $\lambda = 0.25$, and
XENON1T sensitivity has only moderate coverage of the parameter space.

For $\lambda = -0.25$ in \figref{Plot8b_MSMD_yNeg}, the ``no
solution'' region covers $M_S \lesssim 850~\gev$ for $M_S < M_T$.  No
blind spot regions exist for $\lambda = -0.25$, but the relative
strength of the gauge and four point interactions produces regions
with small $\sigmaSI$ along the upper portion of the ``no solution''
boundary and along the high mass $M_S \approx M_T$ line.  LUX has no
constraining power in the plane, and the regions with small $\sigmaSI$
avoid even XENON1T reach despite the lack of true blind spot behavior.
For both $\lambda = 0$ and $\lambda = -0.25$ well-tempering of at
least $\xiwt \lesssim 0.1$ is required to avoid XENON1T projected
bounds except for nearly pure triple DM, although most of the viable
regions have well-tempering $\xiwt \lesssim 0.01$.

As in the singlet-doublet scalar case, for $\lambda \le 0$ the blind
spot fine-tuning is trivial, $\xibs = 1$, and $\xibs$ is only
physically meaningful when $\lambda > 0$.  For $\lambda > 0$, however,
the degree of fine-tuning needed to produce cancellation is
significantly smaller than in the singlet-doublet scalar case.  The
pink region in \figref{Plot8b_MSMD_yPos} has tuning of $\xibs \lesssim
0.1$ and covers approximately half of the area left viable after
XENON1T.  However, the remaining unshaded area avoids XENON1T
sensitivity with minimal tuning, $\xibs \gtrsim 0.1$.  The lower
portion of this allowed area also has minimal tuning from
well-tempering, $\xiwt > 0.1$.  Hence, for $\lambda >0$,
singlet-triplet DM can accommodate viable thermal relic DM with
minimal tuning.

\begin{figure}[tb]
  \includegraphics*[width=0.48\textwidth]{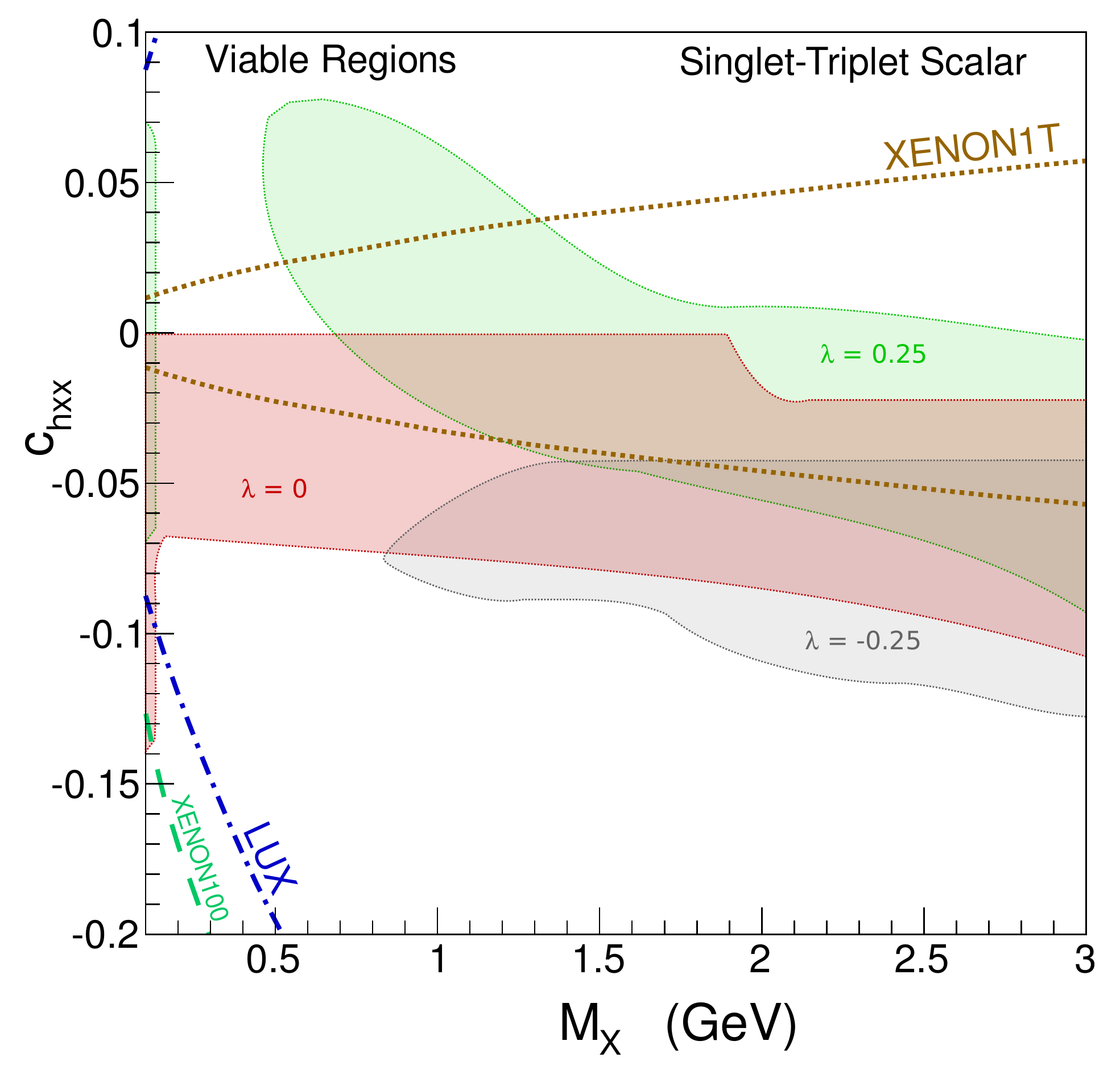}
\vspace*{-.1in}
\caption{\label{fig:PlotViableST} \textit{Viable regions with
    $\Omegachi=\Omega_{\rm DM}$ as a function of $\mchi$ and
    $c_{h\chi\chi}$.}  Contours and regions shown are the same as in
  \figref{Plot6}.}
\end{figure}

The viable parameter space of singlet-triplet DM is depicted in the
physical $(\mchi, \chxx )$ plane in \figref{PlotViableST}.  For
$\lambda = 0$, the behavior is similar to the singlet-doublet case,
except the region is ``stretched'' horizontally and ``squeezed''
vertically -- $\chxx$ can be close to zero up to $\mchi \lesssim
1.9~\tev$, and the increase of $|\chxx|$ along the lower boundary is
more gradual.  A small region extending to $\chxx \approx -0.14$ is
allowed for very low mass, where $\mchi$ is below the Higgs production
threshold and thus $\kappa$ can be significantly larger.  For $\lambda
= 0.25$, most of the viable region is restricted to $\mchi \gtrsim
500~\gev$, while for $\lambda = -0.25$, $\mchi \gtrsim 700~\gev$ is
required.  For $\mchi \gtrsim 500~\gev$ the behavior for $\lambda =
\pm 0.25$ is similar to the singlet-doublet scalar case.  For all of
the choices $\lambda = 0, \pm 0.25$, there are parameter regions which
are beyond the projected reach of XENON1T.  In contrast to the
singlet-doublet scalar case, the sensitivity of direct detection
experiments weakens as $\mchi$ increases.

\section{Conclusions and Future Directions}
\label{sec:conclusion}

Simplified models are a powerful tool for studying the generic
behavior of WIMP DM.  Theories in which DM couples to the SM via the
Higgs are of particular interest because Higgs-mediated DM-nucleon
scattering is just now being probed by the current generation of
direct detection experiments.  In this paper we have constructed and
analyzed simplified models of mixed DM describing a stable particle
composed of a mixture of a singlet and an electroweak doublet or
triplet.  In these models DM undergoes thermal freeze-out through
electroweak interactions to accommodate the observed DM relic
abundance.  Mixing between the singlet and non-singlet states is
induced via DM-Higgs couplings, and is in general correlated with
signals in direct detection.

We have determined the viable parameter space of these models
subject to current LUX limits and the projected reach of XENON1T.
Present experimental constraints from LUX place stringent limits on
mixed DM models, with a DM mass of at least a few hundred GeV in most
cases.  The projected reach of XENON1T is significantly stronger,
extending to masses of at least 1~TeV, and in many cases larger.
Using simplified models of mixed DM, we have identified direct
detection blind spots, which are parameter regions at which $\sigmaSI$
vanishes identically, nullifying experimental limits on spin
independent DM-nucleon scattering.  Finally, we have quantified the
degree of fine-tuning required for mixing angles ($\xiwt$) and for
blind spot cancellations ($\xibs$) required for thermal relic DM which
is experimentally viable.  Our results for each of our simplified
models are summarized in the discussion below.

First, we studied singlet-doublet Majorana fermion DM, which is a
generalization of mixed bino-Higgsino DM in the MSSM or
singlino-Higgsino DM in the NMSSM.  In these models the observed
thermal relic density can be produced for $\mchi < M_D \lesssim
1~\tev$ through mixing with a small Higgs coupling, but requires a
significant degree of well-tempering.  It is also possible to achieve
$\eqomega$ with large Higgs couplings, particularly with $M_D \gtrsim
1~\tev$, but avoiding direct detection constraints in such cases
requires blind spot cancellations.  While most of the parameter space
that avoids the LUX bound requires little tuning, $\xiwt, \xibs >
0.1$, after XENON1T nearly all models either exhibit a significant
degree of blind spot tuning, $\xibs < 0.1$, or must have mixing angles
which are sensitively well-tempered to produce $\eqomega$ through
coannihilation with small Higgs couplings with $\xiwt < 0.1$.  Thus,
XENON1T strongly constrains the parameter space of singlet-doublet DM.

The constraints placed by direct detection on singlet-doublet scalar
models are also substantial.  In such models $\eqomega$ can be
achieved at any mass through mixing for $\mchi < M_D \lesssim
550~\gev$ with small DM-Higgs couplings, requiring significant
well-tempering as in the fermionic case.  For $\mchi \gtrsim
550~\gev$, however, annihilation through Higgs-mediated processes is
required to accommodate $\eqomega$.  Hence, these models require
larger Higgs couplings and are subject to stronger direct detection
bounds.  Current limits from LUX place strong bounds up to large
$\mchi$ and XENON1T constrains almost the entire parameter space
studied.  In the examined parameter space, the few allowed regions
needed significant coannihilation, with fine-tuning of $\xibs < 0.1$
to remain viable.  However, a more comprehensive study of the full
seven-dimensional parameter space might yield regions of lesser
fine-tuning consistent with XENON1T.

The behavior of singlet-triplet scalar DM is qualitatively similar to
that of singlet-doublet DM.  However, both gauge boson and Higgs
mediated annihilation processes are quantitatively stronger.  The
correct relic density can be achieved through mixing alone for $\mchi
< M_T \lesssim 1.9~\tev$ though such processes, though the well-mixed
region still requires significant well-tempering.  Moreover, both the
trilinear $h\chi\chi$ and quartic $hh\chi\chi$ interactions are
stronger than in the singlet-doublet scalar case, enhancing
annihilation even for smaller mixing angles and thus weakening direct
detection bounds.  LUX has little constraining power on the
singlet-triplet parameter space.  XENON1T constraints are strong for
vanishing or negative quartic couplings, allowing for minimal
well-tempering only for nearly pure triplet models with $1.9~\tev
\lesssim \mchi \lesssim 2.1~\tev$.  For positive quartic couplings,
however, a significant region where DM is dominantly singlet and has
minimal tuning of any sort, $\xiwt, \xibs > 0.1$, remains viable.

The present work has focused exclusively on experimental constraints
from spin independent direct detection.  However, many complementary
probes exist.  For example, even at cancellation points with vanishing
DM-Higgs coupling, there will generically be $Z$-mediated spin
dependent DM-nucleon scattering.  Future direct detection
probes~\cite{Ramberg:2010zz, Liu:2012uva, Bruch:2010eq} and both
current and future constraints from neutrino
telescopes~\cite{Aartsen:2012kia, IceCube:2011aj, Abbasi:2011eq} will
place significant constraints on many of these models.  Moreover,
these models will also be constrained by astrophysical probes such as
FERMI~\cite{Ackermann:2011wa} and HESS~\cite{Abramowski:2013ax}.  We
leave these analyses for future work.

\section*{Acknowledgments}

We are grateful to the Kavli Institute for Theoretical Physics at
Santa Barbara and the Aspen Center for Physics, where part of this
work was conducted.  DS is supported in part by U.S.~Department of
Energy grant DE--FG02--92ER40701 and by the Gordon and Betty Moore
Foundation through Grant No.~776 to the Caltech Moore Center for
Theoretical Cosmology and Physics.  CC is supported by a DOE Early
Career Award DE-SC0010255.

\phantomsection
\bibliography{bibmixeddm01}{}

\end{document}